\documentclass[superscriptaddress,preprint,10pt,prx,oneside]{revtex4-2}%
\usepackage{amsmath}
\usepackage{amsfonts}
\usepackage{amssymb}
\usepackage{graphicx}
\usepackage{mathrsfs}
\usepackage[caption=false]{subfig}
\graphicspath{{pics/}}
\usepackage[ruled, linesnumbered]{algorithm2e}
\usepackage{tikz}
\usepackage{comment}
\usepackage{array}
\usepackage{tabularx}
\usepackage{booktabs}
\newcolumntype{C}{>{\centering\arraybackslash}X}

\PassOptionsToPackage{dvipsnames}{xcolor}
\usepackage[shortpaper]{aps_math}
\usepackage{basic_phys}
\usepackage{basic_comment}
\usepackage{qm_my}
\newcommand{\bbcompo}{\blacksquare}
\renewcommand{\proofend}{\hfill$\square$}
\newcommand{\wjy}[1]{{#1}}
\newcommand{\wjycomment}[1]{}
\newcommand{\shhcomment}[1]{}

\usepackage{ifthen}

\def\bibfile{NQPD}
\ifthenelse{\equal{\bibfile}{}}{%
  \def\myprintbibliography{}%
}{%
  \def\myprintbibliography{%
    \bibliographystyle{apsrev4-2}%
    \bibliography{\bibfile}%
  }%
}%
\usepackage[%
]{hyperref}

\def\myprintglossary{%
}

\begin{document}
\title{Entanglement-assisted circuit knitting:\\ Distributed quantum computing using limited entanglement resources}
\author{Shao-Hua Hu}
\email{shhphy@gmail.com}
\affiliation{Department of Physics, National Tsing Hua University, Hsinchu 30013, Taiwan ROC}

\author{Po-Sung Liu}
\affiliation{Department of Physics, National Cheng Kung University, Tainan 701, Taiwan, ROC}

\author{Jun-Yi Wu}
\email{junyiwuphysics@gmail.com}
\affiliation{Department of Physics, Tamkang University, New Taipei 25137, Taiwan, ROC}
\affiliation{Hon Hai Research Institute, Taipei, Taiwan, ROC}
\affiliation{Physics Division, National Center for Theoretical Sciences, Taipei, Taiwan, ROC}

\begin{abstract}
Distributed quantum computing (DQC) provides a promising route toward scalable quantum computation, where entanglement-assisted LOCC and circuit knitting represent two complementary approaches. The former deterministically realizes nonlocal operations but demands extensive entanglement resources, whereas the latter requires no entanglement yet suffers from exponential sampling overhead. Here, we propose a hybrid framework called \emph{entanglement-assisted circuit knitting} that integrates these two paradigms by performing circuit knitting assisted with a limited amount of entanglement.

We establish a general theoretical framework for entanglement-assisted circuit knitting. Optimal sampling overhead is achieved for Choi-stretchable unitaries with general entanglement resources, while for general unitaries we derive both lower and upper bounds for one-Bell-pair-assisted circuit knitting.We further extend the framework to the black-box setting, which can be treated as a class of quantum combs. This extension releases the need for explicit knowledge of the global unitary of a whole quantum circuit, enables a more flexible embedding structure, and broadens its applicability.
Within this framework, we develop constructive protocols utilizing different resources, including entanglement, local operations, and classical communication. We derive the optimal mixed configuration among these protocols and provide an algorithm for its determination.

Under dynamically probabilistic entanglement distribution, we reveal a trade-off between sampling overhead and entanglement cost in entanglement-assisted circuit knitting. This hybrid approach can thus be viewed as a form of hybrid classical–quantum computation, balancing sampling and entanglement efficiency, and enabling more resource-efficient implementations of distributed quantum computing.
\end{abstract}
\clearpage

\maketitle
\newpage

\tableofcontents

\newpage
\section{Introduction}

Quantum computing offers immense potential to outperform classical computation, however, its realization on large-scale quantum devices remains a central challenge.
Distributed quantum computing (DQC)~\cite{Caleffi2024, Barral2025, Knoerzer2025} addresses this challenge by interconnecting multiple local quantum processing units (QPUs) to collectively implement a global unitary operation.
DQC implementations are generally classified into two approaches: entanglement-assisted local operation and classical communication (LOCC) \cite{Gottesman1999, Eisert2000, AndresMartinez2019, Wu2023, AndresMartinez2024} and circuit knitting \cite{Peng2020, Mitarai2020, Piveteau2022, Mitarai2021, Piveteau2024}.

Although entanglement-assisted LOCC approaches for DQC can deterministically implement a global unitary, they are highly resource-intensive in terms of the number of entangled pairs required.
Such fully entanglement-assisted DQC schemes can be categorized into two types, namely quantum state teleportation~\cite{Bennett1993} and quantum telegate~\cite{Eisert2000}, which serve as fundamental building blocks for constructing distributed quantum processes~\cite{AndresMartinez2019, Wu2023, AndresMartinez2024, Caleffi2024, Barral2025, Knoerzer2025}.

On the other hand, circuit knitting was proposed as a method to simulate a large quantum circuit using smaller subcircuits~\cite{Peng2020}. In contrast to entanglement-assisted approaches, circuit knitting requires no entanglement resources. Instead, it relies on a classical postprocessing technique known as \emph{quasi-probability decomposition (QPD)}~\cite{Pashayan2015, Mitarai2020, Mitarai2021}, which statistically reconstructs the outcomes of a quantum circuit rather than physically implementing the full global operation.
In this framework, the global operation is expressed as a quasi-probabilistic mixture of LOCC, with the original statistics recovered via assigning positive or negative weights to measurement outcomes in classical postprocessing. However, to maintain the same estimation accuracy, circuit knitting incurs a \emph{sampling overhead}, which increases the number of circuit executions and the total runtime. The central objective of circuit knitting is therefore to identify LOCC configurations that minimize this overhead.
Notably, classical communication is not required to achieve the optimal sampling overhead for certain classes of unitary operations~\cite{Ufrecht2023, Piveteau2024, Ufrecht2024, Schmitt2025, Harrow2025}.

Beyond the size of circuits, such a virtual simulation of global operation provides an additional advantage of mitigating gate noises in quantum circuit.
It offers a powerful framework for the \emph{virtual distillation} of quantum resources~\cite{Yuan2024, Takagi2024}, particularly for \emph{entanglement purification}~\cite{Yamamoto2024}.
Moreover, such approaches have been shown to enhance algorithms for \emph{clustered Hamiltonian simulation}~\cite{Harrow2025}.

These two paradigms, entanglement-assisted LOCC and circuit knitting, represent the two extreme regimes of DQC.
The entanglement-assisted approach consumes a large amount of high-fidelity entanglement distributed across the quantum network of QPUs but enables fast runtime.
In contrast, circuit knitting requires no entanglement, but incurs significantly longer runtime due to its sampling overhead, which increases exponentially with the number of nonlocal operations~\cite{Jing2025}.
It implies a fundamental trade-off between entanglement consumption and execution time, that bridges these two paradigms.

\begin{figure}[htb]
    \centering
    \includegraphics[width=0.7\linewidth]{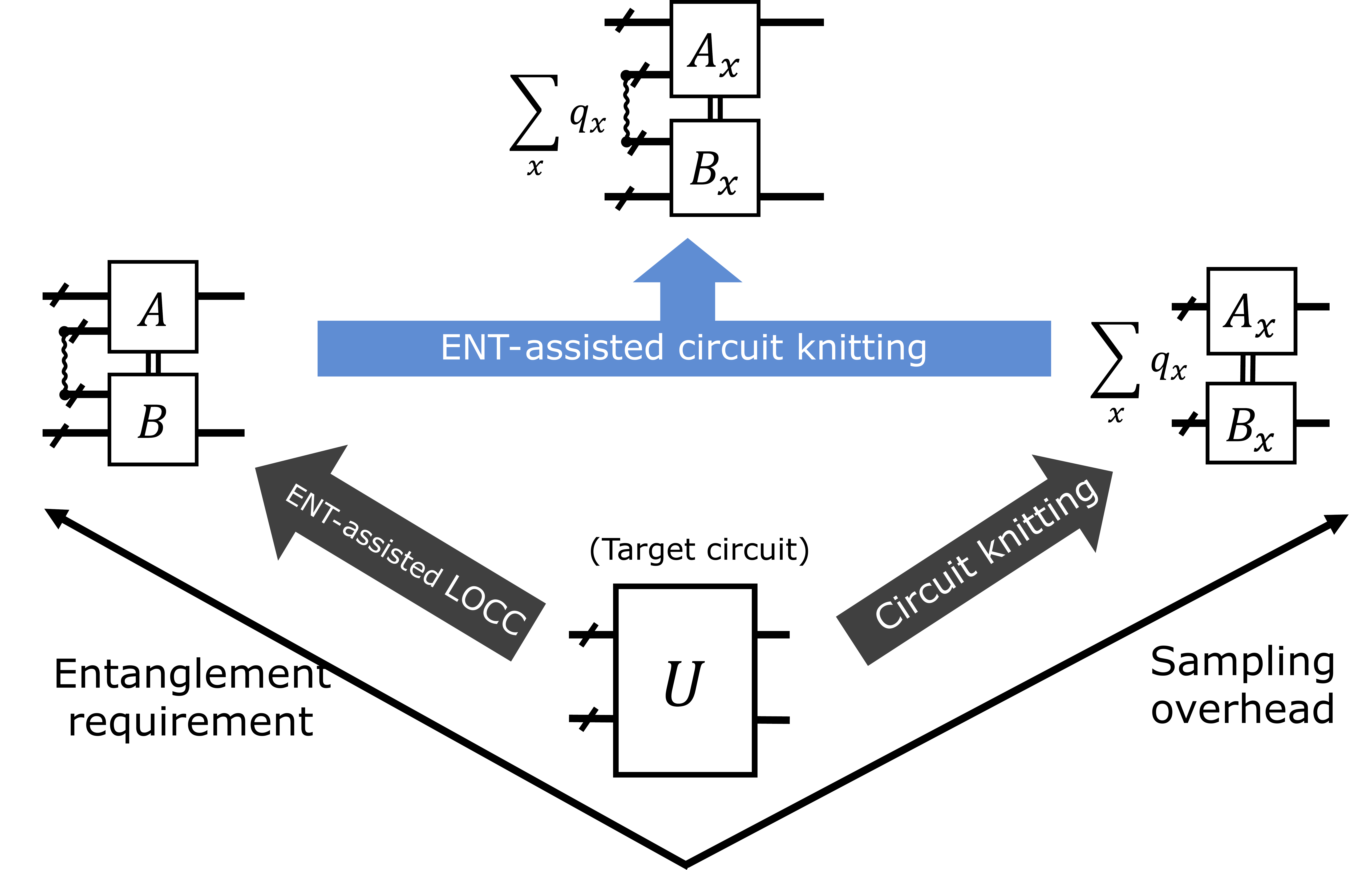}
    \caption{Schematic illustration of entanglement-assisted circuit knitting as a combination of entanglement-assisted LOCC and circuit knitting. Wiggly lines represent shared entanglement resources, while double lines denote classical communication. This framework achieves a balance between the amount of pre-shared entanglement and the resulting sampling overhead.}
    \label{fig_intro}
\end{figure}

In this work, we integrate these two approaches into a \emph{hybrid framework} for DQC. We consider a realistic scenario where local QPUs have access to only a limited number of entangled pairs, which are insufficient for fully entanglement-assisted DQC. Consequently, a sampling overhead must be incurred to simulate the target quantum operation via QPD with limited entanglement resources.
A straightforward approach would be to apply circuit knitting to distill a large amount of virtual entanglement~\cite{Yuan2024, Takagi2024, Yamamoto2024}, which are then be consumed by LOCC to implement DQC. Instead, we directly employ the available physical entanglement to assist circuit knitting. In this approach, the pre-shared physical entanglement is consumed within LOCC as a resource, leading to a unified framework of \emph{entanglement-assisted circuit knitting} as illustrated in Fig.~\ref{fig_intro}, which bridges entanglement-assisted LOCC and circuit knitting.

Entanglement-assisted circuit knitting can be understood as a form of classical–quantum hybrid computation, in which a finite amount of entanglement is introduced to significantly reduce the sampling overhead to a practical level, thereby preserving quantum advantage. We establish a theoretical framework to derive upper and lower bounds on the sampling overhead under limited entanglement resources and identify the optimal overhead under certain conditions. The resulting trade-off between sampling overhead and entanglement provides physical insight into the origin and persistence of quantum advantage in distributed quantum computing.

We further extend the framework to a powerful black-box setting, formulated as an instance of a quantum comb. This extension provides a well-defined embedding structure, enables efficient determination of QPD configurations, and broadens applicability to quantum circuit compilation. Specifically, in this setting, we derive an optimal protocol for one-Bell-pair-assisted circuit knitting under dynamically probabilistic entanglement distribution. Our results reveal a trade-off between sampling overhead and entanglement consumption, which enables efficient utility of probabilistic entanglement distribution in DQC. It therefore facilitates practical implementations of entanglement-assisted distributed quantum computing in realistic quantum networks.

This paper is organized as follows.
Section~\ref{sec::r-assisted_Q_op} establishes the theoretical framework of resource-assisted quasi-probability decomposition over free operations.
In Sec.~\ref{sec::ent-assisted_circknit}, we apply this framework to circuit knitting, taking entanglement as the assisting resource and characterizing the resulting overhead. Section~\ref{sec::ent-assist_bb_circknit} extends the framework to the black-box setting for a broader application with unknown interleaving quantum channels.
Section~\ref{sec::tradeoff_gamma_ENT} derives a protocol for one-Bell-pair-assisted circuit knitting under dynamically probabilistic entanglement distribution and demonstrates the trade-off between entanglement cost and sampling overhead.
Finally, Sec.~\ref{sec::conclusion} concludes our results.

\section{Resource-assisted quasi-probability decomposition}
\label{sec::r-assisted_Q_op}
It is convenient to employ the vectorization formulation \cite{D’Ariano2016} in the Liouville space to describe density operators and their evolution under quantum channels.
Here, we adopt the double ket notation to represent the vectorization of an operator. The vectorization of the identity operator is given by $\kett{I_d} := \sum_{i=0}^{d-1}\ket{i,i}$. Accordingly, we denote the vectorization of a general operator $\hat{O}$ by $\kett{O} := (\hat{O}\otimes \hat{I}_d)\kett{I_d}$. The density operator after the vectorization is then given by
\begin{equation}
  \kett{\rho} = (\hat{\rho}\otimes \hat{I}_d)\kett{I_d}.
\end{equation}

The unitary transformation of a vectorized state is denoted by the tilde mark and written as
\begin{equation}
\label{eq::tilde_U_def}
  \widetilde{U}:=\hat{U}\otimes\hat{U}^{\ast}.
\end{equation}
The operator sum representation of a quantum channel $\widetilde{Q}$ in this formalism can be then conveniently expressed as
\begin{equation}
\label{eq::op_sum}
  \widetilde{Q} = \sum_{i}\widetilde{K}_{i},
\end{equation}
where $\widehat{K}_{i}$ are the Kraus operators of $Q$.
Note that the Choi state $\widehat{J}_{Q}$ is the conventional representation of a quantum channel $\widetilde{Q}$. For a $d$-dimensional unitary channel $U$, the corresponding Choi state is
\begin{equation}
\label{eq::Choi_St_def}
  \widehat{J}_{Q} = (Q\otimes\widehat{\id})[\projector{\Phi_{d}}]
  \;\;
  \text{ with }
  \;\;
  \ket{\Phi_{d}} = \frac{1}{\sqrt{d}}\sum_{i=0}^{d-1}\ket{i,i}.
\end{equation}
The vectorization of the Choi state is then given by
\begin{equation}
  \kett{J_{Q}} := (\widehat{J}_{Q}\otimes\id) \kett{I_{d^{2}}} = \widetilde{Q\otimes\id}\kett{\Phi_{d}}.
\end{equation}
In this paper, both the tilde notation $\widetilde{Q}$ and the vectorized Choi-state representation $\kett{J_{Q}}$ will be used in different contexts.

\bigskip

\begin{figure*}[htb]
    \centering
    \hfill
    \subfloat[]{\includegraphics[width=0.4\textwidth]{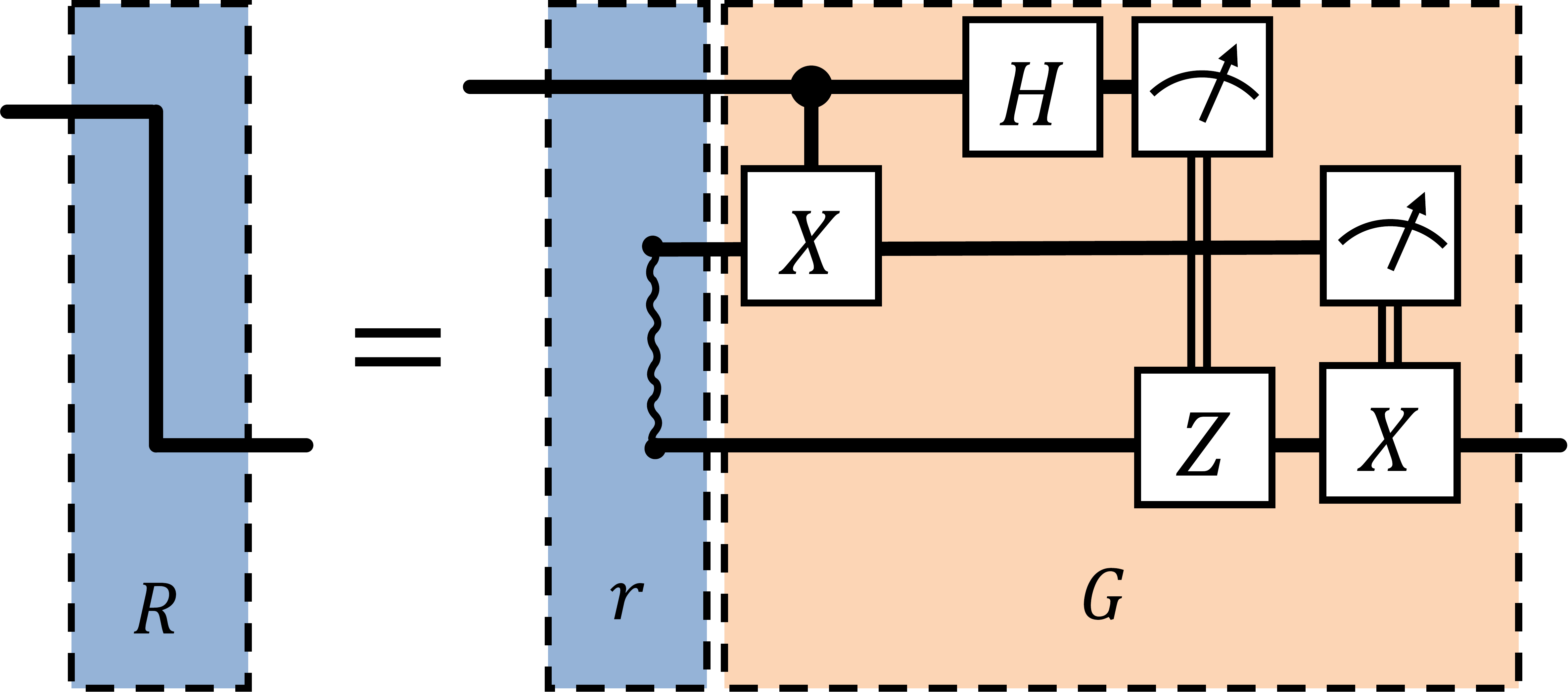}}
    \hfill
    \subfloat[]{\includegraphics[width=0.4\textwidth]{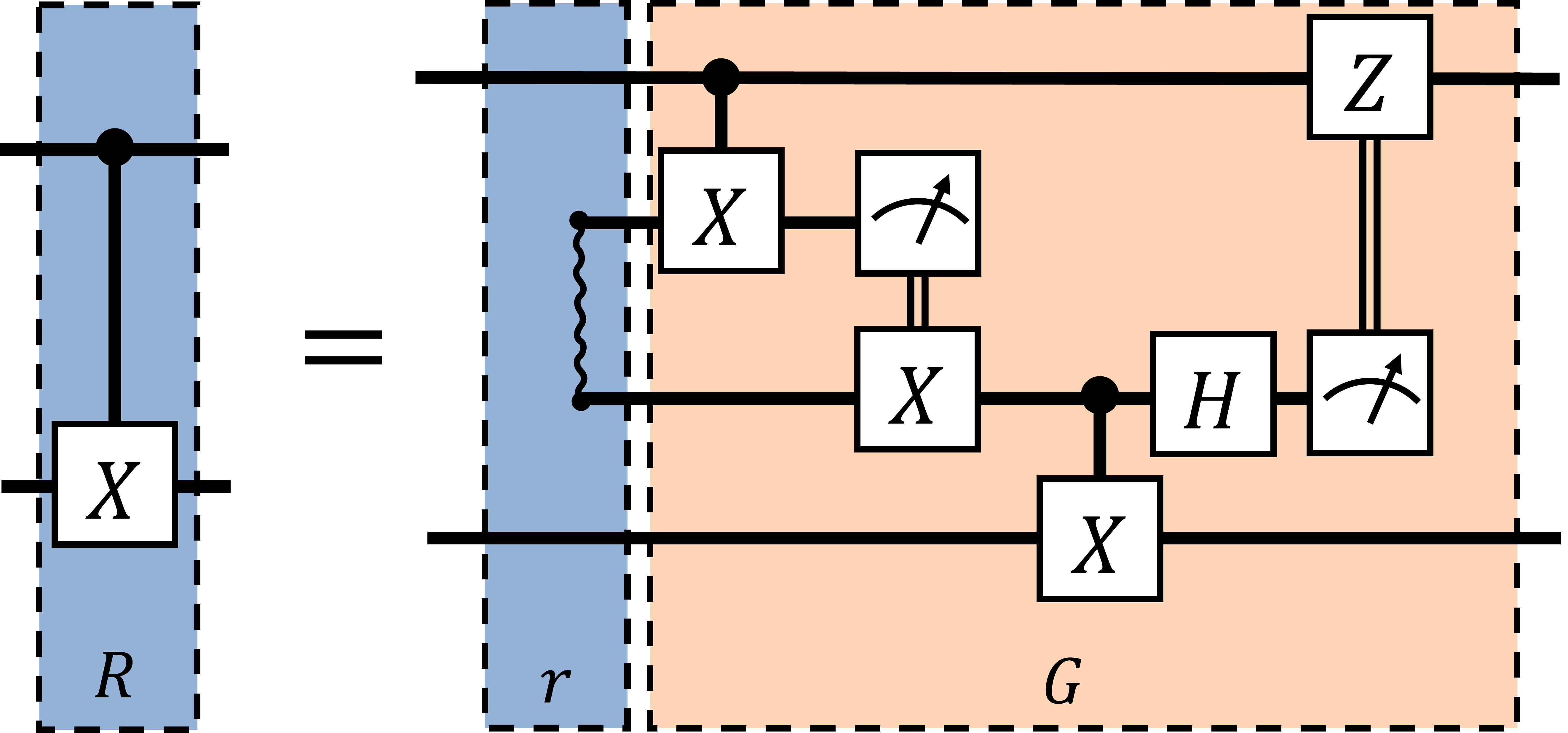}}
    \hfill{ }
    \caption{Diagram of (a) quantum state teleportation and (b) telegate. The blue one indicates the resource in the process, where the orange one indicates the free operation.}
    \label{fig::DQC-diagram}
\end{figure*}

With an ancillary subspace, one can construct an instrument $\{\widetilde{K}'_{m}\}_{m}$ through engineering the quantum operation $\widetilde{Q}$ with an input $\kett{\rho_{anc}}$ and a positive-operator-valued-measure (POVM) measurement $\{\bbra{M_{m}^{(anc)}}\}_{m}$ on the ancillary subspace
\begin{equation}
\label{eq::q_instr_def}
  \widetilde{K}'_{m} =
  \bbra{M_{m}^{(\text{anc.})}}\widetilde{Q}\kett{\rho_{\text{anc.}}}.
\end{equation}
Such a construction of state-assisted quantum instruments has been employed in entanglement-assisted distributed quantum computing \cite{Eisert2000, Wu2023, AndresMartinez2024}, in which maximally entangled states are employed as the ancillary.
Examples of such protocol are shown in Fig. \ref{fig::DQC-diagram}.
We can describe the state preparation of $\kett{\rho}$ as a quantum operation mapping classical information to the ancillary Hilbert space,
\begin{equation}
  \widetilde{r}_{\rho}^{(\text{anc.})} := \kett{\rho_{\text{anc.}}}.
\end{equation}
In general, one can implement a pre-operation $\widetilde{P}$ before the state preparation. As a whole, one can construct a quantum instrument through the following composition of quantum operations
\begin{equation}
\label{eq::q_instr_st_ast_def}
  \widetilde{K}'_{m}
  =
  \bbra{M_{m}^{(\text{anc.})}}\widetilde{Q}\circ\widetilde{r}_{\rho}^{(\text{anc.})}\circ\widetilde{P}.
\end{equation}
In a more general framework, the operation $\widetilde{r}$ is not restricted to the initial preparation of ancillary states. Instead, it may represent any quantum operation implemented with the assistance of a suitable quantum-state preparation.

\subsection{Resource-free quasi-probability decomposition over free operations}

\begin{figure}[htb]
  \centering
  \hfill
  \subfloat[]{\includegraphics[width=0.45\textwidth]{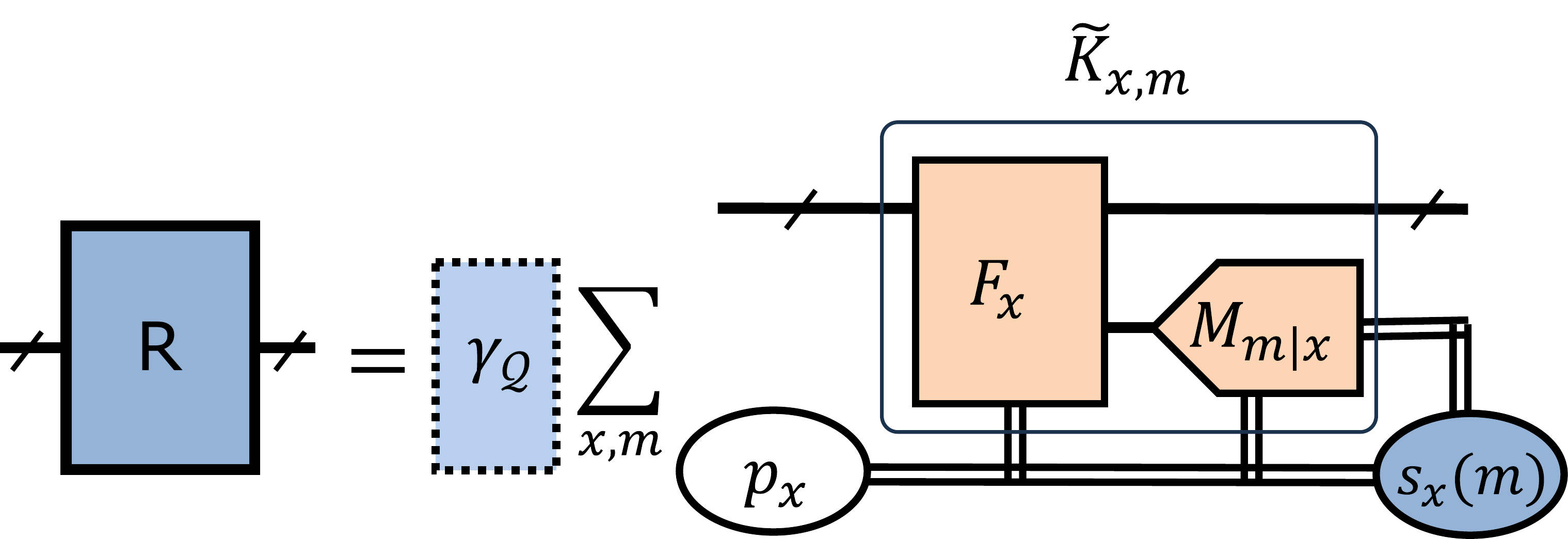}}
  \hfill
  \subfloat[]{\includegraphics[width=0.48\textwidth]{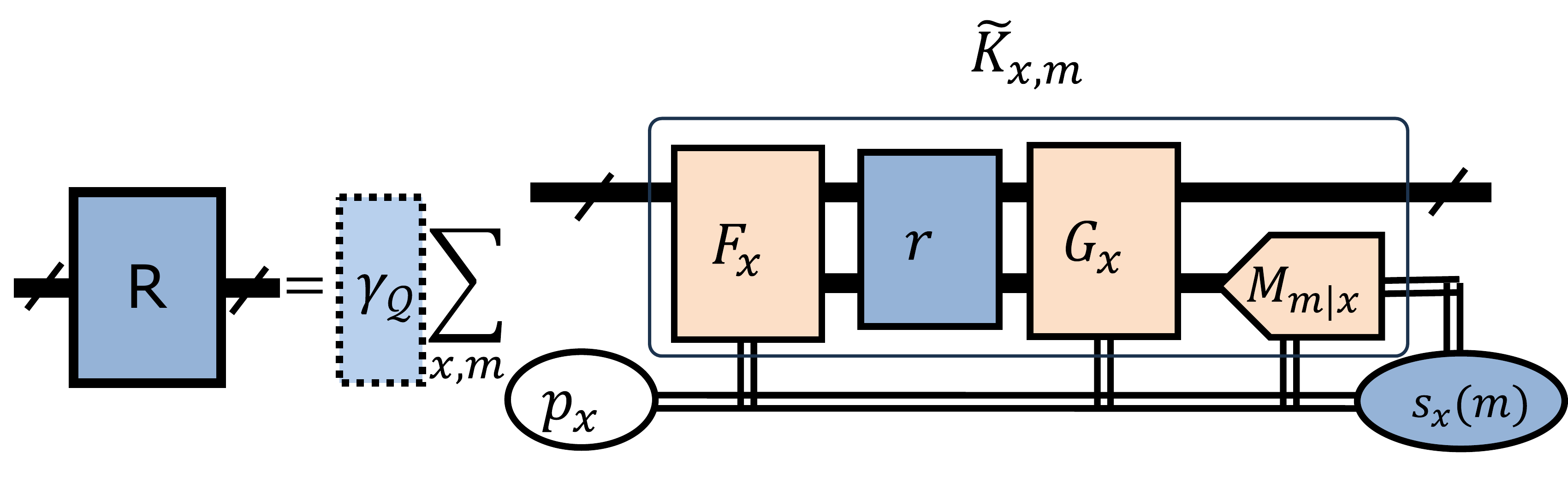}}
  \hfill
  \caption{(a) Resource-free QPD (b) Resource-assisted QPD}
  \label{fig::QPD_types}
\end{figure}

\wjy{
Let $\mathbb{F}$ be a set of free operations in the framework of quantum resource theory \cite{Chitambar2019} that satisfies the following three conditions. (1) It contains the identity map, $\widetilde{\id} \in \mathbb{F}$. (2) It is closed under composition, i.e. $(\widetilde{E}_a\circ \widetilde{E}_b)\in \mathbb{F}$ if $\widetilde{E}_{a}$ and $\widetilde{E}_{b}$ are both free. (3) it is closed under the convex combination, i.e. $((1-p)\widetilde{E}_a + p\widetilde{E}_b)\in \mathbb{F}$ if $\widetilde{E}_{a}$ and $\widetilde{E}_{b}$ are both free and $p \in [0,1]$.
In many tasks of quantum information processing, quantum advantage is achievable only via resourceful operations that lie outside $\mathbb{F}$. However, these resourceful operations can be simulated by quasi-probability decomposition (QPD) using only the free operations in $\mathbb{F}$, paying classical sampling overhead.
}

Let $\mathbb{X}$ denote a set of labels $x$ associated with the available free operations $\widetilde{F}_{x}\in\mathbb{F}$.
A quasi-probability decomposition (QPD) over these free maps shown in Fig. \ref{fig::QPD_types} (a) is constructed as follows.
One samples $x\in\mathbb{X}$ according to a probability distribution $\{p_x\}_{x}$ and implements the corresponding free operation $\widetilde{F}_{x}$.
Conditioned on the classical label $x$, a free POVM measurement $\mathcal{M}_{x} = \{\bbra{M_{m|{x}}^{(\text{anc.})}}\}_{m}$ is performed on an ancillary subspace.
\wjy{Here, we say a POVM measurement $\mathcal{M}_{x}$ is free, if it cannot generate a resourceful state stochastically from free states, i.e.
\begin{equation}
\label{eq::free_POVM}
  \bbrakett{M_{m}^{(\text{anc.})}|\rho} \in \mathbb{F}, \;\;
  \text{for all } m \text{ and } \kett{\rho} \in \mathbb{F}.
\end{equation}
After the free POVM measurement $\mathcal{M}_{x}$, a binary sign function $s_x(m)=\pm1$ is then assigned to each POVM outcome $m$ in the classical post-processing stage.}
Altogether, a QPD over the free maps $\mathbb{F}$ can be formulated as follows.
\begin{definition}[Resource-free QPD over free maps]
Let $\mathbb{X} = \{x\}$ be a set of classical labels associated with a probability distribution $\{p_{x}\}_{x}$. For each label $x$, there are a free operation $\widetilde{F}_x \in \mathbb{F}$ and a $\mathbb{F}$-free POVM $\mathcal{M}_{x}=\{\bbra{M_{m|x}^{(\text{anc.})}}\}_{m}$ available.
The tuples $\mathcal{Q}=\{(p_{x}, \widetilde{F}_{x}; s_{x}, \mathcal{M}_{x})\}_{x\in\mathbb{X}}$ is a \emph{QPD configuration} of $\widetilde{R}$, if the quantum operation $\widetilde{R}$ can be constructed by sampling $\mathbb{F}$ over $\mathbb{X}$ with the assignment $s_{x}$ to each POVM $\mathcal{M}_{x}$,
\begin{align}
\label{eq_QPD_without_r}
  \widetilde{R} = \mathcal{\gamma}_{\mathcal{Q}} \sum_{x\in \mathbb{X}} p_x \,
  \sum_{m} s_{x}(m) \widetilde{K}_{x,m}
  \;\;  \text{ with } \;\;
  \widetilde{K}_{x,m} := \bbra{M_{m|x}^{(\text{anc.})}} \,
  \widetilde{F}_x,
\end{align}
where $\mathcal{\gamma}_{\mathcal{Q}}$ is a normalization factor that normalizes the RHS of the equation to a CPTP map.
\end{definition}

Since $s_{m}$ can be negative, without the normalization factor $\gamma_{\mathcal{Q}}$, the quantum operation is in general a CPTN (complete-positive trace-non-increasing) map.
To ensure that $\widetilde{R}$ is a CPTP, it must fulfill $\frac{1}{d_{in}}\bbrakett{\id_{out}|\widetilde{R}|\id_{in}} = 1$, which determines the normalization factor as
\begin{equation}
  \gamma_{\mathcal{Q}}
  =
  \left(\frac{1}{d_{in}}
  \sum_{x\in \mathbb{X}} p_x \,
  \sum_{m} s_{x}(m)
  \bbrakett{\id_{out}|
    \widetilde{K}_{x,m}
  |\id_{in}}\right)^{-1}
  \ge 1.
\end{equation}
The equality holds, when $s_{x}(m)=1$ for all $x$ and $m$.
Whenever such a decomposition exists, it can be employed to estimate expectation values of the form $\bbra{O} \Tilde{R} \kett{\rho}$, for any input state $\kett{\rho}$ and observable $\bbra{O}$.
Given a QPD of $\widetilde{R}$, the expectation value can be rewritten as
\begin{align}
  \bbra{{O}} \widetilde{R} \kett{\rho}
  =
  \gamma_{\mathcal{Q}} \sum_{x\in \mathbb{X}} p_x \,
  \sum_{m} s_{x}(m) \bbra{O} \,
  \widetilde{K}_{x,m} \kett{\rho}.
\end{align}
Such a QPD configuration enables the estimation of the expectation value of $\widehat{O}$ via Monte Carlo sampling as follows:
\begin{enumerate}
  \item Sample $x$ with probability $p_x$ and apply the corresponding free map $\widetilde{F}_x$ to the input state $\kett{\rho}$.
  \item Conditioned on $x$, perform the measurement $\mathcal{M}_{x}^{(\text{anc.})} = \{\bbra{M_{m|x}^{(\text{anc.})}}\}_{m}$ on the ancillary qubits.
  \item Assign the outcome $m$ with the sign $s_{x}(m)$ in the classical post-processing.
\end{enumerate}

However, one has to pay a price for using free operations to simulate a resourceful operation due to the additional normalization factor $\gamma_{\mathcal{Q}}$.
Since it amplifies the statistical uncertainty, to compensate for the amplified uncertainty, one needs to measure more samples.
More precisely, according to Hoeffding’s inequality, to estimate the outcome with the same accuracy, the total sampling number increases by a factor of $\gamma_{\mathcal{Q}}^{2}$ \cite{Pashayan2015, Peng2020, Mitarai2021}.
The normalization factor $\gamma_{\mathcal{Q}}$ is then referred to as the \emph{sampling overhead} of the QPD configuration $\mathcal{Q}$.

Given a set of free maps $\mathbb{F}$, the \emph{QPD sampling overhead} of a resourceful operation $\widetilde{R}$ over $\mathbb{F}$ is defined as the minimum $\gamma_{\mathcal{Q}}$ achievable among all QPD configurations constructed from free maps in $\mathbb{F}$, which is denoted and determined as follows
\begin{align}
\label{eq::overhead_over_F}
  \gamma_{\mathbb{F}}(\widetilde{R}) :=
  \inf \left\{ \gamma_{\mathcal{Q}} :  \text{$\mathcal{Q}$ is a QPD configuration of $\widetilde{R}$ over the free maps $\mathbb{F}$}.
    \right\}
\end{align}

\subsection{Resources-assisted quasi-probability decomposition over free operations}

To enhance the practical feasibility of circuit knitting, we incorporate entangled-state preparation into the QPD framework to formulate an entanglement-assisted QPD that reduces sampling overhead via entanglement resources. Conversely, from an equivalent perspective, one may incorporate QPD into entanglement-assisted DQC to reduce the required entanglement at the cost of increased sampling overhead.
By treating entangled-state preparation as a resourceful channel, we unify entanglement resources and QPD sampling overhead within the general framework of resource-assisted QPD, implemented via the following sampling procedure:
\begin{enumerate}
  \item Sample a classical label $x\in\mathbb{X}$ with the probability distribution $\{p_x\}_{x\in\mathbb{X}}$.
  \item Conditioned on $x$, implement a $\widetilde{r}$-assisted quantum operations, $\widetilde{G}_{x}\circ\widetilde{r}\circ\widetilde{F}_{x}$, where  $\widetilde{F}_x$ and $\widetilde{G}_x$ are pre- and post-operations, respectively, which are both free maps.
  \item Perform the free POVM measurement $\mathcal{M}_{x}^{(\text{anc.})} = \{\bbra{M_{m|x}^{(\text{anc.})}}\}_{m}$ on the ancillary subsystem to construct a quantum instrument $\{\widetilde{K}_{x,m}\}_{x,m}$.
  \item Assign each outcome $m$ with a sign $s_{x}(m)$ in the classical post-processing.
\end{enumerate}
A resource-assisted QPD is formally defined as follows.
\begin{definition}[Resource-assisted quasi-probability decomposition over free maps]
\label{def::r-assist_QPD}
Let $\mathbb{X} = \{x\}$ be a set of classical labels associated with a probability distribution $\{p_{x}\}_{x}$.
For each label $x$, there are two free operations $\widetilde{F}_x, \widetilde{G}_x$, a free POVM $\mathcal{M}_{x}=\left\{\bbra{M_{m|x}^{(\text{anc.})}}\right\}_{m}$, and additionally a resourceful operation $\widetilde{r}$ available.
The tuples $\mathcal{Q}=\{(p_{x},\widetilde{F}_{x},\widetilde{G}_{x}; s_{x},\mathcal{M}_{x})\}_{x\in\mathbb{X}}$ is a \emph{$\widetilde{r}$-assisted QPD configuration} for a target quantum operation $\widetilde{R}$, if the quantum operation $\widetilde{R}$ can be constructed by sampling $\widetilde{F}_{x}$ and $\widetilde{G}_{x}$ from $\mathbb{F}$ over $\mathbb{X}$ with the assignment $s_{x}$ to each POVM $\mathcal{M}_{x}$,
\begin{align}
\label{eq_QPD_with_r}
    \widetilde{R} = \gamma_\mathcal{Q} \, \sum_{x\in \mathbb{X}}\sum_{m} p_{x}\,s_{x}(m)\,\widetilde{K}_{x,m}
    \;\;\text{ where }\;\;
    \widetilde{K}_{x,m} := \bbra{M_{m|x}^{(\text{anc.})}} \widetilde{G}_{x} \circ \tilde{r} \circ \tilde{F}_{x},
\end{align}
where $ \gamma_\mathcal{Q}$ is the normalization factor that normalizes
the RHS of the equation to a CPTP map.
\end{definition}

The resource-assisted QPD encompasses both resource-free circuit knitting and fully entanglement-assisted DQC as two extremal cases.
These correspond, respectively, to a trivial assisting resource $\widetilde{r}=\kett{\id_{\mathrm{anc.}}}$ (resource-free) and a QPD-free configuration with $s_x=1$ (fully entanglement-assisted).
As an illustrative example, the quantum state teleportation and quantum telegate protocols shown in Fig.~\ref{fig::DQC-diagram} are fully entanglement-assisted DQC schemes implemented via entanglement-assisted LOCC.
Within our framework, taking LOCC as the free maps, they can be viewed as special instances of Bell-state-assisted QPD with the configuration $\mathcal{Q} = \left(p_{x}=1, \widetilde{F}_{x}=\widetilde{\id}, \widetilde{G}_{x} ; s_{x} = 1, \mathcal{M}_{x} = \bbra{\id}\right)$, where the resourceful operations are highlighted in blue and $\widetilde{G}_x$ denotes the LOCC operations highlighted in orange.

\begin{definition}[Sampling overhead]
\label{def::overhead_gamma}
The normalization factor $\gamma_{\mathcal{Q}}$ in Eq.~\eqref{eq_QPD_with_r} is called the \emph{$\widetilde{r}$-assisted $\mathcal{Q}$-sampling overhead for $\widetilde{R}$}, which is denoted and determined by
\begin{equation}
  \gamma_{\mathcal{Q}}(\widetilde{r}\rightarrow\widetilde{R})
  =
  \left(\frac{1}{d_{in}}
  \sum_{x\in \mathbb{X}}\sum_{m} p_x \, s_{x}(m)
    \bbrakett{\id_{\text{out}}| \widetilde{K}_{x,m} |\id_{\text{in}}}
  \right)^{-1}.
\end{equation}
Given a set of free maps $\mathbb{F}$, the \emph{$\widetilde{r}$-assisted $\mathbb{F}$-sampling overhead for a quantum operation $\widetilde{R}$} is defined as the minimum sampling overhead of the $\widetilde{r}$-assisted QPD configurations constructed from the free maps in $\mathbb{F}$,
\begin{align}
  \gamma_{\mathbb{F}}(\widetilde{r}\rightarrow\widetilde{R}) :=
  \inf \left\{ \gamma_{\mathcal{Q}}  :
    \text{$\mathcal{Q}(\widetilde{r}\rightarrow\widetilde{R})$ is a $\widetilde{r}$-assisted QPD for $\widetilde{R}$ with $\widetilde{F}_{x}\in\mathbb{F}$ and $\widetilde{G}_{x}\in\mathbb{F}$}
  \right\}.
\end{align}
\end{definition}
The arrow ``$\rightarrow$'' indicates the conversion of the initial resource operation $\widetilde{r}$ to the target resource operation $\widetilde{R}$. For consistence, we omit the arrow in the QPD overhead in Eq.~\eqref{eq::overhead_over_F} for resource-free QPD, that is
\begin{equation}
  \gamma_{\mathbb{F}}(\Tilde{R}) := \gamma_{\mathbb{F}}(\tilde{f}\rightarrow\Tilde{R}) \; \forall \Tilde{f}\in\mathbb{F}.
\end{equation}

The overheads of resource-assisted $\mathbb{F}$-QPDs have the following properties.
\begin{lemma}[Rules for resource-assisted QPDs]
\label{lemma::gamma_properties}
Let $\widetilde{A}, \widetilde{B}, \widetilde{C}, \widetilde{D}$ be arbitrary quantum operations.
Resource-assisted QPDs over a set of free maps $\mathbb{F}$ follow the following rules:
\begin{description}
    \item[Triangle transition submultiplicativity]
    For a transition $(\widetilde{A}\rightarrow\widetilde{B}\rightarrow\widetilde{C})$, it holds
    \begin{equation}
      \gamma_{\mathbb{F}}(\widetilde{A} \rightarrow \widetilde{C})
      \leq \gamma_{\mathbb{F}}(\widetilde{A} \rightarrow \widetilde{B}) \cdot \gamma_{\mathbb{F}}(\widetilde{B} \rightarrow \widetilde{C}).
    \end{equation}

    \item[Ordering under free operations]
    If $\gamma_{\mathbb{F}}(\widetilde{A} \rightarrow \widetilde{B}) = 1 $, the resource-free overhead of $\widetilde{A}$ and $\widetilde{B}$ is ordered by
    \begin{equation}
      \gamma_{\mathbb{F}}(\widetilde{A}) \geq \gamma_{\mathbb{F}}(\widetilde{B}).
    \end{equation}
%

    \item[Composition submultiplicativity for output channels] The $\widetilde{A}\otimes\widetilde{B}$-assisted QPD of the composition $\widetilde{C}\circ\widetilde{D}$ is upper bound by the product the of the overheads of $(\widetilde{A}\rightarrow\widetilde{C})$ and $(\widetilde{B}\rightarrow\widetilde{D})$,
    \begin{equation}
      \gamma_{\mathbb{F}}(\widetilde{A}\otimes\widetilde{B} \rightarrow \widetilde{C}\circ\widetilde{D})
      \leq \gamma_{\mathbb{F}}(\widetilde{A} \rightarrow \widetilde{C}) \cdot \gamma_{\mathbb{F}}(\widetilde{B} \rightarrow \widetilde{D}).
    \end{equation}

    \item[Parallel-composition submultiplicativity]
    As a result of the composition submultiplicativity for output channels, the parallel implementation $(\widetilde{C}\otimes\widetilde{D})$ has the same upper bound,
    \begin{equation}
      \gamma_{\mathbb{F}}(\widetilde{A}\otimes\widetilde{B} \rightarrow \widetilde{C}\otimes\widetilde{D})
      \leq \gamma_{\mathbb{F}}(\widetilde{A} \rightarrow \widetilde{C}) \cdot \gamma_{\mathbb{F}}(\widetilde{B} \rightarrow \widetilde{D}).
    \end{equation}


%
    \item[Output-channel convexity] A convex mixing of two output channels $\widetilde{B}$ and $\widetilde{C}$ is upper bounded by the convex sum of their overheads,
    \begin{equation}
      \gamma_{\mathbb{F}}(\widetilde{A} \rightarrow p\widetilde{B}+(1-p)\widetilde{C})
      \leq p\gamma_{\mathbb{F}}(\widetilde{A} \rightarrow \widetilde{B})+(1-p)\gamma_{\mathbb{F}}(\widetilde{A} \rightarrow \widetilde{C}),
      \forall 0\le p\le 1.
    \end{equation}
    \item[Input-resource concavity]
    A convex mixing of two input resource channels $\widetilde{A}$ and $\widetilde{B}$ is lower bounded by the convex sum of their overheads,
    \begin{equation}
      \gamma_{\mathbb{F}}(p\widetilde{A}+(1-p) \widetilde{B} \rightarrow \widetilde{C}) \ge p\gamma_{\mathbb{F}}(\widetilde{A}\rightarrow \widetilde{C}) + (1-p)\gamma_{\mathbb{F}}(\widetilde{B} \rightarrow \widetilde{C}),
      \forall 0\le p\le 1.
    \end{equation}
     \item[Free-map monotonicity] For any free map $\widetilde{f}\in \mathbb{F}$, composition with $\widetilde{f}$ on either the input resource or the output channel results in a monotonic behavior,
         \begin{equation}
           \max\left\{ \gamma_{\mathbb{F}}(\widetilde{A}\rightarrow \widetilde{f}\circ \widetilde{B}),\gamma_{\mathbb{F}}(\widetilde{A}\rightarrow \widetilde{B}\circ\widetilde{f}) \right\}
           \le \gamma_{\mathbb{F}}(\widetilde{A}\rightarrow \widetilde{B}) \le
           \min\left\{ \gamma_{\mathbb{F}}(\widetilde{f}\circ\widetilde{A}\rightarrow \widetilde{B}), \gamma_{\mathbb{F}}(\widetilde{A}\circ \tilde{f}\rightarrow \widetilde{B}) \right\}.
         \end{equation}
\end{description}
\begin{proof}
  see Appendix \ref{sec::proof_gamma_properties}
\end{proof}
\end{lemma}

\section{Entanglement-assisted circuit knitting}
\label{sec::ent-assisted_circknit}

In distributed quantum computing, one considers the implementation of a global unitary operation across two QPUs assisted by shared entanglement resources.
Within the resource-assisted QPD framework introduced in Def.~\ref{def::r-assist_QPD}, the assisting resource channel $\widetilde{r}$ corresponds to an entanglement-preparation channel, while the set of free operations $\mathbb{F}$ consists of separable state preparation and local operations with classical communication (LOCC) \cite{Chitambar2014}.
In more restricted scenarios where classical communication is not feasible during local operations (LO), the set of free operations can be further restricted to LO only.
\wjy{Formally, a general local operation $\widetilde{\mathcal{E}}$ is expressed as
\begin{equation}
  \widetilde{\mathcal{E}} = \sum_{i,j}p_{i,j}\widetilde{A}_{i}\otimes\widetilde{B}_{j},
\end{equation}
where $p_{i,j}\ge0$, $\sum_{i,j}p_{i,j}=1$, and $\widetilde{A}_{i}$ and $\widetilde{B}_{j}$ are local CPTP maps.
Under the set of LOCC free operations, the free POVM measurement defined in Eq.~\eqref{eq::free_POVM} is equivalent to a measurement implemented via LOCC followed by a computational-basis measurement, according to the Naimark dilation theorem~\cite{Naimark1943}.
}

The conventional circuit knitting technique \cite{Peng2020, Piveteau2024, Ufrecht2024, Schmitt2025, Harrow2025} can be viewed as an instance of entanglement-free QPD of global unitaries over LOCC. It provides a practical framework for realizing distributed modular quantum computing \cite{AndresMartinez2024}, where large-scale quantum circuits are executed using multiple small-scale QPUs. The QPD of global unitaries over LOCC across local modular QPUs thus offers a natural strategy for entanglement-free distributed quantum computation.

\begin{figure}[htb]
    \centering
    \includegraphics[width=0.9\linewidth]{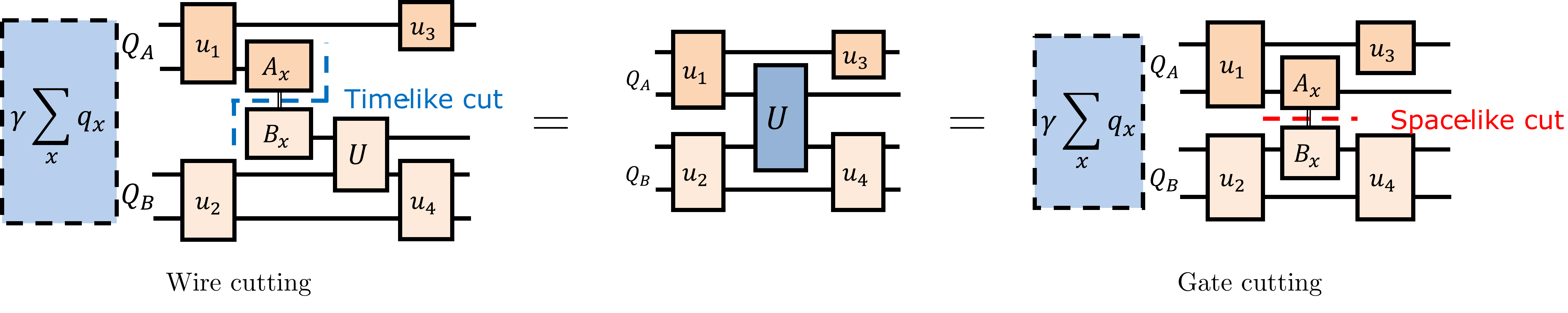}
    \caption{This figure illustrates two types of circuit knitting, namely wire cutting and gate cutting.}
    \label{fig_intro-circuit-knitting}
\end{figure}

As illustrated in Fig.~\ref{fig_intro-circuit-knitting}, circuit knitting typically involves two types of circuit partitioning: wire cutting in the time domain and gate cutting in the spatial domain.
From the perspective of DQC, the former corresponds to a localization of a quantum channel with respect to the bipartition of (input - output), whereas the latter corresponds to a localization with respect to two sets of qubits.
These two forms of circuit cutting can be interpreted as resource-free QPD of state teleportation and gate teleportation, respectively, both serving as fundamental building blocks of entanglement-assisted DQC.

Consider a bipartite system that consists of two QPUs. The target resourceful quantum operations of circuit knitting are global bipartite unitaries. In general, a global bipartite unitary can be written in the \emph{local unitary decomposition (LUD)} with the following definition.
\begin{definition}[Local unitary decomposition]
\label{def::LUD}
  Let $\widehat{U}$ be a unitary acting on two subsystems $A|B$. A \emph{local unitary decomposition (LUD)} of $\widehat{U}$ is given by
  \begin{align}
      \widehat{U} = \sum_i \lambda_i \widehat{A}_i \otimes \widehat{B}_i,
  \end{align}
  where $\hat{A}_i$ and $\hat{B}_i$ are all unitary and the coefficient  $\lambda_{i}$ is a positive real number, which is called a \emph{LUD coefficient}.
  We call a unitary operator is \emph{KAK-like}, if it has a LUD, which is also an operator-Schmidt decomposition \cite{Tyson2003}.
  In such a LUD, the LUD coefficients coincide with the \emph{operator-Schmidt coefficients}.
\end{definition}
Note that the LUD is referred to as the \emph{local decomposition} in Ref.~\cite{Harrow2025}.
In a LUD of a general unitary, the local operators $\{\hat{A}_i\}_i$ and $\{\hat{B}_i\}_i$ are not necessarily orthogonal with respect to the Hilbert–Schmidt inner product.
For KAK-like unitaries, however, one can construct an LUD in which the local operators $\{\hat{A}_i\}_i$ and $\{\hat{B}_i\}_i$ are orthogonal unitaries. In this case, the decomposition simultaneously constitutes an operator–Schmidt decomposition.
As important examples, all the two-qubit gates (i.e., SU(4) operators) are KAK-like \cite{Harrow2025, Schmitt2025}.

From a practical perspective, circuit knitting relies on the LUD rather than a general operator-Schmidt decomposition, since the local unitaries appearing in the LUD directly enable the construction of the local controlled-unitary operations that are required for implementing circuit knitting.
By contrast, the operator–Schmidt decomposition provides a natural connection to the Choi-state representation.
A KAK-like unitary that admits both an LUD and an operator-Schmidt decomposition therefore possesses a particularly well-structured decomposition for QPD, since the LUD enables a constructive QPD protocol, while the operator-Schmidt decomposition yields a tight lower bound on the resource-free sampling overhead via its Choi state.
Together, these properties give rise to the optimal resource-free QPD \cite{Schmitt2025, Harrow2025}, which we summarize in the following proposition.
\begin{proposition}[Entanglement-free circuit knitting \cite{Schmitt2025, Harrow2025}]
\label{prop::ent-free_CK}
Let $\hat{U}$ be a bipartite unitary with the LUD $\hat{U} = \sum_i \lambda_i \hat{A}_i\otimes \hat{B}_i$ with a coefficient $\boldvec{\lambda} = (\lambda_{1},...,\lambda_{K})$.
It has an entanglement-free LO-QPD overhead upper bounded by $\gamma_{\mathrm{LO}}(\widetilde{U} ) \leq  2|| \bm{ \lambda}||_{1}^2 - || \bm{ \lambda}||_{2}^2$.
Moreover, when $\hat{U}$ is KAK-like, its optimum overhead is explicitly determined by its operator Schmidt coefficients $\boldvec{\lambda}$,
\begin{align}\label{s5}
    \gamma_{\mathrm{LOCC}}(\Tilde{U} )
    = \gamma_{\mathrm{LO}}(\Tilde{U} )
    =  2|| \bm{ \lambda}||_{1}^2 - 1.
\end{align}
\end{proposition}%

\bigskip

In this section, beyond the conventional entanglement-free circuit knitting, we incorporate entanglement assistance into the QPD framework and investigate the relationship between available entanglement resources and the resulting reduction in QPD overhead.

It is worth noting that the overhead of a $(A|B)$-bipartite quantum channel $\widetilde{\Lambda}^{(A|B)}$ over LOCC is always lower bounded by the overhead of its Choi state $\kett{J_{\Lambda}^{(A|B)}}$.
This follows directly as a corollary of the triangle submultiplicativity property (Lemma~\ref{lemma::gamma_properties}) for the transition
\wjy{$(\Tilde{r}\rightarrow\widetilde{\Lambda}^{(A|B)}\rightarrow\kett{J_{\Lambda}^{(A|B)}})$, together with the fact that preparing the Choi state $\widehat{J}_{\Lambda}^{(A|B)}$ from the quantum channel $\widetilde{\Lambda}^{(A|B)}$ is free using the free-state preparation of $\ket{\Phi_{d_{A}}^{(A)}}\otimes\ket{\Phi_{d_{B}}^{(B}}$, where $\ket{\Phi_{d_{A}}^{(A)}}$ and $\ket{\Phi_{d_{B}}^{(B)}}$ are the $d_{A,B}$-dimensional maximally entangled state on the local systems $A$ and $B$, respectively,
\begin{equation}
  \kett{J_{\Lambda}^{(A|B)}} =
  \left(\widetilde{\Lambda}^{(A|B)}\otimes\widetilde{\id}^{(A|B)}\right)
  \kett{\Phi_{d_{A}d_{B}}^{(A|B)}}
  \;\;\text{ with }\;\;
  \ket{\Phi_{d_{A}d_{B}}^{(A|B)}} := \ket{\Phi_{d_{A}}^{(A)}}\otimes \ket{\Phi_{d_{B}}^{(B)}}.
\end{equation}
}
\begin{corollary}[Choi-state preparation as a lower bound]
\label{coro::Choi_state_lower_bound}
  The $\tilde{r}$-assisted LOCC-QPD of a quantum channel $\widetilde{\Lambda}$ has a sampling overhead lower bounded by its Choi-state preparation $\kett{J_{\Lambda}}$,
  \begin{equation}
    \gamma_{\mathrm{LOCC}}(\tilde{r}\rightarrow \kett{J_{\Lambda}^{(A|B)}})
    \le
    \gamma_{\mathrm{LOCC}}(\tilde{r}\rightarrow \widetilde{\Lambda}^{(A|B)}).
  \end{equation}
\end{corollary}
For notational simplicity, we omit the superscript $(A|B)$ indicating the bipartition throughout the remainder of the paper.
This corollary leads to a general reduction of the lower bound on the QPD overhead via the assistance of entanglement resources.
\begin{proposition}
\label{prop::r-qpd_lower_bound_by_ent}
Let $\hat{U}$ be a bipartite unitary, for any pure state $\kett{\psi}$ with the entanglement entropy $\mathscr{E}({J_U}) \geq \mathscr{E}({\psi} )$, the $\psi$-assisted LOCC-QPD overhead is lower bounded by
\begin{align}
    2^{\mathscr{E}({J_U})-\mathscr{E}(\psi )} \le \gamma_{\mathrm{LOCC}}(\kett{\psi}\rightarrow \Tilde{U}).
\end{align}

\begin{proof}
See Appendix \ref{sec::proof_EACK}.
\end{proof}
\end{proposition}
Proposition~\ref{prop::r-qpd_lower_bound_by_ent} establishes a trade-off between the lower bound on the LOCC-QPD overhead and the entanglement of the input resource. However, this lower bound is in general not tight.

\bigskip

Note that for a sufficiently powerful entanglement resource $\widetilde{r}_{\text{ent}}$, whose entanglement rank exceeds the operator rank of the target unitary, the unitary is possible to be deterministically implemented via $\widetilde{r}_{\text{ent}}$-assisted LOCC without sampling overhead~\cite{Stahlke2011}, for example using the EJPP telegate protocol~\cite{Eisert2000}.
Here, we instead consider partially entangled resources whose entanglement rank is smaller than the operator rank required to implement the target unitary.

In the remainder of this section, we will derive a tighter lower and upper bounds for entanglement-assisted circuit knitting. Our main results consist of two parts. First, we determine the optimal overhead for Choi-state-stretchable unitaries.
\begin{equation}
  \gamma_{LOCC}\left(\widetilde{r} \rightarrow \widetilde{U}\right) =
  \gamma_{LOCC}\left(\widetilde{r} \rightarrow \kett{J_{U}}\right).
\end{equation}
Second, we establish universal lower and upper bounds on one-Bell-state-assisted QPD for general unitaries.
\begin{equation}
  \gamma_{min}\le\gamma_{LOCC}\left(\kett{\Phi_{2}}\rightarrow \widetilde{U}\right)\le\gamma_{max}.
\end{equation}



\subsection{Entanglement-assisted circuit knitting for Choi-stretchable unitaries}

Corollary~\ref{coro::Choi_state_lower_bound} implies that if the Choi state $\kett{J_{\Lambda}}$ of a quantum channel $\widetilde{\Lambda}$ can be used to implement the channel $\widetilde{\Lambda}$ via LOCC, i.e. $\gamma_{\mathrm{LOCC}}(\kett{J_{\Lambda}} \rightarrow \widetilde{\Lambda}) = 1$,
then the optimal QPD overhead of the Choi state coincides with that of the quantum channel itself. This follows directly from the submultiplicativity of the transition $(\widetilde{r} \rightarrow \widetilde{\Lambda} \overset{\text{free}}{\rightarrow}\kett{J_{\Lambda}} \overset{\text{free}}{\rightarrow} \widetilde{\Lambda})$.
A channel satisfying $\gamma_{\mathrm{LOCC}}(\kett{J_{\Lambda}} \rightarrow \widetilde{\Lambda}) = 1$ is referred to as \emph{Choi-stretchable} \cite{Pirandola2017}. For such a unitary channel, the optimal $\widetilde{r}$-assisted QPD can therefore be characterized by the following theorem.
\begin{theorem}[Resource-assisted circuit knitting for Choi-stretchable unitaries]\label{thm::gamma-LOCC-Choi-stretchable}
\wjy{%
For a Choi-stretchable unitary, i.e. $\gamma_{\mathrm{LOCC}}(\kett{J_{\Lambda}}\rightarrow\widetilde{\Lambda})$, the optimum overhead of a $\widetilde{r}$-assisted QPD for implementing the unitary $\widetilde{U}$ coincides with the optimum overhead for preparing its Choi state $\kett{J_{U}}$,
\begin{align}
\label{eq::opt_gamma_Choi-stretchable}
    \gamma_{\mathrm{LOCC}}(\Tilde{r} \rightarrow \Tilde{U})
    = \gamma_{\mathrm{LOCC}}(\Tilde{r} \rightarrow \kett{J_U}).
\end{align}
If $\tilde{r} = \kett{\rho}$ is a state preparation, the overhead is given by its $d_{U}$-dimensional fully entangled fraction $F_{d_{U}}(\rho)$,
\begin{align}
    \gamma_{\mathrm{LOCC}}(\kett{{\rho}} \rightarrow \Tilde{U})
    = \frac{2}{F_{d_{U}}( {\kett{\rho}} )}-1,
\end{align}
where $d_{U}$ is the operator Schmidt rank of $\widetilde{U}$, and $F_{d}( \rho )$ is the fully entangled fraction of the state $\rho$ for the $d$-dimensional maximally entangled state $\ket{\Phi_d}=\frac{1}{\sqrt{d}}\sum_{i=1}^{d}\ket{i,i}$,
\begin{align}
  F_{d}(\rho) := \max_{\Tilde{\epsilon}\in\mathrm{LOCC}}\left\{\bbra{\Phi_d}\Tilde{\epsilon}\kett{{\rho} }\right\}.
\end{align}}
\begin{proof}
The equality in Eq.~\eqref{eq::opt_gamma_Choi-stretchable} follows direct from the submultiplicativity of the transition $(\widetilde{r} \rightarrow \widetilde{U} \overset{\text{free}}{\rightarrow}\kett{J_{U}} \overset{\text{free}}{\rightarrow} \widetilde{U})$.

For a space-like cutting, according to Theorem 1 in \cite{Stahlke2011}, $\gamma_{\mathrm{LOCC}}(\kett{J_U} \rightarrow \Tilde{U})=1$ implies that $\kett{J_U}$ has uniform Schmidt coefficients, which means that $\kett{J_U} $ is a maximally entangled state with a Schmidt rank of $d_U$. For a time-like bipartition of the identity channel $\widehat{\id}^{(A\rightarrow B)}$ with $A$ and $B$ as the input and output, the Schmidt coefficients of $\kett{J_{\id^{(A\rightarrow B)}}}$ are uniform by itself.

According to \cite{Yuan2024, Takagi2024}, the overhead of virtual distillation of $\rho$ for a $d_U$-dimensional maximally entangled state $\ket{\Phi_{d}}$ is determined by the fully entangled fraction of $\rho$,
  \begin{equation}
    \gamma_{\mathrm{LOCC}}(\kett{\rho}\rightarrow\kett{J_{U}}) =
    \gamma_{\mathrm{LOCC}}(\kett{\rho}\rightarrow\kett{\Phi_{d_U}}) = \frac{2}{F_{d_U}( \rho )}-1.
  \end{equation}
This completes the proof.
\end{proof}
\end{theorem}
As a result of Theorem \ref{thm::gamma-LOCC-Choi-stretchable}, one can show an exponential reduction of overhead for Choi-stretchable unitaries.
\begin{corollary}[Exponential reduction of overhead for Choi-stretchable unitaries]
  For a Choi-stretchable unitary $\widetilde{U}$, the $n$-Bell-state-assisted circuit knitting over LOCC reduces the overhead from $m$-Bell-state-assisted circuit knitting ($m<n$) exponential to
  \begin{equation}
    \frac{
      \gamma_{\mathrm{LOCC}}\left(\kett{\Phi_{2}}^{\otimes n}\rightarrow \widetilde{U}\right)+1
    }{
      \gamma_{\mathrm{LOCC}}\left(\kett{\Phi_{2}}^{\otimes m}\rightarrow \widetilde{U}\right)+1
    }
    =
    2^{-(n-m)}.
  \end{equation}
\end{corollary}

The Choi-stretchable unitaries considered in Theorem~\ref{thm::gamma-LOCC-Choi-stretchable} admit a general scheme applicable to both wire cutting and gate cutting, which correspond to the bipartitions with respect to $(\mathrm{input}_{A} \mid \mathrm{output}_{B})$ in time and two qubit subsets $(\mathbb{Q}_{A} \mid \mathbb{Q}_{B})$ in space, respectively. It should be noted that the relevant free LOCC operations in these two scenarios are $(\mathrm{input}_{A} \mid \mathrm{output}_{B})$-LOCC and $(\mathbb{Q}_{A} \mid \mathbb{Q}_{B})$-LOCC, respectively. Accordingly, the Choi state preparation of the corresponding unitary must also be performed using the LOCC with respect to the same corresponding partition.
\wjy{Note that for a spacial bipartition $(\mathbb{Q}_{A} \mid \mathbb{Q}_{B})$, a $SU(4)$ Clifford unitary is always Choi-stretchable \cite{Pirandola2017}.}

The notion of Choi-stretchable unitaries in~\cite{Pirandola2017} primarily emphasizes quantum channels $\mathcal{E}^{(\mathrm{in}_{A} \rightarrow \mathrm{out}_{B})}$ for state teleportation, \wjy{where the corresponding LOCC operations are defined with respect to the $(\mathrm{input}_{A}\mid\mathrm{output}_{B})$ bipartition.} Note that the Choi state $\kett{\Phi_{d}}$ of the identity channel $\widetilde{\id}^{(A \rightarrow B)}$ from $A$ to $B$ can be obtained via sending half of a locally prepared maximally entangled state from $A$ to $B$, i.e. $\gamma_{LOCC}(\widetilde{\id}^{(A \rightarrow B)}\rightarrow\kett{\Phi_{d}^{A|B}})=1$. On the other hand, the identity channel $\widetilde{\id}^{(A \rightarrow B)}$ is Choi-stretchable. This directly yields a corollary for the optimal overhead of resource-assisted wire cutting, which constitutes a natural extension of the results in \cite{Bechtold2024, Bechtold2025}.
\begin{corollary}[Resource-assisted wire cutting] \label{cor_optimal_wire_cutting}
    The optimum $\tilde{r}$-assisted QPD overhead over LOCC for a $d$-dimensional identity channel is given by the $\tilde{r}$-assisted QPD overhead of the $d$-dimensional maximally entangled state,
    \begin{align}
        \gamma_{\mathrm{LOCC}}\left( \tilde{r} \rightarrow \Tilde{\id}_d^{(A\rightarrow B)}  \right) = \gamma_{\mathrm{LOCC}}\left( \tilde{r} \rightarrow \kett{\Phi_{d}}\right)
        \;\;\text{ with }\;\;
        \ket{\Phi_{d}} = \frac{1}{\sqrt{d}}\sum_{i=0}^{d-1}\ket{i,i}.
    \end{align}
    If $\tilde{r} = \kett{\rho}$ is a state preparation, the optimal overhead is then determined by the $d$-dimensional fully entangled fraction $F_{d}(\rho)$,
  \begin{align}
  \label{eq::gamma_wire_cut}
    \gamma_{\mathrm{LOCC}}\left( \kett{\rho}\rightarrow \Tilde{\id}_d^{(A\rightarrow B)}  \right) = \frac{2}{F_{d}( \rho )}-1.
  \end{align}
\end{corollary}
Note that if $\kett{\rho}$ is a separable state, which has the fully entangled fraction $F_d(\kett{\rho}) = \frac{1}{d}$, the formula in Eq.~\eqref{eq::gamma_wire_cut} reproduces the resource-free QPD overhead $\gamma_{\text{LOCC}} = 2d-1$ for wire cutting derived in \cite{Brenner2025, Pednault2023, Harada2024}.


\subsection{Universal one-Bell-pair-assisted circuit knitting}
\label{sec::one-Bell_CK}
In this section, we establish universal lower and upper bounds on the QPD of general unitaries over LOCC with the assistance of one Bell pair $\ket{\Phi_{2}}$.
Without loss of generality, the assisting Bell state is expressed as $\ket{\Phi_{2}} = (\ket{00}+\ket{11})/\sqrt{2}$ in the computational basis.

As a consequence of Corollary~\ref{coro::Choi_state_lower_bound}, the $\Phi_{2}$-assisted QPD overhead for implementing a target unitary $\widetilde{U}$ is lower bounded by the $\Phi_{2}$-assisted QPD overhead for its Choi state $\kett{J_{U}}$. A general lower bound can be therefore characterized by the Schmidt-rank robustness~\cite{Johnston2018} of the Choi state.
\begin{theorem}[Lower bound on one-Bell-pair-assisted circuit knitting over LOCC]\label{thm::lBound_one-Bell_CK}
The overhead of a one-Bell-state-assisted QPD of a unitary over LOCC is lower bounded by
\begin{align}
    \gamma_{\mathrm{LOCC}}(\kett{ {\Phi}_2} \rightarrow \widetilde{U}) \geq ||\boldvec{s}||_{1}^2 - 1 + \left(\max\{0, 2s_1 - ||\boldvec{s}||_{1}\}\right)^2,
\end{align}
where $s_i$ is the Schmidt coefficient of the Choi state $\kett{J_{U}}$ with decreasing order.
\begin{proof}
See Appendix \ref{sec::proof_EACK}.
\end{proof}
\end{theorem}

\bigskip

\begin{figure}[htb]
    \centering
    \includegraphics[width=0.9\textwidth]{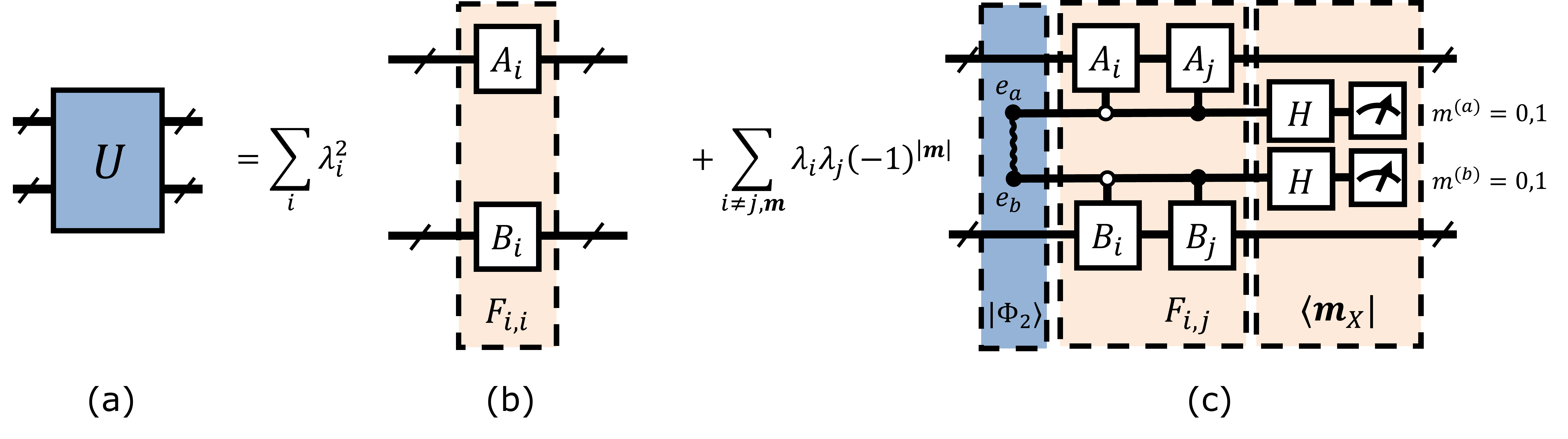}
    \caption{$\Phi_{2}$-assisted QPD over LO. (a) A target unitary. (b) Diagonal terms that are already LO. (c) Cross terms, which are implemented with $\Phi_{2}$-assisted LO. The wiggle line denotes the Bell state $|\Phi_2\rangle = \frac{1}{\sqrt{2}}(|0,0\rangle +|1,1\rangle)$.
    The solid dot denotes the control unitary operates at $\projector{1}$, and the white dot denotes the control unitary operates at $\projector{0}$.}
    \label{fig::circuit-2LO}
\end{figure}

To obtain an upper bound, we construct an explicit $\Phi_{2}$-assisted QPD for the target unitary.
Given an LUD of the target unitary $\widehat{U}$ (Definition \ref{def::LUD}), one can decompose $\widetilde{U}$ into a diagonal term $\widetilde{D}_{U}$ and a cross term $\widetilde{C}_{U}$,
\begin{equation}
  \widetilde{U} = \widetilde{D}_{U} + \widetilde{C}_{U},
  \text{ where }
  \widetilde{D}_{U} = \sum_{i}\lambda_{i}^{2}\widetilde{\Lambda}_{i}
  \;\;\text{ and }\;\;
  \widetilde{C}_{U} = \sum_{i\neq j}\lambda_{i}\lambda_{j}\widehat{\Lambda}_{i}\otimes\widehat{\Lambda}_{j}^{\ast}.
\end{equation}
As shown in Fig.~\ref{fig::circuit-2LO}(b), the diagonal terms can be implemented using local operations (LO) alone.
For the cross term $i\neq j$, as shown in Fig.~\ref{fig::circuit-2LO} (c), one can employ a LO $\widetilde{F}_{i,j}$ acting on the input resource $\ket{\Phi_{2}}$ on the ancillary qubits $(e_{a},e_{b})$ followed by a local $X$-basis measurement $\{\bbra{\boldvec{m}_{X}^{(anc.)}}: (m^{(a)},m^{(b)}) \in\{0,1\}^{\otimes 2}\}$ without classical communication.
The LO $\widetilde{F}_{i,j}$ is a local control unitary given by
\begin{equation}
\label{eq::free_LO_for_Bell_QPD}
  \widetilde{F}_{i,j} =
  \left(
    \widehat{\pi}_{0}^{(e_{a})}\otimes\widehat{A}_{i}
    +
    \widehat{\pi}_{1}^{(e_{a})}\otimes\widehat{A}_{j}
  \right)
  \otimes
  \left(
    \widehat{\pi}_{0}^{(e_{b})}\otimes\widehat{B}_{i}
    +
    \widehat{\pi}_{1}^{(e_{b})}\otimes\widehat{B}_{j}
  \right).
\end{equation}
The resulting $\Phi_{2}$-assisted LO-QPD configuration is summarized in the following theorem, which provides a well-characterized upper bound on the overhead for a general unitary.
\begin{theorem}[Upper bound on one-Bell-pair-assisted circuit knitting over LO]\label{thm::one-Bell_CK_LO}
Let $\widehat{U}$ be a bipartite unitary with the LUD $\widehat{U} = \sum_i \lambda_i \widehat{A}_i\otimes \widehat{B}_i$, and $\kett{\Phi_2}$ be a two-qubit maximally entangled state.
There exists a $\Phi_{2}$-assisted QPD for $\widetilde{U}$ over LO
\begin{align}
\label{eq::LUD_2_QPD_config}
  \widetilde{U} =
  \gamma_{\mathcal{Q}} \left(
    \sum_{i}p_{i,i}(\widetilde{A}_{i}\otimes\widetilde{B}_{i})
    +
    \sum_{i,j} p_{i,j}\sum_{\bm{m} \in \{0,1\}^{\otimes2}} (-1)^{|\bm{m}|} \bbra{\boldvec{m}_{X}^{(anc.)}}\tilde{F}_{i,j}  \kett{\Phi_2},
  \right)
\end{align}
where $p_{i,j} = \frac{\lambda_i\lambda_j}{||\boldvec{\lambda}||_{1}^2}$, and $\widetilde{F}_{i,j}$ is given in Eq.~\eqref{eq::free_LO_for_Bell_QPD} and illustrated in Fig. \ref{fig::circuit-2LO}.
The overhead of a $\Phi_2$-assisted LO-QPD is then upper bounded by this explicit QPD configuration,
\begin{align}
  \gamma_{\mathrm{LO}}(\kett{{\Phi}_2} \rightarrow \Tilde{U} )
  \leq
  \gamma_{\mathcal{Q}}
  =
  \|\boldvec{\lambda}\|_{1}^2.
\end{align}
\begin{proof}
See Appendix \ref{sec::proof_EACK}.
\end{proof}
\end{theorem}

The explicit implementation of the QPD in Eq.~\eqref{eq::LUD_2_QPD_config} is illustrated in Fig. \ref{fig::circuit-2LO}.
It is worth noting that since the one-norm is multiplicative under the tensor product of the vector, $||{\boldvec{\lambda}\otimes \boldvec{\lambda}'}||_{1} = ||\boldvec{\lambda}||_{1}\cdot ||\boldvec{\lambda}'||_{1}$, the overhead of the LO-QPD in Eq.~\eqref{eq::LUD_2_QPD_config} is multiplicative with respect to parallel QPDs,
\begin{equation}
  \gamma_{\mathrm{LO}}\left(\kett{\Phi_{2}}\rightarrow\widetilde{U}\otimes\widetilde{U}'\right)
  \le
  ||\boldvec{\lambda}||_{1}^{2}\cdot ||\boldvec{\lambda}'||_{1}^{2}.
\end{equation}
For a KAK-like unitary, it leads to an upper bound on the $\Phi_{2}$-assisted LO-QPD overhead $\gamma_{\mathrm{LO}}(\kett{{\Phi}_2} \rightarrow \Tilde{U} ) $ that coincides with the regularized optimal entanglement-free LOCC-QPD overhead $\gamma_{\mathrm{LOCC}}^{\infty}(\Tilde{U})$ \cite{Piveteau2024, Schmitt2025},
\begin{align}
  \gamma_{\mathrm{LO}}\left(\kett{{\Phi}_2} \rightarrow \widetilde{U} \right)
  \leq
  ||\boldvec{\lambda}||_{1}^{2}
  =
  \lim_{n\rightarrow \infty} \sqrt[n]{\gamma_{\mathrm{LOCC}}(\widetilde{U}^{\otimes n})}
  =:\gamma_{\mathrm{LOCC}}^{\infty}(\Tilde{U}).
\end{align}

The result for a one-Bell-state-assisted QPD can be extended to the case where the pre-shared entanglement is not maximally entangled in the following corollary.
\begin{corollary}
\label{coro::non-max_ent_QPD}
Let $\ket{\psi(r)} = \sqrt{\frac{1+r}{2}}\ket{0,0}+\sqrt{\frac{1-r}{2}}\ket{1,1}$ be a partially entangled state with $r\in [0,1]$.
For an arbitrary target unitary $\widehat{U}$ with a LUD $\widehat{U} = \sum_i \lambda_i \widehat{A}_i\otimes \widehat{B}_i$, one can adopt the QPD configuration in Fig.~\ref{fig::circuit-2LO} via replacing the Bell state $\ket{\Phi_{2}}$ in Fig.~\ref{fig::circuit-2LO} by $\ket{\psi(r)}$. The corresponding overhead of this configuration determines an upper bound on the overhead
$\gamma_{\mathrm{LO}}(\kett{ {\psi}(r)} \rightarrow \widetilde{U})$,
\begin{align}
\label{eq::partial_ent_LO_CK}
  \gamma_{\mathrm{LO}}(\kett{ {\psi}(r)} \rightarrow \widetilde{U} ) 
  \leq ||\boldvec{\lambda}||_{2}^2+
  \frac{1}{\sqrt{1-r^2}}(||\boldvec{\lambda}||_{1}^2 - ||\boldvec{\lambda}||_2^2).
\end{align}
For $r\ge3/4$, entanglement-free circuit knitting provides smaller overhead.
\end{corollary}
\begin{proof}
With the same configuration in Fig.~\ref{fig::circuit-2LO} (c) with a replacement of the Bell state by $\ket{\psi(r)}$, one obtains the following QPD for the cross terms $i\neq j$,
\begin{align}
    \sum_{\bm{m}\in \{0,1\}^2}(-1)^{|\bm{m}|}\bbra{\bm{m}}\tilde{F}_{i,j} |\Phi_2\rangle\rangle
    = \frac{1}{\sqrt{1-r^2}}\sum_{\bm{m}\in \{0,1\}^2}(-1)^{|\bm{m}|}\bbra{\bm{m}}\tilde{F}_{i,j} |\psi(r)\rangle\rangle ,
    \;\;\forall i\neq j,
\end{align}
As a result, the overhead in Eq.~\eqref{eq::partial_ent_LO_CK} can be provided by the following the QPD configuration,
\begin{align}
    \Tilde{U} = \sum_i \lambda_i^2 \tilde{A}_i\otimes \tilde{B}_i + \frac{2}{\sqrt{1-r^2}}\sum_{i>j}\lambda_i \lambda_j \sum_{\bm{m}\in \{0,1\}^2}(-1)^{|\bm{m}|}\bbra{\bm{m}}\tilde{F}_{i,j} |\psi(r)\rangle\rangle \label{eq_QPD-with-NME}.
\end{align}
\end{proof}

\begin{figure}[htb]
    \centering
    \def\offsetleft{-0.7}
    \begin{tikzpicture}
      \node at (0,0) {\includegraphics[trim={1cm 0cm 1cm 0cm }, clip, width= 0.9\linewidth]{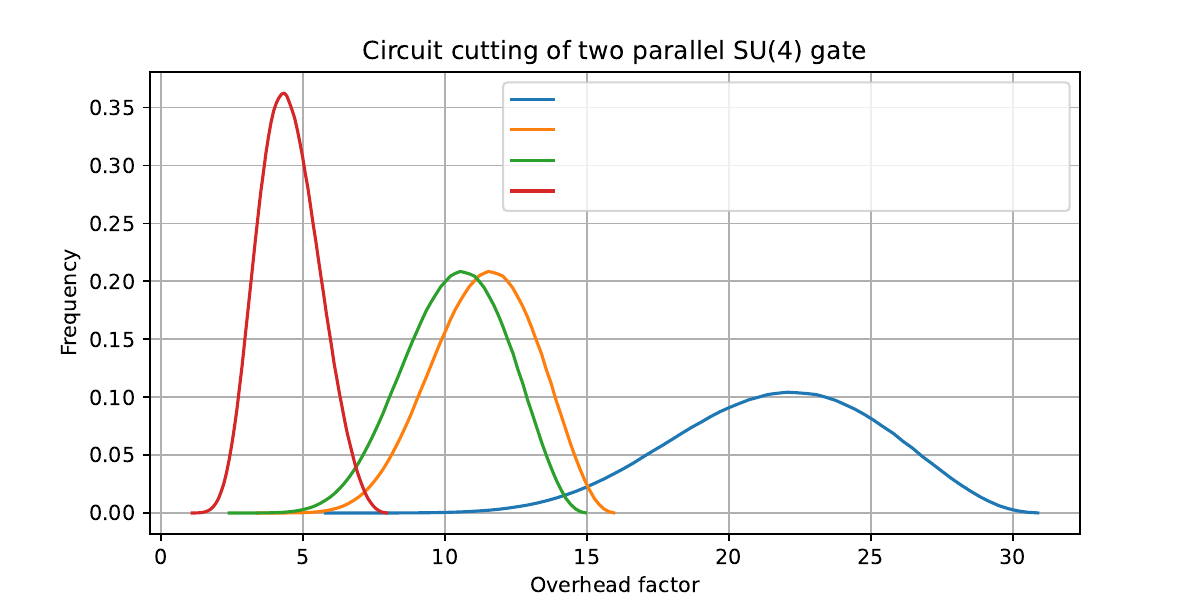}}; 
      \node[anchor=south west] at (\offsetleft, 2.75) {$\gamma_{\text{LOCC}}(\widetilde{U})$: entanglement-free, exact value, Prop \ref{prop::ent-free_CK}};
      \node[anchor=south west] at (\offsetleft, 2.3) {$\gamma_{\text{LO}}(\Phi_{2}\rightarrow\widetilde{U})$: $\Phi_{2}$-assisted, upper bound, Thm. \ref{thm::one-Bell_CK_LO}};
      \node[anchor=south west] at (\offsetleft, 1.85) {$\gamma_{\text{LOCC}}(\Phi_{2}\rightarrow\widetilde{U})$: $\Phi_{2}$-assisted, lower bound, Thm. \ref{thm::lBound_one-Bell_CK}};
      \node[anchor=south west] at (\offsetleft, 1.4) {$\gamma_{\text{LOCC}}(\Phi_{2}\rightarrow\widetilde{U})$: $\Phi_{2}$-assisted, lower bound, Prop. \ref{prop::r-qpd_lower_bound_by_ent}};
    \end{tikzpicture}
    \caption{Distribution of QPD overhead for $10^{8}$ samples of two parallel two-qubit gates, $\widehat{U}^{\otimes 2}$, drawn from Haar-random $\mathrm{SU}(4)$ unitaries. With the assistance of one Bell state, the sampling overhead is reduced approximately by a factor of two compared to the entanglement-free circuit knitting.}
    \label{fig_pdf-overhead}
\end{figure}

For illustration, we numerically generate two parallel two-qubit gates sampled from Haar-random unitaries and plot the probability distribution of their sampling overheads under different bounds in Fig.~\ref{fig_pdf-overhead}. The blue curve shows the distribution of the overhead for entanglement-free circuit knitting (Proposition~\ref{prop::ent-free_CK}).
The red curve indicates the lower bound derived from the entanglement entropy of the Choi states, which is not tight in most instances.
The orange and green curves represent the upper bound (Theorem~\ref{thm::one-Bell_CK_LO}) and lower bound (Theorem~\ref{thm::lBound_one-Bell_CK}) on the overhead of one-Bell-state-assisted circuit knitting, respectively.
We observe that the sampling overhead of one-Bell-pair-assisted circuit knitting is approximately reduced by a factor of two compared to the entanglement-free case.

\bigskip

In Fig.~\ref{fig_pdf-overhead}, one can also observe a clear gap between the lower and upper bound on the $\Phi_{2}$-assisted circuit knitting.
For a KAK-like unitary, the Schmidt coefficients in Theorem~\ref{thm::lBound_one-Bell_CK} and the LUD coefficients in Theorem~\ref{thm::one-Bell_CK_LO} coincides, $\boldvec{s} =\boldvec{\lambda}$. It leads to a fixed gap $\Delta$ between the lower and upper bounds given as follows.
\begin{equation}
\label{eq::LO-LOCC_gap_KAK-like}
  \|\boldvec{\lambda}\|_{1}^{2}-\Delta
  \le
  \gamma_{\mathrm{LOCC}}(\kett{\Phi_{2}}\rightarrow\widetilde{U})
  \le
  \|\boldvec{\lambda}\|_{1}^{2}
  \;\;\text{ with }\;
  \Delta = 1-\left(\max(\{0,2\lambda_{1}-\|\boldvec{\lambda}\|_{1}\})\right)^{2}.
\end{equation}
To reduce this gap, classical communication (CC) should be incorporated into the QPD configuration in Theorem~\ref{thm::one-Bell_CK_LO} to further reduce the overhead.

Suppose the target unitary $\widehat{U}$ has the LUD $\widehat{U} = \sum_{i}\lambda_{i}\widehat{\Lambda}_{i}$. Let $\{\widehat{\Lambda}_{i}\}_{i=1,...,d_{U}}$ be the set of local unitaries in the LUD.
There can be a subset of local-unitary indices $L=\{i_{j}\}_{j=1}^{k}$ appearing in the LUD of $\widetilde{U}$, which can be combined to construct a rank-$k$ Choi-stretchable unitary by taking their uniform operator sum.
\begin{equation}
\label{eq::CS_construction}
  \widehat{S}_{L} := \frac{1}{\sqrt{|L|}}\sum_{j\in L}\widehat{\Lambda}_{j}.
\end{equation}
The $\Phi_{2}$-assisted LO-QPD of $\widetilde{S}_{L}$ is given by
\(
\gamma_{\mathrm{LO}}(\ket{\Phi_{2}} \rightarrow \widetilde{S}_{L}) = |L|.
\)
Classical communication enables the implementation of the optimal LOCC-QPD for this Choi-stretchable unitary according to Theorem~\ref{thm::gamma-LOCC-Choi-stretchable} with an optimal $\Phi_{2}$-assisted LOCC-QPD overhead
\(
\gamma_{\mathrm{LOCC}}(\kett{\Phi_{2}} \rightarrow \widetilde{S}_{L}) = |L|-1.
\)
Incorporating the LOCC-QPD in Theorem~\ref{thm::gamma-LOCC-Choi-stretchable} for such a Choi-stretchable unitary therefore reduces the overhead by one.

One can then construct a quantum channel from a set Choi-stretchable unitaries,
\begin{equation}
  \widetilde{\mathcal{E}}_{S} = \sum_{L\in\mathbb{S}} p_{L}\, \widetilde{S}_{L}
  \;\;\text{ with }\;\;
  \mathbb{S}:=\{L: \widetilde{S}_{L} \text{ is Choi-stretchable}\},
\end{equation}\wjy{
such that the target unitary admits a convex decomposition into a Choi-stretchable component $\widetilde{\mathcal{E}}_{S}$ and a complementary part $\widetilde{\mathcal{E}}_{\mathrm{ELO}}$, which can be implemented with $\Phi_{2}$-assisted LOCC-QPD (Theorem \ref{thm::gamma-LOCC-Choi-stretchable}) and $\Phi_{2}$-assisted LO-QPD (Theorem \ref{thm::one-Bell_CK_LO}), respectively,}
\begin{equation}
\label{eq::CS_Non-CS_decomposition}
  \widetilde{U}
  =
  c_{S}\,\widetilde{\mathcal{E}}_{S}
  +
  (1-c_{S})\,\widetilde{\mathcal{E}}_{\mathrm{ELO}}.
\end{equation}
Here, $0 \le c_{S} \le 1$ denotes the mixing weight of $\widetilde{\mathcal{E}}_{S}$.
Since each Choi-stretchable unitary contributes a unit reduction in overhead,
incorporating the LOCC-QPD (Theorem~\ref{thm::gamma-LOCC-Choi-stretchable}) for $\widetilde{\mathcal{E}}_{S}$ reduces the total overhead by $c_{S}$.
\begin{equation}
  \gamma_{\mathrm{LOCC}}(\kett{\Phi_{2}}\rightarrow\widetilde{U})
  \le
  \|\boldvec{\lambda}\|_{1}^{2}-c_{S}
  \;\;\text{ with }\;\;
  0\le c_{S}\le 1.
\end{equation}

Note that the fraction $c_{S}$ has an upper limit conditioned by the diagonal terms $\widetilde{D}_{U}$ of the target unitary.
Each components $\widetilde{S}_{L}$ in $\widetilde{\mathcal{E}}_{S}$ can also be decomposed into the diagonal and cross terms
\begin{equation}
\label{eq::CS_UL}
  \widetilde{U}_{L} = \widetilde{D}_{L} + \widetilde{C}_{L}
  \;\;\text{ with }\;\;
  \widetilde{D}_{L} := \frac{1}{|L|}\sum_{i\in L}\widetilde{\Lambda}_{i}
  \,\,\text{ and }\,\,
  \widetilde{C}_{L} := \frac{1}{|L|}\sum_{(i\neq j)\in L}\widehat{\Lambda}_{i}\otimes\widehat{\Lambda}_{j}^{\ast}.
\end{equation}
The entanglement-assisted-LO term $\widetilde{\mathcal{E}}_{\mathrm{ELO}}$ is explicitly given by
\begin{equation}
  \widetilde{\mathcal{E}}_{\mathrm{ELO}}
  =
  \widetilde{D}_{U} - c_{S}\sum_{L\in\mathbb{S}}p_{L}\widetilde{D}_{L}
  +
  \widetilde{C}_{U} - c_{S}\sum_{L\in\mathbb{S}}p_{L}\widetilde{C}_{L}
\end{equation}
To ensure that the decomposition in Eq.~\eqref{eq::CS_Non-CS_decomposition} reduces the overhead to the greatest extent, one can impose the condition that the diagonal terms of the complementary LO-QPD component $\widetilde{\mathcal{E}}_{\mathrm{ELO}}$ are positive semidefinite,
\begin{equation}
  \widetilde{D}_{U} - c_{S}\sum_{L\in\mathbb{S}}p_{L}\widetilde{D}_{L}\succcurlyeq 0.
\end{equation}
The maximum $c_{S}$ that satisfies this condition provides the largest overhead reduction via incorporating the LOCC-QPD (Theorem~\ref{thm::gamma-LOCC-Choi-stretchable}) into the LO-QPD (Theorem~\ref{thm::one-Bell_CK_LO}), which allows us to further reduce the overhead as follows.
\begin{corollary}[Upper bound on one-Bell-pair-assisted circuit knitting over LOCC]
\label{coro::one-Bell_CK_LOCC}
Let $\widehat{U}$ be a bipartite unitary with the LUD $\widehat{U} = \sum_i \lambda_i \widehat{\Lambda}_{i}$, where $\widehat{\Lambda}_{i} = \widehat{A}_i\otimes \widehat{B}_i$ are local unitaries, and $\kett{\Phi_2}$ be a two-qubit maximally entangled state.
Suppose $\mathbb{S}$ is the set of index subset associated with all Choi-stretchable unitaries $\widetilde{S}_{L}$ constructed in Eq.~\eqref{eq::CS_construction},
\begin{equation}
  \mathbb{S} := \{L: \widetilde{S}_{L}=\frac{1}{|L|}\sum_{j\in L}\widehat{\Lambda}_{j} \text{ is Choi-stretchable}\}.
\end{equation}
One can then construct a $\Phi_{2}$-assisted QPD over LOCC given in Eq.~\eqref{eq::CS_Non-CS_decomposition}. Such a QPD upper bounds the optimal LOCC-QPD overhead by
\begin{equation}
  \gamma_{\mathrm{LOCC}}(\kett{{\Phi}_2} \rightarrow \Tilde{U} ) \leq
  ||\boldvec{\lambda}||_{1}^2-c_{S},
\end{equation}
where $c_{S}$ denotes the maximum fraction of the Choi-stretchable quantum channel $\widetilde{\mathcal{E}}_{S}$ in $\widetilde{U}$ that satisfies the following condition,
\begin{align}
\label{eq::one-Bell_LOCC-QPD_cond}
  c_{S} = \max_{\{p_{L}\}_{L\in\mathbb{S}}} c,
  \;\;\text{ s.t. }\;\;
  c \sum_{L}\frac{p_{L}}{|L|}\delta_{l\in L}
  \le
  \lambda_{l}^{2}.
\end{align}
\begin{proof}
  See Appendix \ref{sec::proof_EACK}.
\end{proof}
\end{corollary}\wjy{
The choice of the mixing weight $c_{S}$ for the Choi-stretchable component is not necessarily restricted to its maximum value. Any weight satisfying $c \leq c_{S}$ is allowed.}
It is worth noting that the Choi-stretchable component $\widetilde{\mathcal{E}}_{S}$ can also be implemented using more general entanglement resources beyond a single Bell pair, including multiple Bell pairs or higher-dimensional entangled states.

\section{Entanglement-assisted black-box circuit knitting}
\label{sec::ent-assist_bb_circknit}

Many quantum circuits consist of two-qubit gates interleaved with variable single-qubit gates, such as in parameterized quantum circuits. The single-qubit gates between two-qubit gates are tunable and may vary depending on the specific tasks. To construct an efficient resource-assisted QPD configuration that is independent of these task-dependent interleaving quantum channels, it is necessary to treat the intervening variable channels as \emph{black boxes} and develop a resource-assisted QPD scheme that remains agnostic to their internal structure.

As illustrated in the upper part of Fig.~\ref{fig::blackbox_QPD}, we consider a sequence of resource operations $\{\widetilde{R}^{(t)}\}_{t=1}^{T}$ interleaved with unknown quantum channels $\{\widetilde{\mathcal{B}}^{(t)}\}_{t=1}^{T}$. We refer to this composition as the \emph{black-box composition} of the sequence $\{\widetilde{R}^{(t)}\}_{t}$ and denote it by a black-square symbol,
\begin{equation}
  \bbcompo_{t=1}^{T}\; \widetilde{R}^{(t)}
  :=
  \widetilde{\mathcal{B}}^{(T)}\circ\widetilde{R}^{(T)} \circ  \cdots \circ
  \widetilde{\mathcal{B}}^{(1)}\circ\widetilde{R}^{(1)}.
\end{equation}
In this configuration, the target channels form an instance of a quantum comb~\cite{Chiribella2008a,Chiribella2008}, while the black-box channels form an unknown quantum comb.
Note that the additional Hilbert space shared by the black-box channels and untouched by the target channels can also be removed, if the black-box channels act as the identity on this space.

A straightforward QPD for this scheme is to simply allocate the input resource $\Tilde{r}$ to one individual target channel $\widetilde{R}_{t'}$, while the others employ the resource-free black-box circuit knitting.
This QPD configuration follows the results derived in Section \ref{sec::ent-assisted_circknit}, which returns a trivial upper bound on the overhead of resource-assisted black-box circuit knitting according to ``\emph{composition submultiplicativity for output channels}'' in Lemma \ref{lemma::gamma_properties}
\begin{equation}
  \gamma_{\mathbb{F}}\left(
    \Tilde{r}\rightarrow\bbcompo_{t}\widetilde{R}^{(t)}
  \right)
  \le
  \gamma_{\mathbb{F}}\left(
    \Tilde{r}\rightarrow\widetilde{R}^{(t')}
  \right)
  \gamma_{\mathbb{F}}\left(
    \bbcompo_{t\neq t'}\widetilde{R}^{(t)}
  \right).
\end{equation}
It means that a collective $\Tilde{r}$-assisted black-box QPD protocol provides advantage over a $\Tilde{r}$-assisted QPD protocol derived in Section \ref{sec::ent-assisted_circknit}.
To benefit from this advantage, we extend the theoretical framework of resource-assisted QPDs to the black-box setting in Section~\ref{sec::r-bb-QPD}, and develop an advanced collective QPD for this scheme over the free operations of LO and LOCC in Section~\ref{sec::EntBBQPD_CS} and \ref{sec::EntBBQPD_Bell}.

\subsection{Resource-assisted black-box quasi-probability decomposition}
\label{sec::r-bb-QPD}

The resource-assisted QPD can be further generalized to the \emph{black-box setting} introduced in Ref.~\cite{Schmitt2025}. A straightforward QPD approach is to apply an independent $\widetilde{r}^{(t)}$-assisted QPD to each target operation $\widetilde{R}^{(t)}$.
However, this strategy is inefficient in terms of resource consumption.
Instead, it is possible to employ a single resourceful operation $\widetilde{r}$ jointly across multiple non-sequential free operation $\widetilde{F}_{x}^{(t)}$.
Such joint utility of the same resource can reduce both the overall entanglement consumption and the sampling overhead.

\begin{figure}[htb]
  \centering
  \includegraphics[width=0.8\textwidth]{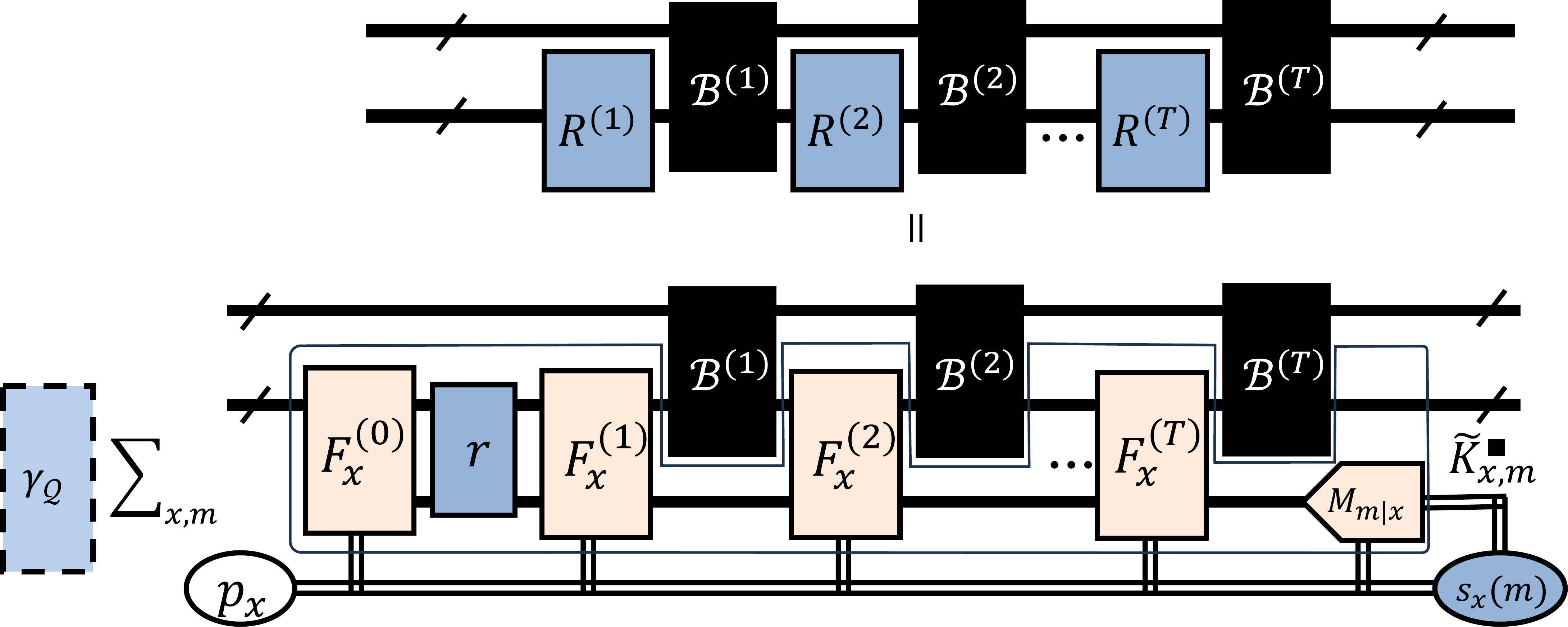}
  \caption{Resource-assisted black-box QPD}
  \label{fig::blackbox_QPD}
\end{figure}

The corresponding \emph{resource-assisted black-box QPD} for this class of target operations is illustrated in Fig.~\ref{fig::blackbox_QPD}. Its implementation proceeds via the following sampling procedure:
\begin{enumerate}
  \item Sample a classical label $x \in \mathbb{X}$ according to the probability distribution $\{p_x\}_{x \in \mathbb{X}}$.

  \item Conditioned on $x$, implement a $\widetilde{r}$-assisted quantum operation
  \[
  ( \bbcompo_{t=1}^{T} \widetilde{F}_{x}^{(t)} )
  \circ \widetilde{r} \circ \widetilde{F}_{x}^{(0)},
  \]
  which begins with a pre-operation $\widetilde{F}_{x}^{(0)}$, followed by the application of the resourceful operation $\widetilde{r}$. After the resourceful operation $\widetilde{r}$, each target operation $\widetilde{R}^{(t)}$ is then replaced by a free operation $\widetilde{F}_{x}^{(t)}$, while the internal black-box channels remain untouched and are embedded within the QPD configuration.

  \item Perform the free POVM measurement
  $\mathcal{M}_{x}^{(\mathrm{anc.})} = \{\bra{M_{m|x}^{(\mathrm{anc.})}}\}_{m}$
  on the ancillary subsystem to construct a quantum instrument
  $\{\widetilde{K}_{x,m}\}_{x,m}$.

  \item Assign each outcome $m$ a sign $s_{x}(m)$ in the classical post-processing.
\end{enumerate}
This sampling procedure implements a \emph{resource-assisted black-box QPD} for a black-box interleaved resourceful operations, which is formally defined as follows.

\begin{definition}[Resource-assisted black-box QPD]
\label{def::r-assist_bb_QPD}
Suppose a resource operation $\Tilde{r}$ is available.
Let $\mathbb{X} = \{x\}$ be a set of classical labels associated with a probability distribution $\{p_{x}\}_{x}$.
As shown in Fig.~\ref{fig::blackbox_QPD}, for each $x$, a free pre-operation $\widetilde{F}_{x}^{(0)}$ is implemented before $\Tilde{r}$, followed by a sequence of free operations $\{\widetilde{F}_x^{(t)}\}_{1 \leq t \leq T}$ ordered in time, interleaved by a sequence of unknown quantum channels $\{\widetilde{\mathcal{B}}^{(t)}\}_{1 \leq t \leq T}$.
In the end, a free POVM $\mathcal{M}_{x}=\{\bbra{M_{m|x}^{(\text{anc.})}}\}_{m}$ is implemented to construct a black-box quantum instrument $\{\widetilde{K}_{x,m}^{\bbcompo}\}_{x,m}$.
Each POVM $\mathcal{M}_{x}$ is assigned with a sign function $s_{x}$.
The tuples $\mathcal{Q}=\{(p_{x},\widetilde{F}_{x}^{(0)},...,\widetilde{F}_{x}^{(T)}; s_{x},\mathcal{M}_{x})\}_{x\in\mathbb{X}}$ is a \emph{$\widetilde{r}$-assisted black-box QPD configuration} for the black-box composition $\bbcompo_{t=1}^{T} \; \Tilde{R}^{(t)}$, if it can be constructed by
\begin{align}
\label{eq_QPD_with_r_blackbox}
  \bbcompo_{t=1}^{T} \; \Tilde{R}^{(t)} 
  =
  \gamma_{\mathcal{Q}}\;\sum_{x\in \mathbb{X}}p_x \sum_{m} s_{x}(m) \widetilde{K}_{x,m}^{\bbcompo}
  \;\;\text{ with }\;\;
  \widetilde{K}_{x,m}^{\bbcompo}
  :=
  \bbra{M_{m|x}}  (\bbcompo_{t=1}^{T}\;\Tilde{F}_{x}^{(t)}  ) \circ\tilde{r}\circ\Tilde{F}^{(0)}_{x},
\end{align}
where $\gamma_{\mathcal{Q}}$ is the \emph{sampling overhead} of $\mathcal{Q}$ that normalizes the RHS of Eq~\eqref{eq_QPD_with_r_blackbox} to a CPTP map. The optimal $\gamma_{\mathcal{Q}}$ over the free maps is denoted by
\wjy{%
\begin{align}
  \gamma_{\mathbb{F}}\left(
    \Tilde{r}\rightarrow
    \bbcompo_{t}\widetilde{R}^{(t)}
  \right) :=
  \inf \left\{ \gamma_{\mathcal{Q}}  :
    \text{$\mathcal{Q}(\widetilde{r}\rightarrow \bbcompo_{t}\widetilde{R}^{(t)})$ is a $\widetilde{r}$-assisted QPD for $\bbcompo_{t}\widetilde{R}^{(t)}$ with $\widetilde{F}_{x}^{(t)}\in\mathbb{F},\;\forall x,t$}
  \right\}.
\end{align}
}%
\end{definition}

A general lower bound can be established by observing that a resource-assisted black-box QPD can simulate the sequential composition of quantum channels
\(
\bigcirc_{t} \bigl( \widetilde{\mathcal{E}}^{(t)}\widetilde{R}^{(t)} \bigr)
\)
for any given set of interleaving channels $\{\widetilde{\mathcal{E}}^{(t)}\}_{t}$. Consequently, this yields a lower bound on the overhead of a resource-assisted black-box QPD over the free operations $\mathbb{F}$, stated as follows.
\begin{proposition}
\label{prop::lower_bound_black-box_QPD}
Suppose $\{\widetilde{\mathcal{R}}^{(t)}\}_{t=1,...,T}$ be a sequence of target resourceful operations.
The overhead of a resource-assisted black-box QPD is always lower-bounded by that of a fixed interleaving channel setting $\{\widetilde{\mathcal{E}}^{(t)}\}_{1 \leq t \leq T}$,
  \begin{equation}
    \gamma_{\mathbb{F}}\left(
      \widetilde{r}\rightarrow\bigcirc_{t=1}^{T}(\widetilde{\mathcal{E}}^{(t)}\widetilde{R}^{(t)})
    \right)
    \le
    \gamma_{\mathbb{F}}(\widetilde{r}\rightarrow\bbcompo_{t=1}^{T}\widetilde{R}^{(t)}),
  \end{equation}
if each channel $\widetilde{\mathcal{E}}^{(t)}$ is a free map, i.e. $\{\widetilde{\mathcal{E}}^{(t)}\}_{t} \subset \mathbb{F}$.
\begin{proof}
  This is a result of the triangle transition submultiplicativity in Lemma \ref{lemma::gamma_properties} for the transition
  \(
    \Tilde{r}
    \rightarrow
    \bbcompo_{t}\widetilde{R}^{(t)}
    \rightarrow
    \bigcirc_{t} (\widetilde{\mathcal{E}}^{(t)} \widetilde{R}^{(t)})
  \),
  \begin{equation}
    \gamma_{\mathbb{F}}\left(
      \widetilde{r}\rightarrow\bigcirc_{t}(\widetilde{\mathcal{E}}^{(t)}\widetilde{R}^{(t)})
    \right)
    \le
    \gamma_{\mathbb{F}}\left(\widetilde{r}\rightarrow\bbcompo_{t}\widetilde{R}^{(t)}\right)
    \gamma_{\mathbb{F}}\left(\bbcompo_{t}\widetilde{R}^{(t)}\rightarrow \bigcirc_{t}(\widetilde{\mathcal{E}}^{(t)}\widetilde{R}^{(t)})\right).
  \end{equation}
  Since $\widetilde{\mathcal{E}}^{(t)}$ is a free operation for all $t$, it holds that $\gamma_{\mathbb{F}}(\bbcompo_{t}\widetilde{R}^{(t)} \rightarrow \bigcirc_{t} (\widetilde{\mathcal{E}}^{(t)} \widetilde{R}^{(t)}) )=1$.
  This completes the proof.
\end{proof}
\end{proposition}

\begin{figure}[htb]
    \centering
    \includegraphics[width= 0.7\linewidth]{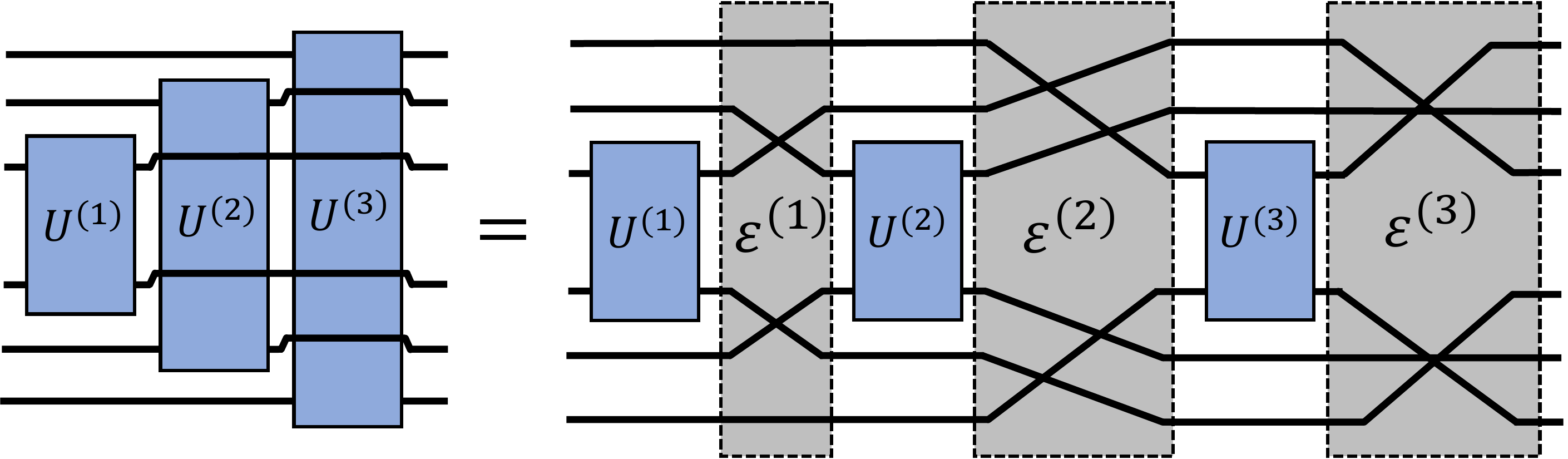}
    \caption{Parallelization of the black-box composition of unitaries via local swapping. The example illustrates the case of $T=3$, while the same construction applies to an arbitrary number of unitaries.}
    \label{fig_swap-trick}
\end{figure}

In the remaining subsections, we derive lower and upper bounds on the sampling overhead of resource-assisted black-box QPDs for unitaries, using entanglement preparation as the resource and LO or LOCC as the free operations.

\subsection{Entanglement-assisted black-box circuit knitting for Choi-stretchable unitaries}
\label{sec::EntBBQPD_CS}

Consider a sequence of unitaries $\{\widetilde{U}^{(t)}\}_{t=1}^{T}$ as the target black-box operations.
One can obtain the parallel implementation $\bigotimes_{t} \widetilde{U}^{(t)}$ via choosing local swapping operations as the interleaving quantum channels of the black-box composition $\bbcompo_{t} \widetilde{U}^{(t)}$.
The overhead of a black-box QPD for $\bbcompo_{t} \widetilde{U}^{(t)}$ can be therefore lower bounded by that of the parallel implementation $\bigotimes_{t} \widetilde{U}^{(t)}$.
Explicitly, the local operations are a sequence of qubit permutations,   \begin{equation}
  \{\widetilde{\mathcal{E}}^{(t)}\}_{t=1,...T} = \{
  \widetilde{\mathcal{E}}_{\text{swap}}^{(1,2)},\,
  \widetilde{\mathcal{E}}_{\text{swap}}^{(1,3)}\circ\widetilde{\mathcal{E}}_{\text{swap}}^{(1,2)}, \,
  ..., \, \widetilde{\mathcal{E}}_{\text{swap}}^{(1,t+1)}\circ\widetilde{\mathcal{E}}_{\text{swap}}^{(1,t)}, \, ..., \,
  \widetilde{\mathcal{E}}_{\text{swap}}^{(1,T)}
  \}.
\end{equation}
As illustrated in Fig.~\ref{fig_swap-trick} for $T=3$, the composition of operations $\{\widetilde{\mathcal{E}}^{(t)} \widetilde{U}^{(t)}\}_{t}$ parallelizes the composition of the unitaries, i.e. $\bigotimes_{t} \widetilde{U}^{(t)} = \bigcirc_{t} (\widetilde{\mathcal{E}}^{(t)} \tilde{U}^{(t)})$.
As a result, an entanglement-assisted black-box QPD is lower bounded as follows.
\begin{corollary}\label{coro::lower-bound-BB}
  Let $ \{\tilde{U}^{(t)}\}_t$ be a set of bipartite unitaries and $\Tilde{r}$ be a quantum operation as the input resource. The overhead of $\widetilde{r}$-assisted QPD is lower bounded by
  \begin{align}
    \gamma_{\mathbb{F}}\left(\Tilde{r}\rightarrow \bigotimes_{t} \kett{J_{U^{(t)}}} \right)
    \le
    \gamma_{\mathbb{F}}\left(\Tilde{r}\rightarrow \bigotimes_{t} \Tilde{U}^{(t)} \right)
    \leq
    \gamma_{\mathbb{F}}\left(\Tilde{r}\rightarrow \bbcompo_{t} \widetilde{U}^{(t)}\right).
  \end{align}
  The set of free operations can be LO or LOCC.
\end{corollary}

This corollary allows us to extend Theorem~\ref{thm::gamma-LOCC-Choi-stretchable} to entanglement-assisted QPDs for the black-box composition of Choi-stretchable unitaries.
Consider the transition
\(
\left( \widetilde{r}
\rightarrow
\bigotimes_{t} \kett{J_{U}^{(t)}}
\rightarrow
\bbcompo_{t} \widetilde{U}^{(t)} \right).
\)
One can show that the overhead of the black-box QPD of Choi-stretchable unitaries,
\(
\gamma_{\mathbb{F}}(\widetilde{r} \rightarrow \bbcompo_{t} \widetilde{U}^{(t)}),
\)
is upper bounded by the overhead required for preparing the corresponding Choi states. This overhead is optimal and coincides with the lower bound established in Corollary~\ref{coro::lower-bound-BB}.
\begin{theorem}[Resource-assisted black-box circuit knitting for Choi-stretchable unitaries]\label{thm_BB-with-Clifford}
Let $ \{\tilde{U}^{(t)}\}_t$ be a sequence of Choi-stretchable unitaries, i.e. $\gamma_{\mathrm{LOCC}}\left(\kett{J_{U^{(t)}}}\rightarrow \widetilde{U}^{(t)}\right) = 1$ for all $t$.
For any initial resourceful operation $\Tilde{r}$, the optimal black-box sampling overhead over LOCC is equivalent to the overhead of the parallel preparation of their Choi states,
\begin{align}
\label{eq::BB_QPD_Choi-stretchable}
  \gamma_{\mathrm{LOCC}} \left(\Tilde{r}\rightarrow \bigotimes_{t}\kett{J_{U^{(t)}}}  \right)
  =
  \gamma_{ \mathrm{LOCC}}\left( \Tilde{r} \rightarrow \bbcompo_{t}\Tilde{U}^{(t)} \right).
\end{align}
If $\Tilde{r} = \kett{\rho}$ is a state preparation, the optimal overhead is given by the fully entangled fraction $F_{D}(\rho)$ of $\rho$, where $D = \prod_{t} d_{U^{(t)}}$ and $d_{U^{(t)}}$ is the operator Schmidt rank of $\widetilde{U}^{(t)}$,
\begin{align}
    \gamma_{ \mathrm{LOCC} }\left( \kett{\rho}\rightarrow \bbcompo_{t}\widetilde{U}^{(t)} \right)
    = \frac{2}{F_D(\kett{\rho})} - 1.
\end{align}
\begin{proof}
See Appendix \ref{sec::proof_BB-EACK}.
\end{proof}
\end{theorem}

\subsection{Universal one-Bell-pair-assisted black-box circuit knitting}
\label{sec::EntBBQPD_Bell}

\begin{figure}[htb]
  \centering
  \includegraphics[width= 0.9 \textwidth]{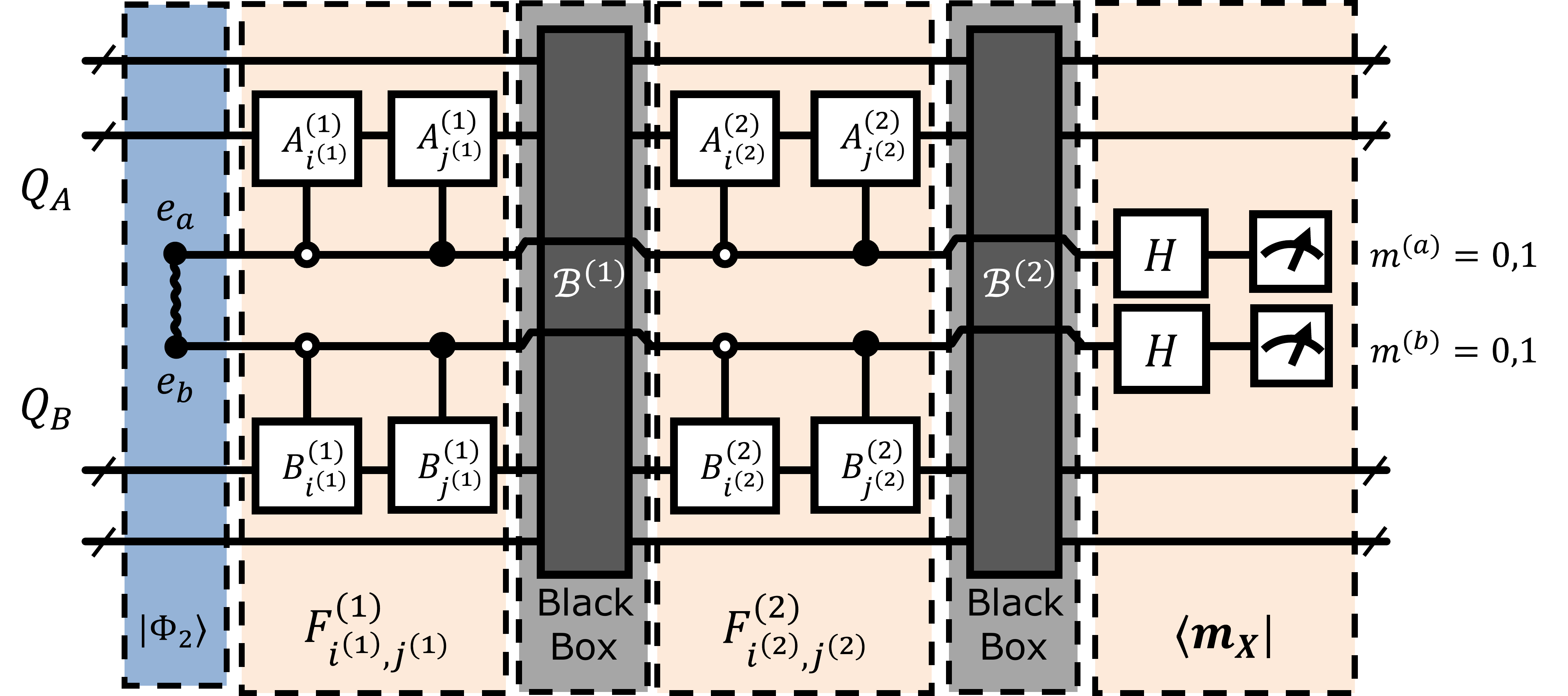}
  \caption{A black-box QPD configuration for the cross terms assisted by a Bell state. The wiggly line denotes a Bell state.}
  \label{fig::black-box-with-Bell}
\end{figure}

Suppose that at most only one two-qubit Bell state is available for each measurement shot. We can derive a universal lower bound and an upper bound on the one-Bell-pair-assisted black-box QPD for general unitaries via an extension of Section \ref{sec::one-Bell_CK} to the black-box scenario.

First, a lower bound can be derived from the Choi-state preparation of the parallel composition $\bigotimes_{t}\widetilde{U}_{t}$ according to Corollary~\ref{coro::lower-bound-BB}, which is determined by the $\Phi_{2}$-assisted QPD overhead for the Choi state $\bigotimes_{t}\kett{J_{U^{(t)}}}$.
Similar to Theorem~\ref{thm::lBound_one-Bell_CK}, one can then show that a general lower bound is characterized by the Schmidt-rank robustness~\cite{Johnston2018}.
\begin{corollary}[Lower bound on one-Bell-pair-assisted BB circuit knitting over LOCC]\label{coro::lBound_one-Bell_BB_CK}
The overhead of a one-Bell-state-assisted black-box circuit knitting of a sequence of  unitaries $\{\widetilde{U}_{t}\}_{t}$ over LOCC is lower bounded by
\begin{align}\label{eq::lBound_one-Bell_BB_CK}
  \gamma_{\mathrm{LOCC}}\left(
    \kett{ {\Phi}_2} \rightarrow \bbcompo_{t}\widetilde{U}^{(t)}
  \right)
  \geq
  \prod_{t}||\boldvec{s}^{(t)}||_{1}^2 - 1 + \left(\max\{0, 2\prod_{t}s_1^{(t)} - \prod_{t}||\boldvec{s}^{(t)}||_{1}\}\right)^2,
\end{align}
where $s_i^{(t)}$ is the Schmidt coefficient of the Choi state $\kett{J_{U^{(t)}}}$ with decreasing order.
\begin{proof}
According to Theorem \ref{thm::lBound_one-Bell_CK},
\begin{equation}
  \gamma_{\mathrm{LOCC}}\left(
    \kett{ {\Phi}_2} \rightarrow \bigotimes_{t}\kett{J_{U^{(t)}}}
  \right)
  \ge
  ||\boldvec{S}||_{1}^2 - 1 + \left(\max\{0, 2S_1 - ||\boldvec{S}^{(t)}||_{1}\}\right)^2,
\end{equation}
where $\boldvec{S} = (S_{1},S_{2},...)$ are the Schmidt coefficients of the Choi-state $\bigotimes_{t}\kett{J_{U^{(t)}}}$ sorted in decreasing order.
Since the first-norm $\|\boldvec{S}\|_{1}$ and the maximum Schmidt coefficient $S_{1}$ are both multiplicative with respect to the tensor product over the $t$, it holds
\begin{equation}
  \gamma_{\mathrm{LOCC}}\left(
    \kett{ {\Phi}_2} \rightarrow \bigotimes_{t}\kett{J_{U^{(t)}}}
  \right)
  \ge
  \prod_{t}||\boldvec{s}^{(t)}||_{1}^2 - 1 + \left(\max\{0, 2\prod_{t}s_1^{(t)} - \prod_{t}||\boldvec{s}^{(t)}||_{1}\}\right)^2,
\end{equation}
As a result of Corollary \ref{coro::lower-bound-BB}, one arrives at Eq.~\eqref{eq::lBound_one-Bell_BB_CK}.
\end{proof}
\end{corollary}

Similar to Theorem \ref{thm::one-Bell_CK_LO}, an upper bound on the overhead of the black-box QPD over LO can be derived from an explicit construction.
The black-box QPD starts with the LUD of each unitary,
\begin{equation}
  \widehat{U}^{(t)} = \sum_{i}\lambda_{i}^{(t)} \widehat{\Lambda}_{i}^{(t)}
  \;\;\text{ with }\;\;
  \widehat{\Lambda}_{i}^{(t)} = \widehat{A}_{i}^{(t)}\otimes \widehat{B}_{i}^{(t)}.
\end{equation}
Taking the example of $T =2$, the black-box composition of $\{\widetilde{U}_{1},\widetilde{U}_{2}\}$ can also be decomposed into the sum of diagonal and cross terms
\begin{equation}
  \bbcompo_{t}\widetilde{U}^{(t)}
  =
  \widetilde{\mathcal{B}}_{2}\widetilde{U}_{2}\widetilde{\mathcal{B}}_{1}\widetilde{U}_{1}
  =
  \widetilde{D}_{\mathrm{BB}} + \widetilde{C}_{\mathrm{BB}},
\end{equation}
where $\widetilde{D}_{\mathrm{BB}}$ and $\widetilde{C}_{\mathrm{BB}}$ are the diagonal and cross terms, respectively
\begin{align}
  \widetilde{D}_{\mathrm{BB}} :=
  \sum_{i^{(1)},i^{(2)}}\lambda_{i^{(1)}}^{2}\lambda_{i^{(2)}}^{2}\;
  (\widetilde{\mathcal{B}}_{2}\,
  \widetilde{\Lambda}_{i^{(2)}}^{(2)}\,
  \widetilde{\mathcal{B}}_{1}\,
  \widetilde{\Lambda}_{i^{(1)}}^{(1)})
  \;\;\text{ and }\;\;
  \widetilde{C}_{\mathrm{BB}} :=
  \sum_{(i^{(1)},i^{(2)})\neq(j^{(1)},j^{(2)})}
  \lambda_{i^{(1)}}\lambda_{i^{(2)}}\lambda_{j^{(1)}}\lambda_{j^{(2)}}\;
  \widetilde{C}_{ij},
\end{align}
where each cross term is given by
\begin{equation}
  \widetilde{C}_{\boldvec{i},\boldvec{j}} =
  \frac{1}{2}
    \left(
    \widetilde{\mathcal{B}}^{(2)}
    \circ
    (\widehat{\Lambda}_{i^{(2)}}^{(2)}\otimes\widehat{\Lambda}_{j^{(2)}}^{(2)\ast})
    \circ
    \widetilde{\mathcal{B}}^{(1)}
    \circ
    (\widehat{\Lambda}_{i^{(1)}}^{(1)}\otimes\widehat{\Lambda}_{j^{(1)}}^{(1)\ast})
    +
    \widetilde{\mathcal{B}}^{(2)}
    \circ
    (\widehat{\Lambda}_{j^{(2)}}^{(2)}\otimes\widehat{\Lambda}_{i^{(2)}}^{(2)\ast})
    \circ
    \widetilde{\mathcal{B}}^{(1)}
    \circ
    (\widehat{\Lambda}_{j^{(1)}}^{(1)}\otimes\widehat{\Lambda}_{i^{(1)}}^{(1)\ast})
  \right).
\end{equation}
The diagonal term can be implemented using LO, while the cross term $\widetilde{C}_{ij} $ can be implemented using the configuration illustrated in Fig.~\ref{fig::black-box-with-Bell}.
With the assistance of the ancillary two-qubit Bell state, one implements local control unitaries $\widetilde{F}_{i^{(t)},j^{(t)}}^{(t)}$ for each $t$.
\begin{equation}
\label{eq::free_LO_for_Bell_BB_QPD}
  \widetilde{F}_{i^{(t)},j^{(t)}}^{(t)} =
  \left(
    \widehat{\pi}_{0}^{(e_{a})}\otimes\widehat{A}_{i^{(t)}}
    +
    \widehat{\pi}_{1}^{(e_{a})}\otimes\widehat{A}_{j^{(t)}}
  \right)
  \otimes
  \left(
    \widehat{\pi}_{0}^{(e_{b})}\otimes\widehat{B}_{i^{(t)}}
    +
    \widehat{\pi}_{1}^{(e_{b})}\otimes\widehat{B}_{j^{(t)}}
  \right).
\end{equation}
In the end, one implements local $X$-basis measurements $\bbra{\boldvec{m}_{X}}$ and assign a sign function $(-1)^{|\boldvec{m}|}$ to construct the cross terms,
\begin{equation}
\label{eq::cross_term_BB_QPD}
  \widetilde{C}_{\boldvec{i},\boldvec{j}} =
  \sum_{\boldvec{m}}(-1)^{|\boldvec{m}|}
  \bbra{\boldvec{m}_{X}}
  \widetilde{\mathcal{B}}^{(2)}\,\widetilde{F}_{i^{(2)},j^{(2)}}\,
  \widetilde{\mathcal{B}}^{(1)}\,\widetilde{F}_{i^{(1)},j^{(1)}}\,
  \kett{\Phi_{2}}.
\end{equation}
The resulting QPD does not depend on the black-box channel $\{\widetilde{\mathcal{B}}^{(t)}\}_{t}$.
The configuration for $\widetilde{C}_{\boldvec{i},\boldvec{j}}$ can be directly extended to arbitrary $T$, where $\boldvec{i}=(i^{(1)}, i^{(2)},...,i^{(T)})$ and $\boldvec{j}=(j^{(1)}, j^{(2)},...,j^{(T)})$ are two vectors of classical labels with a length of $T$
\begin{equation}
\label{eq::cl_labels_BB_QPD}
  (\boldvec{i},\boldvec{j})\in
  \left(
    \bigotimes_{t=1}^{T}\mathbb{Z}_{\mathrm{rank}(U^{(t)})}, \bigotimes_{t=1}^{T}\mathbb{Z}_{\mathrm{rank}(U^{(t)})}
  \right).
\end{equation}
The associated quasi-probabilities are determined by the LUD coefficients $\{\lambda_i\}$, which yield an upper bound on the overhead of $\Phi_{2}$-assisted black-box QPD over LO, as stated in the following theorem.
\begin{theorem}[Upper bound on one-Bell-pair-assisted black-box circuit knitting over LO]\label{thm::Bell_BBCK_LO}
Let $\{\tilde{U}^{(t)}\}_{t=1}^{T}$ be a sequence of untiaries, which have the LUD coefficients $\{\boldvec{\lambda}^{(t)}\}_{t}$.
Suppose $\ket{\Phi_{2}}$ is a two-qubit Bell state.
One can construct a $\Phi_{2}$-assisted black-box QPD over LO for $\{\tilde{U}^{(t)}\}_t$ as follows,
\begin{equation}
  \bbcompo_{t}\widetilde{U}^{(t)}
  =
  \gamma_{\mathcal{Q}}
  \left(
    \sum_{\boldvec{i}}p_{\boldvec{i},\boldvec{i}}
    \;\bbcompo_{t}(\widetilde{A}_{i^{(t)}}^{(t)}\otimes\widetilde{B}_{i^{(t)}}^{(t)})
    +
    \sum_{\boldvec{i}\neq\boldvec{j}}p_{\boldvec{i},\boldvec{j}}
    \sum_{\boldvec{m}\in\mathbb{Z}_{2}^{\otimes 2}}
    (-1)^{|\boldvec{m}|}
    \bbra{\boldvec{m}_{X}} (\bbcompo_{t} \widetilde{F}_{i^{(t)},j^{(t)}}^{(t)}) \kett{\Phi_{2}}
  \right),
\end{equation}
where $\widetilde{F}_{i^{(t)},j^{(t)}}^{(t)}$ is given in Eq.~\eqref{eq::free_LO_for_Bell_BB_QPD} and illustrated in Fig.~\eqref{fig::black-box-with-Bell}, and the classical labels $(\boldvec{i},\boldvec{j})$ are sampled from the set in Eq.~\eqref{eq::cl_labels_BB_QPD} with the probability
\begin{equation}
  p_{\boldvec{i},\boldvec{j}} =
  \prod_{t=1}^{T}   \frac{\lambda_{i^{(t)}}^{(t)}\lambda_{j^{(t)}}^{(t)}}{\|\boldvec{\lambda}^{(t)}\|_{1}^{2}}.
\end{equation}
The optimum overhead over LO is upper bounded by $\gamma_{\mathcal{Q}}$, which is determined by the 1-norm of the LUD coefficients,
\begin{equation}
  \gamma_{ \mathrm{LO} }\left(
    \kett{\Phi_2} \rightarrow \bbcompo_{t}\widetilde{U}^{(t)}
  \right)
  \le
  \gamma_{\mathcal{Q}}
  =
  \prod_{t=1}^{T}\|\boldvec{\lambda}^{(t)}\|_{1}^{2}.
\end{equation}
\begin{proof}
See Appendix \ref{sec::proof_BB-EACK}.
\end{proof}
\end{theorem}

As a remark, the construction of black-box QPD through diagonal and cross terms can be also employed to derive the optimal entanglement-free black-box QPD over LO for KAK-like unitaries~\cite{Schmitt2025, Harrow2025}, in which one employ the so-called ``double Hadamard test''~\cite{Harrow2025} and similar technique~\cite{Schmitt2025} to implement the each cross term $\widetilde{C}_{\boldvec{i},\boldvec{j}}$, which requires an overhead of $2$. Replacing the $\Phi_{2}$-assisted LO-QPD for the cross terms with the entanglement-free LO-QPD, one arrives at the optimal entanglement-free black-box QPD over LO for KAK-like unitaries
\begin{equation}
\label{eq::E0-BBCK_LO}
  \gamma_{\mathrm{LOCC}}( \bbcompo_{t} \widetilde{U}^{(t)})
   =
   \gamma_{\mathrm{LO}}( \bbcompo_{t} \widetilde{U}^{(t)})
   =  2\prod_t ||\bm{\lambda}^{(t)}||_{1}^2 - 1.
\end{equation}

\subsection{Effectiveness of black-box circuit knitting}

\begin{figure}
  \centering
  \includegraphics[width=\textwidth]{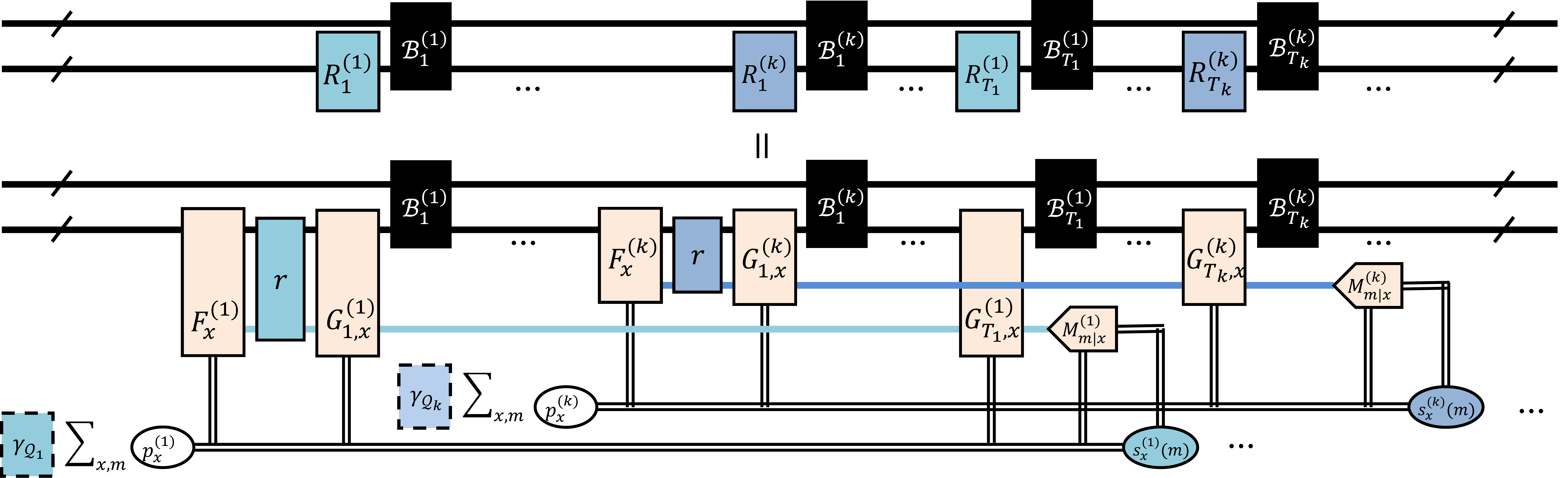}
  \caption{Multiple-copy $\Tilde{r}$-assisted black-box QPD with nested embedding construction.}
  \label{fig::embedding_BB_QPD}
\end{figure}

In general, as a result of the ``composition sub-multiplicativity'' in Lemma \ref{lemma::gamma_properties}, the black-box QPD assisted by $n$-copy resources can provide an overhead lower than the one assisted by one-copy resource,
\begin{equation}
  \gamma_{\mathbb{F}}(\Tilde{r}^{\otimes n}\rightarrow\bbcompo_{t}\widetilde{R}^{(t)})
  \le
  \gamma_{\mathbb{F}}(\Tilde{r}\rightarrow\bbcompo_{t}\widetilde{R}^{(t)})
\end{equation}
A straightforward configuration of $n$-copy $\widetilde{r}$-assisted black-box QPDs can be constructed by allocating each $\widetilde{r}$-assisted black-box QPD to a subset $\mathbb{T}_{i}$ of the target quantum channels, with the resulting overhead given as follows.
\begin{equation}
  \gamma_{\mathcal{Q}}\left(\Tilde{r}\rightarrow\bbcompo_{t\in
    \mathbb{T}_{i}}\widetilde{R}^{(t)}
  \right)
  \text{ with }
  \mathbb{T}_{i}\subseteq\{\widetilde{R}^{(i)}\}_{i=1,...,T}.
\end{equation}
By definition, the black-box configuration admits an embedding construction, originally introduced in~\cite{Wu2023} for fully entanglement-assisted DQC. As illustrated in Fig.~\ref{fig::embedding_BB_QPD}, this configuration allows each target channel to be treated as a black box, enabling a $\widetilde{r}$-assisted black-box QPD to be nested and embedded within another black-box QPD. Such an intrinsic nested embedding structure releases the requirement on the restricted embedding rules used in fully entanglement-assisted DQC~\cite{Wu2023}, thereby offering a broader range of applications.

With such a nested embedding configuration, the total overhead of an $n$-copy $\widetilde{r}$-assisted black-box QPD is upper bounded by
\begin{equation}
  \gamma_{\mathbb{F}}\left(
    \Tilde{r}^{\otimes n}\rightarrow\bbcompo_{t}\widetilde{R}^{(t)}
  \right)
  \le
  \gamma_{\mathbb{F}}\left(\bbcompo_{t\in \mathbb{T}_{0}}\widetilde{R}^{(t)} \right)
  \prod_{i=1}^{n}\gamma_{\mathcal{Q}}\left(
    \Tilde{r}\rightarrow\bbcompo_{t\in \mathbb{T}_{i}}\widetilde{R}^{(t)}
  \right)
  \;\;\text{ with }\;\;
  \bigcup_{i=0}^{n}\mathbb{T}_{i} = \{1,...,T\},
\end{equation}
where $\mathbb{T}_{0}$ is the set without allocation of resources.
Note that more advanced QPD constructions with lower overhead may jointly exploit the tensor-product resource $\widetilde{r}^{\otimes n}$, rather than using each $\widetilde{r}$ individually.

\bigskip

The effectiveness of a particular black-box QPD $\mathcal{Q}$ for a unitary $\widetilde{U}$ can be characterized via the $T$-copy regularized overhead $\bar{\gamma}_{\mathcal{Q}}^{(T)}(\widetilde{r}\rightarrow \widetilde{U})$,
\begin{equation}\label{eq::regularized_overhead_def}
  \bar{\gamma}_{\mathcal{Q}}^{(T)}(\widetilde{r}\rightarrow \widetilde{U})
  :=
  \sqrt[T]{\gamma_{\mathcal{Q}}(\widetilde{r}\rightarrow \widetilde{U}^{\bbcompo T})},
\end{equation}
\wjy{%
where $\widetilde{U}^{\bbcompo T}$ denotes the black-box composition of $T$-copy identical unitary channel $\widetilde{U}$.
\begin{equation}
  \widetilde{U}^{\bbcompo T} :=\bbcompo_{t=1}^{T}\; \widetilde{U}.
\end{equation}
}%
Taking all the channel $\widetilde{R}_{t}^{(i)} = \widetilde{U}$, for $N$ copies of the target $\widetilde{U}$, one allocates the $i$-th resource channel $\Tilde{r}$ to $T_{i}$ copies of $\widetilde{U}$.
The average overhead for each $\widetilde{U}$ under the allocation $\{T_{0}, T_{1},...,T_{n}\}$ with $N = \sum_{i}T_{i}$ is given as follows
\begin{equation}
  \bar{\gamma}_{\mathcal{Q}}
  =
  \sqrt[N]{
    \gamma_{\mathbb{F}}\left(
      \widetilde{U}^{\bbcompo T_{0}}
    \right)
    \prod_{i=1}^{n} \gamma_{\mathcal{Q}}\left(
      \Tilde{r}\rightarrow\widetilde{U}^{\bbcompo T_{i}}
    \right)
  }
  \text{ with }
  \sum_{i}T_{i} = N.
\end{equation}
The logarithm of $\bar{\gamma}_{\mathcal{Q}}$ is then the average of the log of $T_{i}$-copy regularized overhead over the allocation of $n$-copy resourceful channels $\Tilde{r}^{\otimes n}$ to $\{T_{1},...,T_{n}\}$-copy unitaries,
\begin{equation}
  \log\bar{\gamma}_{\mathcal{Q}}
  =
  \sum_{i=1}^{n} \frac{T_{i}}{N}
  \log
  \bar{\gamma}_{\mathcal{Q}}^{(T_{i})}
  ( \Tilde{r}\rightarrow\widetilde{U} )
  +
  \frac{T_{0}}{N}
  \bar{\gamma}_{\mathcal{Q}}^{(T_{0})}
  (\widetilde{U})
\end{equation}
The effectiveness of black-box QPD, characterized by the $T$-copy regularized overhead, provides a guidance on efficient allocation of the multiple copies of resources.

Here, we consider the Bell state $\ket{\Phi_{2}}$ as our resource. The $\Phi_{2}$-assisted black-box overhead in Theorem~\ref{thm::Bell_BBCK_LO} and the entanglement-free black-box overhead in Eq.~\eqref{eq::E0-BBCK_LO} both demonstrate an advantage of collectively applying a black-box QPD to the entire set $\{\widetilde{U}^{(t)}\}_{t}$, rather than implementing separate QPDs for each $\widetilde{U}^{(t)}$ sequentially.
\begin{equation}
  \bar{\gamma}_{\mathcal{Q}}^{(T)}\left(
    \kett{\Phi_{2}} \rightarrow \widetilde{U}
  \right)
  \ge
  \bar{\gamma}_{\mathcal{Q}}^{(T+1)}\left(
    \kett{\Phi_{2}} \rightarrow \widetilde{U}
  \right)
  \ge
  \|\boldvec{\lambda}\|_{1}^{2}
  \text{ for all } T.
\end{equation}
Note that the equality holds for the $\Phi_{2}$-assisted black-box QPD over LO in Theorem~\ref{thm::Bell_BBCK_LO}.
This implies that, within this construction, a single Bell state is as efficient, in terms of sampling overhead, as repeatedly using $\ket{\Phi_{2}}$ to implement each unitary individually.
It means that achieving further overhead reduction via multiple uses of Bell states would require surpassing the threshold overhead established by the $\Phi_{2}$-assisted LO-QPD in Theorems~\ref{thm::one-Bell_CK_LO} and~\ref{thm::Bell_BBCK_LO}, which is determined by the $\ell_{1}$-norm of the LUD coefficients
\begin{equation}
  \bar{\gamma}_{\Phi_{2}-\mathrm{LO}}^{(N)}\left(
    \kett{\Phi_{2}}\rightarrow\widetilde{U}
  \right)
  =
  \gamma_{\Phi_{2}-\mathrm{LO}}\left(\kett{\Phi_{2}}\rightarrow\widetilde{U}\right)
  =
  \|\boldvec{\lambda}\|_{1}^{2}
\end{equation}

It is worth noting that, in the large-$N$ limit, the $T$-copy regularized overhead of the entanglement-free QPD over LOCC in Eq.~\eqref{eq::E0-BBCK_LO} asymptotically approaches that of the $\Phi_{2}$-assisted QPD over LO.
\begin{equation}
  \lim_{T\rightarrow\infty}\Bar{\gamma}_{\mathrm{LOCC}}^{(T)}(\widetilde{U})
  =
  \gamma_{\Phi_{2}-\mathrm{LO}}\left(\kett{\Phi_{2}}\rightarrow\widetilde{U}\right).
\end{equation}

On the other hand, if the $T$-copy regularized overhead of a QPD configuration $\mathcal{Q}$ satisfies
\begin{equation}
  \bar{\gamma}_{\mathcal{Q}}^{(T)}\left(
    \kett{\Phi_{2}} \rightarrow \widetilde{U}
  \right)
  <
  \bar{\gamma}_{\mathcal{Q}}^{(T+1)}\left(
    \kett{\Phi_{2}} \rightarrow \widetilde{U}
  \right)
  \le \|\boldvec{\lambda}\|_{1}^{2},
\end{equation}
then multiple uses of the $\Phi_{2}$-assisted QPD configuration yield a strictly smaller overhead.
To achieve such a reduction, one can combine the Choi-stretchable QPD and the $\Phi_{2}$-LO QPD as Corollary \ref{coro::one-Bell_CK_LOCC}.

\begin{figure}[htbp]
  \centering
  \def\offsetleft{-0.8}
  \begin{tikzpicture}
      \node at (0,0) {\includegraphics[trim={1cm 0cm 1cm 0cm }, clip , width= 0.99\linewidth]{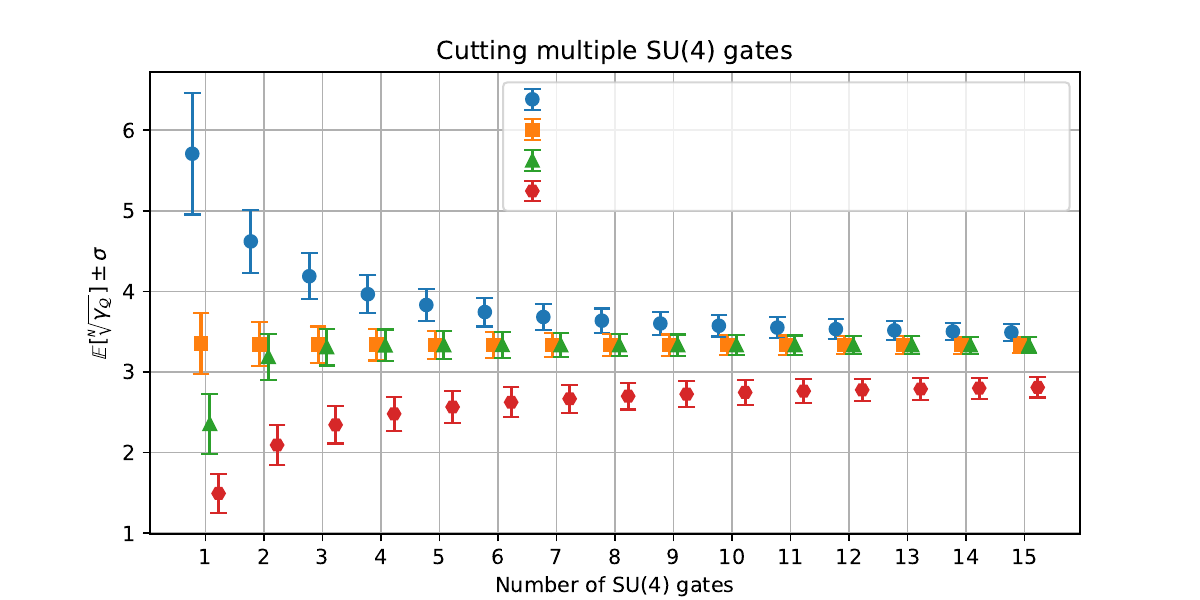}}; 
      \node[anchor=south west] at (\offsetleft, 3) {$\bar{\gamma}_{\text{LOCC}}^{(n)}(\widetilde{U})$: ENT-free, exact value, Eq. \ref{eq::E0-BBCK_LO} \cite{Schmitt2025, Harrow2025}};
      \node[anchor=south west] at (\offsetleft, 2.5) {$\bar{\gamma}_{\text{LO}}^{(n)}(\Phi_{2}\rightarrow\widetilde{U})$: $\Phi_{2}$-assisted, upper bound, Thm. \ref{thm::Bell_BBCK_LO}};
      \node[anchor=south west] at (\offsetleft, 2) {$\bar{\gamma}_{\text{LOCC}}^{(n)}(\Phi_{2}\rightarrow\widetilde{U})$: $\Phi_{2}$-assisted, lower bound, Coro. \ref{coro::lBound_one-Bell_BB_CK}};
      \node[anchor=south west] at (\offsetleft, 1.5) {$\bar{\gamma}_{\text{LOCC}}^{(n)}(\Phi_{2}\rightarrow\widetilde{U})$: $\Phi_{2}$-assisted, lower bound, Prop. \ref{prop::r-qpd_lower_bound_by_ent} \;};
  \end{tikzpicture}
  \caption{$N$-copy regularized overhead $\bar{\gamma}_{\mathcal{Q}}^{(N)}$ evaluated from $10^6$ randomly generated samples of $N$ two-qubit gates, with $N$ ranging from $1$ to $15$. The results compare the averaged overhead values obtained from different upper and lower bounds in the black-box cutting scenario. }
  \label{fig::bb_QPD_overhead}
\end{figure}

We numerically evaluate the $N$-copy regularized overhead of the resource-assisted black-box QPD by randomly generating $N$ two-qubit unitaries $\widetilde{U}$ as the target operations in the black-box composition. Figure~\ref{fig::bb_QPD_overhead} compares the corresponding overheads obtained from different upper and lower bounds.
A significant reduction in the regularized overhead is observed when a single Bell-pair resource (orange squares) is introduced, compared to the entanglement-free circuit knitting case (blue dots). The QPD in Theorem~\ref{thm::gamma-LOCC-Choi-stretchable} and Corollary \ref{coro::one-Bell_CK_LOCC} yields an overhead that lies between that of the $\Phi_{2}$-assisted QPD over LO (Theorem~\ref{thm::one-Bell_CK_LO}, orange squares) and the lower bound established in Theorem~\ref{thm::lBound_one-Bell_CK} (green triangles).
Since two-qubit unitaries are all KAK-like, the upper and lower bounds have a gap given in Eq.~\eqref{eq::LO-LOCC_gap_KAK-like}.
As the number of gate copies increases, the gap in regularized overheads diminishes, and all three QPD configurations converge toward that of the $\Phi_{2}$-assisted QPD over LO.


\bigskip


\section{The tradeoff between entanglement cost and sampling overhead in DQC}
\label{sec::tradeoff_gamma_ENT}

\wjycomment{Add literatures for stochastic distillation.}

From another perspective, beyond reducing sampling overhead, our QPD framework also improves entanglement efficiency in distributed quantum computing (DQC). In practical DQC architectures, entanglement distribution is inherently noisy. \wjy{With the so-called repeat-until-success entanglement generation technique~\cite{Lim2005, Lim2006}, or via stochastic distillation~\cite{Horodecki1999, Rozpedek2018, Li2021, Ku2022, Fang2025}, noisy entanglement distribution can be filtered or distilled into a near-perfect Bell state with a heralded success probability.}
Suppose that an entanglement distribution channel successfully distributes a perfect Bell pair with a probability $p_{\mathrm{ENT}}$. Then, for an entanglement distribution channel equipped with $M$ quantum memories, the probability of distributing exactly $e$ entangled pairs is
\begin{equation}
  p_{e} = (1-p_{ENT})^{M-e}p_{ENT}^{e}.
\end{equation}
Such a DQC architecture naturally allows adaptive mixing of $e$-Bell-pair-assisted circuit-knitting protocols, conditioned on the heralded entanglement preparation. In each round of entanglement distribution, regardless of how many entangled pairs are successfully generated, one can always employ our constructive entanglement-assisted circuit-knitting protocols in Theorems~\ref{thm_BB-with-Clifford} and~\ref{thm::Bell_BBCK_LO} to implement the target global unitary.

Suppose that $\mathbb{Q}=\{\mathcal{Q}_{1}, \ldots, \mathcal{Q}_{n}\}$ denotes the set of QPD configurations that can be implemented in a laboratory, given the practical limitations of the experimental setting, such as the number of quantum memories available for storing $e$ entangled pairs, or the ability to implement classical communication (CC) between two QPUs.
For each configuration $\mathcal{Q}$, the contribution to the sampling overhead arising from the use of $e$ Bell pairs is denoted by $\gamma_{e\mid\mathcal{Q}}$, which contributes to a part of the total overhead $\gamma_{\mathcal{Q}}$,
\begin{equation}
  \gamma_{\mathcal{Q}} = \sum_{e}\gamma_{e|\mathcal{Q}}.
\end{equation}
Accordingly, the frequency of using $e$ entangled pairs in the mixed QPD configuration $\mathcal{Q}$ is given by $(\gamma_{e\mid\mathcal{Q}} / \gamma_{\mathcal{Q}})$.

For a given entanglement distribution channel, the entanglement distribution probabilities $\{p_{e}\}_{e}$ are fixed. One can then mix all available QPD configurations with probabilistic weights $\{p_{\mathcal{Q}}\}_{\mathcal{Q}\in\mathbb{Q}}$ to minimize the total overhead of DQC. It results in an effective QPD configuration with the following total overhead
\begin{equation}
  \gamma_{\mathrm{tot}}
  =
  \min_{\{p_{\mathcal{Q}}:\mathcal{Q}\in\mathbb{Q}\}}
  \sum_{\mathcal{Q}\in\mathbb{Q}} p_{\mathcal{Q}}\,\gamma_{\mathcal{Q}},
\end{equation}
subject to the constraint that the mixed QPD configuration is implementable under the given entanglement distribution statistics.
Specifically, let $q_{e}$ denote the fraction of shots in the total effective QPD configuration that require $e$ Bell pairs, given by
\begin{equation}
  q_{e}
  =
  \sum_{\mathcal{Q}\in\mathbb{Q}}
  \frac{\gamma_{e\mid\mathcal{Q}}}{\gamma_{\mathrm{tot}}}\, p_{\mathcal{Q}}.
\end{equation}

\wjy{
Let us sort $(p_{e})_{e=M,...,1,0}$ and $(q_{e})_{e=M,...,1,0}$ in descending order.
The difference $\sum_{e'=e}^{M} (p_{e'}-q_{e'})$ quantifies the amount of entanglement resource that can be effectively supplied by, or degraded from, higher-entanglement resources with $e' \ge e$. To ensure that the total QPD configuration is implementable, these differences must be nonnegative for all $e$.
The implementability condition is therefore expressed as a majorization relation,
\begin{equation}
  (\ldots, q_{e}, \ldots, q_{0}) \preccurlyeq (\ldots, p_{e}, \ldots, p_{0}),
\end{equation}
which enforces that the cumulative probability of having at least $e$ Bell pairs available, $\sum_{e'=e}^{M} p_{e'}$, is larger or equal to the fraction of subcircuits that require at least $e$ Bell pairs, $\sum_{e'=e}^{M} q_{e'}$, for all $e$.}
This condition ensures that the available entanglement resources are sufficient to support the required QPD configuration.


Since the entanglement distribution is fixed, the average number of available entangled pairs is fixed and given by
\(
\bar{E} = \sum_{e} e\, p_{e}.
\)
The effective entanglement cost $E_{\mathrm{eff}}$, which quantifies the number of entangled pairs required per effective shot, is then solely determined by the total sampling overhead,
\begin{equation}
  E_{\mathrm{eff}} = \gamma_{\mathrm{tot}}^{2}\, \bar{E}.
\end{equation}
Consequently, the effective entanglement cost is minimized over the available QPD strategies $\mathbb{Q}$, whenever the total overhead $\gamma_{\mathrm{tot}}$ is minimized.
This mixing strategy enhances the efficiency of DQC under probabilistic entanglement distribution, as it eliminates the need to wait until sufficient entanglement has been accumulated to perform fully-entanglement-assisted telegates. Moreover, it also relaxes the requirements on quantum memory for storing entangled states.

\begin{figure}
  \centering
  \includegraphics[width=0.9\textwidth]{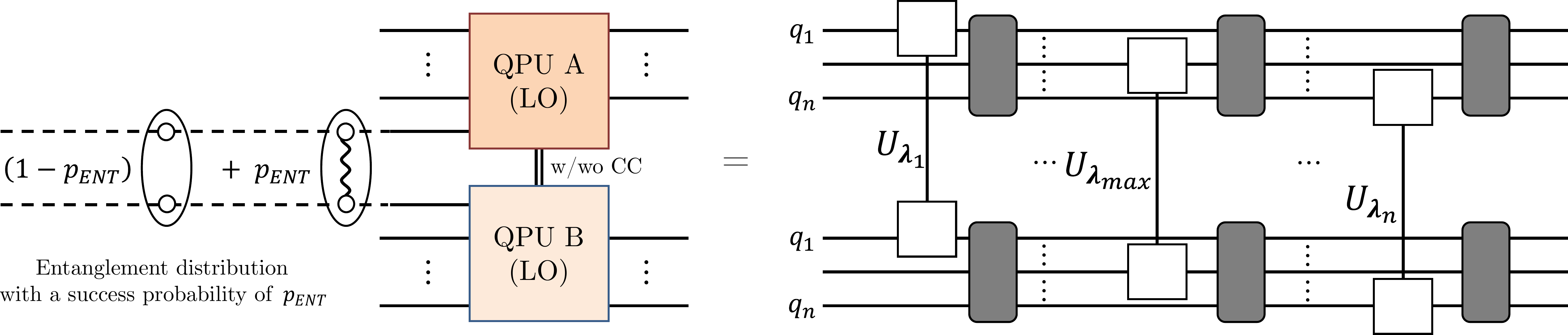}
  \caption{Distributed quantum computing (DQC) architecture with a single-qubit auxiliary entanglement-sharing channel for implementing parallel rank-2 unitary gates $\{U_i\}_{i=1}^{n}$. Each gate $U_i$ is characterized by a Schmidt coefficient vector $\boldsymbol{\lambda}_i$. The vector $\boldsymbol{\lambda}_{\max}$ denotes the Schmidt coefficient with the largest 1-norm among all gates.}\label{fig::DQC_architecture}
\end{figure}

\bigskip

As an illustrative example, consider the black-box implementation of $n$ parallel rank-2 two-qubit gates $\{U_{i}\}_{i=1,...,n}$, each characterized by Schmidt coefficients $\boldsymbol{\lambda}_{i} = (\lambda_{i,1}, \lambda_{i,2})$. As shown in Fig.~\ref{fig::DQC_architecture}, we consider two $(n+1)$-qubit QPUs, each equipped with a single auxiliary qubit for sharing entanglement between them. In each shot of the distributed implementation of the black-box composite unitary $\bbcompo_{i} \widetilde{U}_{i}$, a Bell pair is successfully established between the QPUs with probability $p_{\mathrm{ENT}}$, while with probability $1 - p_{\mathrm{ENT}}$ no entanglement is generated.
The QPD configurations for this setting should use at most one Bell pair per shot, which include the following explicit configurations:
\begin{enumerate}
  \item (LO) The QPD without entanglement and CC, which corresponds to conventional circuit knitting using LO. Its sampling overhead is determined according to Refs.~\cite{Harrow2025, Piveteau2024}.
  \item ($\Phi_{2}$-LO) The QPD with $\Phi_{2}$-assisted LO but without CC proposed in Theorem~\ref{thm::Bell_BBCK_LO}.
  \item ($\Phi_{2}$-LOCC) The QPD with $\Phi_{2}$-assisted LOCC, which is proposed in Corollary~\ref{coro::one-Bell_CK_LOCC} as a combination of $\Phi_{2}$-LO QPD in Theorem~(\ref{thm::one-Bell_CK_LO},\ref{thm::Bell_BBCK_LO}) and Choi-stretchable QPD in Theorem~(\ref{thm::gamma-LOCC-Choi-stretchable},\ref{thm_BB-with-Clifford}).
  \item ($\Phi_{2}$-Telegate) The quantum telegate with entanglement and CC employed for each instance, such as EJPP telegating~\cite{Eisert2000}. One can consume one Bell pair to implement the unitary $\widetilde{U}_{\max}$, which has the largest 1-norm Schmidt coefficient $\lVert \boldsymbol{\lambda}_{\max} \rVert_{1}^{2}$ among the target unitaries. The remaining $(n-1)$ gates are then decomposed via entanglement-free QPD. The resulting $\Phi_{2}$-assisted-telegate QPD therefore has a sampling overhead equivalent to that of entanglement-free circuit knitting applied to the remaining $(n-1)$ unitaries.
\end{enumerate}

\begin{table}
  \centering
  \begin{tabular}{|c|c|c|c|}
     \hline
     QPD Configuration $\mathcal{Q}$ & $\gamma_{0|\mathcal{Q}}$ & $\gamma_{1|\mathcal{Q}}$ & $\gamma_{\mathcal{Q}}$
     \\\hline
     LO ~\cite{Harrow2025, Piveteau2024}& $2L-1$ & $0$ & $2L-1$
     \\\hline
     $\Phi_{2}$-LO (Theorem \ref{thm::Bell_BBCK_LO}) & 1 & $L-1$ & $L$
     \\\hline
     $\Phi_{2}$-LOCC (Corollary \ref{coro::one-Bell_CK_LOCC}) & $1-c_{S}$ &
     $L-1$
     & $L-c_{S}$
     \\\hline
     $\Phi_{2}$-Telegate \cite{Eisert2000}& $0$ & $2\frac{1}{||\boldvec{\lambda}_{max}||_{1}^{2}}L-1$ & $2\frac{1}{||\boldvec{\lambda}_{max}||_{1}^{2}}L-1$       \\\hline
   \end{tabular}
  \caption{Overheads of different QPD configurations. The quantity $L:=\prod_{i}||\boldvec{\lambda}_{i}||_{1}^{2}$ is the product of the 1-norm of LUD coefficients of all target unitaries.}
  \label{tab::gamma_1QPD}
\end{table}

\begin{figure}[htbp]
  \centering
  \includegraphics[width=0.9\textwidth]{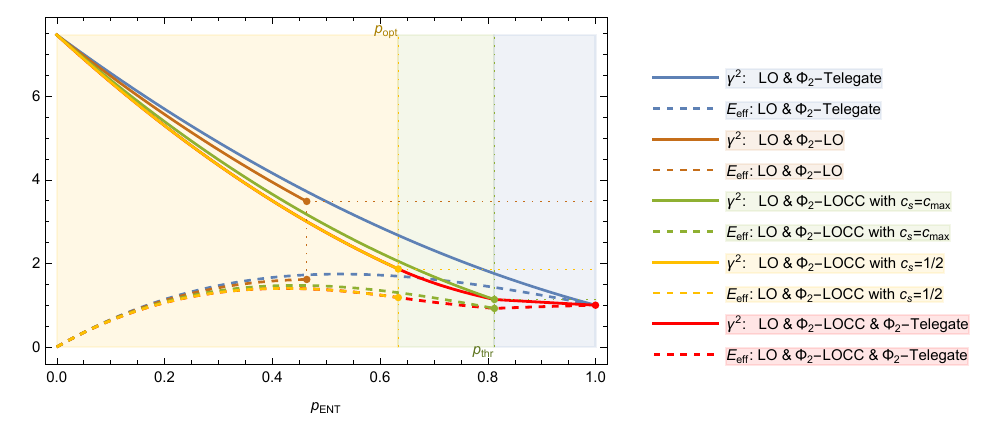}
  \caption{Trade of overhand $\gamma$ and the effective entanglement cost $E_{\mathrm{eff}}$ in the mixed QPD configuration of Algorithm \ref{algo:QPD_mixing}.}\label{fig::Mixed_QPD_statregy}
\end{figure}

The sampling overheads of these four QPD configurations are summarized in Table~\ref{tab::gamma_1QPD}, where $L := \prod_{i} \|\boldvec{\lambda}_{i}\|_{1}^{2}$ denotes the $1$-norm factor associated with the LUD coefficients in the black-box composition of the target unitaries. A naive QPD configuration under probabilistic entanglement distribution can be constructed as a mixture of the LO-QPD and the $\Phi_{2}$-Telegate QPD, with $p_{\Phi_{2}\mathrm{-Tel}} = p_{\mathrm{ENT}}$.
\begin{equation}
\label{eq::gamma_LO-ETel}
  \gamma_{tot} = (1-p_{\mathrm{ENT}})\gamma_{LO} + p_{\mathrm{ENT}}\gamma_{\Phi_{2}-\mathrm{Tel}}
  =
  (2L-1) - 2\left(1-\frac{1}{\|\boldvec{\lambda}_{\max}\|_{1}^{2}}\right)L \, p_{\mathrm{ENT}}.
\end{equation}
In Fig. \ref{fig::Mixed_QPD_statregy}, the blue solid and dashed lines plots the squared overhead $\gamma_{tot}^{2}$ and the effective entanglement cost $E_{\mathrm{eff}}$ of the (LO,$\Phi_{2}$-Tel)-mixed QPD for different $p_{\mathrm{ENT}}$, respectively.

A better configuration can be obtained from the mixture of the LO-QPD and the $\Phi_{2}$-LO-QPD.
The mixing weight for the $\Phi_{2}$-LO-QPD should be chosen such that all the successfully distributed entanglement are used for the $\Phi_{2}$-LO-QPD, which requires a mixing weight $p_{\Phi_{2}\mathrm{-LO}} = p_{\mathrm{ENT}}/p_{\mathrm{thr}}$, where $p_{\mathrm{thr}} = \gamma_{1|\mathcal{Q}}/\gamma_{\mathcal{Q}}=(L-1)/L$. It leads to a total overhead of the (LO,$\Phi_{2}$-LO)-mixed QPD
\begin{equation}
\label{eq::gamma_LO-ELO}
  \gamma_{tot} = (1-p_{\Phi_{2}\mathrm{-LO}})(2L-1) + p_{\Phi_{2}\mathrm{-LO}}\,L
  =
  (2L-1)-L\,p_{\mathrm{ENT}}.
\end{equation}
Compared with Eq.~\eqref{eq::gamma_LO-ETel}, the (LO,$\Phi_{2}$-LO)-mixed QPD always provide lower overhead, whenever $\|\boldvec{\lambda}_{\max}\|<2$.
An advantage of the (LO,$\Phi_{2}$-LO)-mixed QPD is that no classical communication is required, which simplifies its implementation.
In Fig. \ref{fig::Mixed_QPD_statregy}, the orange solid and dashed lines plot the squared overhead $\gamma_{tot}^{2}$ and the effective entanglement cost $E_{\mathrm{eff}}$ of the (LO,$\Phi_{2}$-LO)-mixed QPD for different $p_{\mathrm{ENT}}$, respectively.
Although this configuration provides lower overhead, for a success probability of entanglement greater than the threshold probability, $p_{\mathrm{ENT}}>p_{\mathrm{thr}}$, one can not improve the overhead anymore with the only LO protocol without CC.

Allowing classical communication, an advanced configuration can be obtained from the mixture of the LO-QPD and the $\Phi_{2}$-LOCC-QPD, where the Choi-stretchable weight $c_{s}$ can be chosen as any value below its maximum $c_{\max}$ obtained from Corollary \ref{coro::one-Bell_CK_LOCC}.
In general, for a Choi-stretchable weight $c_{s}$, there exist a threshold probability $p_{\mathrm{thr}}$ given by the probability of entanglement pair used in the QPD configuration, $p_{\mathrm{thr}} = \gamma_{1|\mathcal{Q}}/\gamma_{\mathcal{Q}} =(L-1)/(L-c_{s})$ above which the efficiency of entanglement is saturated and one can not reduce the overhead any more by the (LO, $\Phi_{2}$-LOCC)-mixed QPD.
For an entanglement success probability below the threshold, $p_{\mathrm{ENt}} \le p_{\mathrm{thr}}$, the (LO, $\Phi_{2}$-LOCC)-mixed QPD is mixed via $p_{\Phi_{2}\mathrm{-LOCC}} = p_{\mathrm{ENT}}/p_{\mathrm{thr}}$ probability of $\Phi_{2}$-LOCC-QPD with the Choi-stretchable weight $c_{s}$, while $(1-p_{\Phi_{2}\mathrm{-LOCC}})$ probability of $\Phi_{2}$-LOCC QPD. It leads to a total overhead
\begin{equation}
  \gamma_{tot} = (1-p_{\Phi_{2}\mathrm{-LOCC}})\gamma_{\mathrm{LO}} + p_{\Phi_{2}\mathrm{-LOCC}} \gamma_{\Phi_{2}\mathrm{-LOCC}}
  =
  (2L-1) - \left(L + \frac{(1-c_{s})c_{s}}{L-1}\right)\,p_{\mathrm{ENT}}.
\end{equation}
In Fig. \ref{fig::Mixed_QPD_statregy}, the green solid and dashed lines plot the overhead and effective entanglement cost of the (LO, $\Phi_{2}$-LOCC)-mixed QPD with the Choi-stretchable weight $c_{s} = c_{\max}$.

For a general demonstration, the $c_{\max}$ in Fig. \ref{fig::Mixed_QPD_statregy} is chosen as $c_{\max}>1/2$. One can see that for $c_{\max}>1/2$, it is possible to choose $c_{s}=1/2$ to achieve the lowest overhead up to a threshold probability $p_{\mathrm{opt}} = (L-1)/(L-1/2)$ via mixing $\Phi_{2}$-LOCC QPD and LO-QPD with the weight $p_{\Phi_{2}\mathrm{-LOCC}} = p_{\mathrm{ENT}}/p_{\mathrm{opt}}$ and $1-p_{\Phi_{2}\mathrm{-LOCC}}$, respectively
\begin{equation}
  \gamma_{tot} = (1-p_{\Phi_{2}\mathrm{-LOCC}})\gamma_{\mathrm{LO}} + p_{\Phi_{2}\mathrm{-LOCC}} \gamma_{\Phi_{2}\mathrm{-LOCC}}
  =
  (2L-1) - \left(L + \frac{1}{4(L-1)}\right)\,p_{\mathrm{ENT}}.
\end{equation}
The yellow solid and dashed lines show the overhead and effective entanglement cost of this QPD configuration up to $p_{\mathrm{opt}}$.
This region is highlighted with yellow background.

For $p_{\mathrm{opt}}<p_{\mathrm{ENT}}\le p_{\mathrm{thr}}$, one can further reduce the overhead via the $\Phi_{2}$-LOCC QPD choosing the Choi-stretchable weight according to the success probability, $c_{s} = L-(L-1)/p_{\mathrm{ENT}}$, up to the threshold probability $p_{\mathrm{thr}} = (L-1)/(L-c_{\max})$. It leads to an overhead
\begin{equation}
  \gamma_{tot} = L-c_{s} = \frac{1}{p_{\mathrm{ENT}}}(L-1).
\end{equation}
This region is highlighted with green background in Fig. \ref{fig::Mixed_QPD_statregy}.

For $p_{\mathrm{ENT}} > p_{\mathrm{thr}}$, one can not use the additional entanglement resource to improve the overhead using the $\Phi_{2}$-LOCC QPD.
If the $\Phi_{2}$-Telegate QPD provides an overhead for $p_{\mathrm{ENT}}=1$ lower than the one of $\Phi_{2}$-Telegate QPD,
i.e. $L(2/\|\boldvec{\lambda}_{\max}\|_{1}^{2}-1)<(1-c_{\max})$,
one can then further combine $\Phi_{2}$-LOCC QPD and $\Phi_{2}$-Telegate QPD with the mixing weight $p_{\Phi_{2}\mathrm{-LOCC}} = (1-p_{\mathrm{ENT}})/(1-p_{\mathrm{thr}})$ and $(1-p_{\Phi_{2}\mathrm{-LOCC}})$, respectively,
\begin{align}
  \gamma_{tot} & =
  p_{\Phi_{2}\mathrm{-LOCC}} \gamma_{\Phi_{2}\mathrm{-LOCC}}
  +
  (1-p_{\Phi_{2}\mathrm{-LOCC}}) \gamma_{\Phi_{2}\mathrm{-Tel}}
  \nonumber\\
  & =
  (1-p_{\mathrm{ENT}})\frac{L-c_{\max}}{1-c_{\max}}\Delta_{\gamma}
  +
  \left(
    \frac{2L}{\|\boldvec{\lambda}_{\max}\|^{2}}-1
  \right),
\end{align}
where $\Delta_{\gamma} = (1-c_{s})-(\frac{2}{\|\boldvec{\lambda}_{\max}\|_{1}^{2}}-1)L$ is the overhead difference between $\gamma_{\Phi_{2}\mathrm{-LOCC}}$ and $\gamma_{\Phi_{2}\mathrm{-Tel}}$.
This region is highlighted with blue background in Fig. \ref{fig::Mixed_QPD_statregy}.

In general, the optimal QPD configuration is a combination of (LO, $\Phi_{2}$-LOCC, $\Phi_{2}$-Telgate)-mixed QPD, which can be determined via Algorithm \ref{algo:QPD_mixing}.
In Fig. \ref{fig::Mixed_QPD_statregy}, the red solid and dashed lines show the overhead and effective entanglement cost of the optimal combination of these three QPDs.
Overall, one can see a clear tradeoff relation between the overhead and the effective entanglement cost.
The two extremum DQC protocols, circuit knitting and fully-entanglement-assisted telegating, are the two extremum cases for $p_{\mathrm{ENT}}=0$ and $p_{\mathrm{ENT}}=1$.
Our approach successfully bridges these two extremum protocols by trading off the entanglement resource and sampling overhead.

\bigskip

More concrete examples are shown in Fig. \ref{fig::ent_gamma_trade-off} for black-box composition of multiple control-rotation gates $C_{R}(\theta)^{\bbcompo n}$, with different $\theta \in \{\pi/2, 5\pi/6, \pi\}$ and gate number $n\in\{1,2,3\}$. The control-rotation is expressed as
\begin{equation}
  C_{R}(\theta) = \widehat{\pi}_{0}\otimes\widehat{\id}+\widehat{\pi}_{1}\otimes \widehat{R}(\theta),
\end{equation}
where $\widehat{R}(\theta) = \projector{0}+e^{\imI\theta}\projector{1}$ is a phase gate.
In these examples, one can observe that for small rotation angles the (LO,$\Phi_{2}$-LO)-mixed and (LO,$\Phi_{2}$-LOCC)-mixed QPD provide significant improvement of overhead and entanglement cost over the naive (LO,$\Phi_{2}$-Telegate)-mixed QPD.
For a small gate number, the classical communication significant reduces the overhead and the entanglement cost; while for a large gate number, this improvement diminishes.
This observation suggests that (LO,$\Phi_{2}$-LO)-mixed QPD is powerful enough when the entanglement resource is limited, meanwhile classical communication play a significant role, when sufficient entanglement resource is available.

\bigskip

\section{Conclusion and discussion}
\label{sec::conclusion}

This work introduces a unified framework for \emph{entanglement-assisted circuit knitting}, integrating the principles of entanglement-assisted distributed quantum computing and circuit knitting. The framework is formulated in terms of \emph{resource-assisted quasi-probability decomposition}, as defined in Definitions~\ref{def::r-assist_QPD} and~\ref{def::overhead_gamma}.
Within this setting, LOCC (or LO) are treated as free operations, while entangled-state preparation serves as assisting resources.
Several key properties of the resource-assisted QPD overhead are presented in Lemma~\ref{lemma::gamma_properties}, which form the basis for analyzing both the sampling overhead and the entanglement cost in entanglement-assisted circuit knitting.

For the circuit knitting of a general target unitary assisted with general bipartite entanglement, we obtain the lower bound on the overhead (Corollary \ref{coro::Choi_state_lower_bound} and Proposition \ref{prop::r-qpd_lower_bound_by_ent}). For Choi-stretchable unitaries, the optimal protocol is provided by the entanglement-assisted QPD of the corresponding Choi state, which leads to the optimal overhead characterized by its fully entangled fraction (Theorem~\ref{thm::gamma-LOCC-Choi-stretchable}).
This directly yields the optimal entanglement-assisted wire cutting (Corollary~\ref{cor_optimal_wire_cutting}).

For general unitaries, we have derived a lower (Theorem~\ref{thm::lBound_one-Bell_CK}) and an constructive upper bound (Theorem~\ref{thm::one-Bell_CK_LO}) on the one-Bell-pair-assisted circuit knitting.
The upper bound is constructed over LO operations and determined by the 1-norm of the LUD coefficients of the target unitary.
Although this upper bound is not tight, the gap between the upper and lower bounds is fixed and becomes negligible as the overhead increases. Remarkably, this upper bound coincides with the regularized overhead of standard entanglement-free circuit knitting.
Moreover, the one–Bell-pair-assisted LO QPD requires no classical communication, which enhances its experimental feasibility.

We further reduce the overhead by incorporating classical communication to implement Theorem~\ref{thm::gamma-LOCC-Choi-stretchable} for the Choi-stretchable components within a general unitary, integrated into the framework of Theorem~\ref{thm::one-Bell_CK_LO}. The resulting reduction is determined by the fraction (\emph{Choi-stretchable weight}) of the Choi-stretchable component (Corollary~\ref{coro::one-Bell_CK_LOCC}).

\begin{table}[htbp]
\centering
\renewcommand{\arraystretch}{1.3}
\setlength{\tabcolsep}{6pt}
\begin{tabularx}{\textwidth}{CCCCC}
\toprule
\textbf{Target Resource} & \textbf{Auxiliary Resource} & \textbf{Free Operation} & \textbf{Lower Bound} & \textbf{Upper Bound} \\
\midrule
General unitary & General bipartite entanglement & LOCC & Corollary~\ref{coro::Choi_state_lower_bound}, Proposition~\ref{prop::r-qpd_lower_bound_by_ent} & \\
\midrule
Choi-stretchable unitary & General bipartite entanglement & LOCC & Theorem~\ref{thm::gamma-LOCC-Choi-stretchable} & Theorem~\ref{thm::gamma-LOCC-Choi-stretchable} \\
\midrule
Identity channel (wire cutting) & General bipartite entanglement & LOCC & Corollary~\ref{cor_optimal_wire_cutting} & Corollary~\ref{cor_optimal_wire_cutting} \\
\midrule
General unitary & One Bell pair $\ket{\Phi_2}$ & LOCC & Theorem~\ref{thm::lBound_one-Bell_CK} & \\
\midrule
General unitary & One Bell pair $\ket{\Phi_2}$ & LO & & Theorem~\ref{thm::one-Bell_CK_LO} \\
\midrule
General unitary & Rank-2 partial entanglement & LO & & Corollary~\ref{coro::non-max_ent_QPD} \\
\midrule
General unitary & One Bell pair $\ket{\Phi_2}$ & LOCC & & Corollary~\ref{coro::one-Bell_CK_LOCC} \\
\bottomrule
\end{tabularx}
\caption{Summary of results for entanglement-assisted circuit knitting.}
\label{tab::results_nonBB}
\end{table}

\begin{table}[htbp]
\centering
\renewcommand{\arraystretch}{1.3}
\setlength{\tabcolsep}{6pt}
\begin{tabularx}{\textwidth}{CCCCC}
\toprule
\textbf{Target Resource} & \textbf{Auxiliary Resource} & \textbf{Free Operation} & \textbf{Lower Bound} & \textbf{Upper Bound} \\
\midrule
General unitaries & General bipartite entanglement & LOCC & Proposition~\ref{prop::lower_bound_black-box_QPD}, Corollary~\ref{coro::lower-bound-BB} & \\
\midrule
Choi-stretchable unitaries & General bipartite entanglement & LOCC & Theorem~\ref{thm_BB-with-Clifford} & Theorem~\ref{thm_BB-with-Clifford} \\
\midrule
General unitaries & One Bell pair $\ket{\Phi_2}$ & LOCC & Corollary~\ref{coro::lBound_one-Bell_BB_CK} & \\
\midrule
General unitaries & One Bell pair $\ket{\Phi_2}$ & LO & & Theorem~\ref{thm::Bell_BBCK_LO} \\
\bottomrule
\end{tabularx}
\caption{Summary of results for entanglement-assisted black-box circuit knitting}
\label{tab::results_BB}
\end{table}

We extend the resource-assisted circuit-knitting framework to the \emph{black-box setting} (Definition~\ref{def::r-assist_bb_QPD}), which can be viewed as a special class of quantum combs in which black-box channels are interleaved between a sequence of target resource unitaries. The results for the non–black-box setting are summarized in Table~\ref{tab::results_nonBB}, while those for the black-box setting are summarized in Table~\ref{tab::results_BB}.
A black-box QPD does not depend on the interleaving gates, which enables broader applicability to general quantum algorithms compiled over gate sets. Moreover, a black-box QPD can be applied without explicitly computing the overall unitary, which is generally intractable.

We show that the overhead of such a resource-assisted black-box QPD is lower-bounded by the overhead of the corresponding parallel composition of unitaries (Corollary~\ref{coro::lower-bound-BB}).
If all target unitaries are Choi-stretchable, the optimal overhead for resource-assisted black-box QPD is determined by the fully entangled fraction of the resource states (Theorem~\ref{thm_BB-with-Clifford}).
For general unitaries, we derive a lower bound (Theorem~\ref{coro::lBound_one-Bell_BB_CK}) and an upper bound (Theorem~\ref{thm::Bell_BBCK_LO}) on the overhead of one-Bell-pair-assisted black-box circuit knitting.
The black-box setting possesses an intrinsic embedding structure that enables arbitrary nesting and embedding of entanglement-assisted circuit-knitting blocks. This intrinsic embeddability enhances the effectiveness of entanglement-assisted black-box circuit knitting.

\begin{figure}[htbp]
  \centering
  \includegraphics[width=1\textwidth]{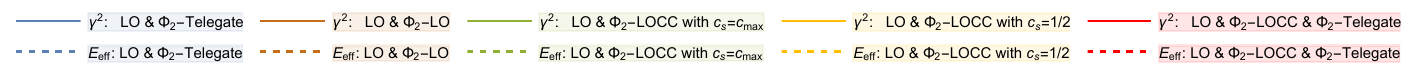}
  \\
  \subfloat[]{\includegraphics[width=1\textwidth]{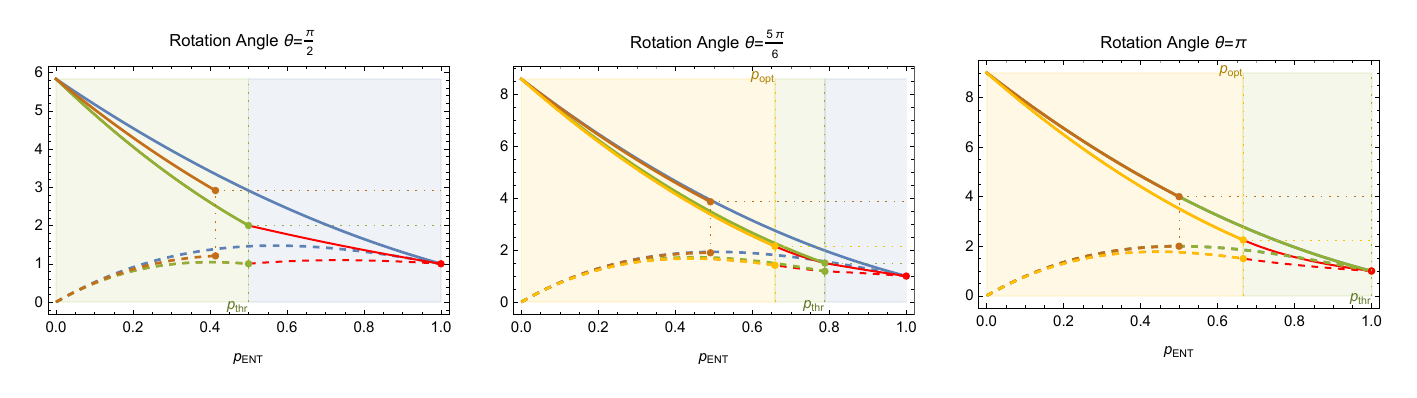}}
  \\
  \subfloat[]{\includegraphics[width=1\textwidth]{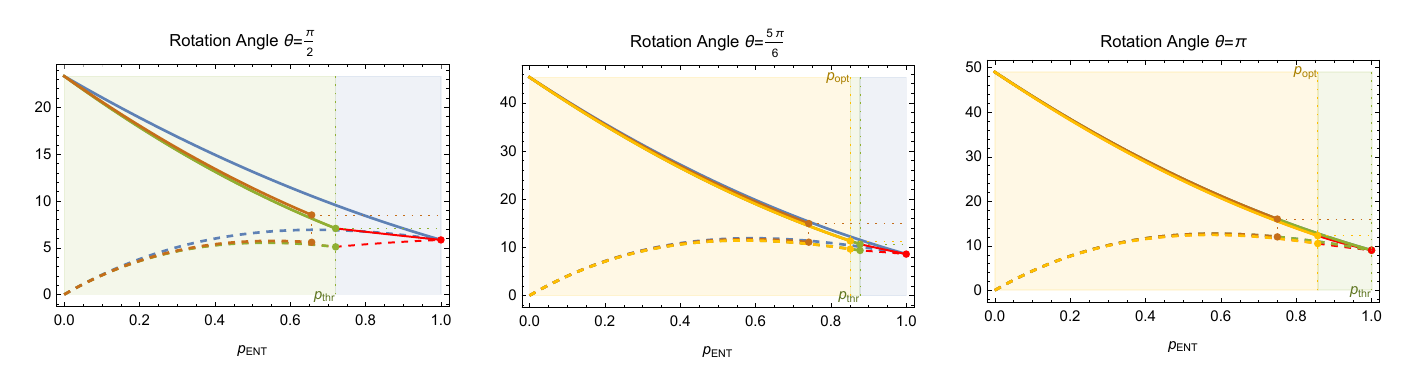}}
  \\
  \subfloat[]{\includegraphics[width=1\textwidth]{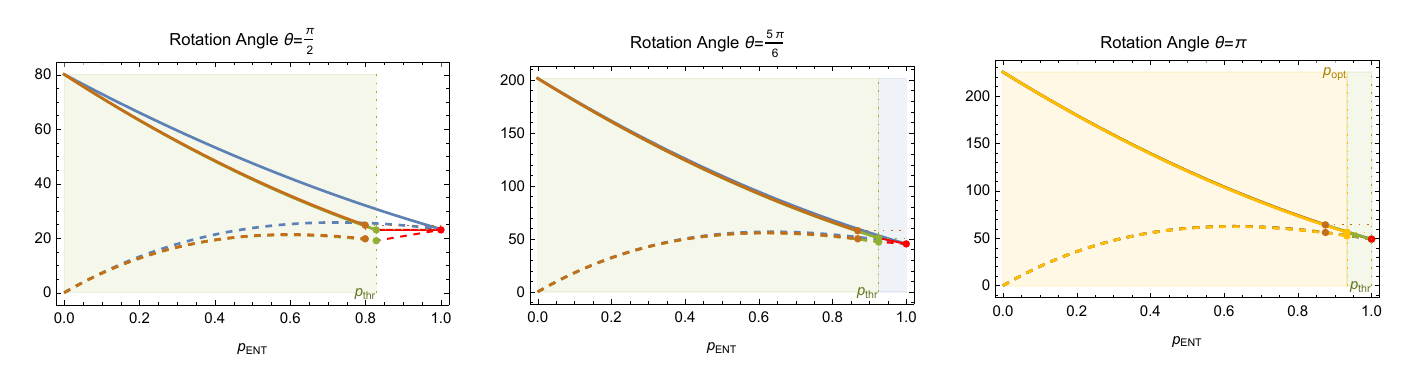}}
  \caption{Trade off between entanglement cost and sampling overhead. (a) One control-phase gate $C_{R}(\theta)$. (b) Two black-box composition of control-phase gates $C_{R}(\theta)^{\bbcompo 2}$. (c) Three black-box composition of control-phase gates $C_{R}(\theta)^{\bbcompo 3}$.}
  \label{fig::ent_gamma_trade-off}
\end{figure}

Consider $\Phi_{2}$-assisted circuit knitting, together with entanglement-free circuit knitting and fully entanglement-assisted telegating, our results identify four types of QPDs, namely LO, $\Phi_{2}$-LO, $\Phi_{2}$-LOCC, and $\Phi_{2}$-Telegate. We develop a protocol that combines these four QPDs and derive an algorithm (Algorithm~\ref{algo:QPD_mixing}) to determine the optimal mixed-QPD configuration for distributed quantum computing under dynamically probabilistic entanglement distribution.
For entanglement distribution success probabilities ranging from zero to one, we reveal a clear trade-off between entanglement cost and sampling overhead, smoothly bridging the entanglement-free circuit knitting and fully entanglement-assisted telegating (Fig.~\ref{fig::Mixed_QPD_statregy}). This behavior is demonstrated through entanglement-assisted black-box circuit knitting of multiple controlled-phase gates (Fig.~\ref{fig::ent_gamma_trade-off}).

\bigskip

These results provide a systematic framework for optimizing resource trade-offs, enabling a balanced use of entanglement cost and sampling overhead, while offering physical insight and practical guidance for scalable distributed quantum computing architectures.
Our present analysis is not yet fully general, as it is restricted to specific classes of unitaries and entanglement resources. Extending the framework to arbitrary bipartite unitaries and more general forms of entanglement remains an important direction for future work, with both conceptual and practical significance.
Another important open question concerns the potential advantages afforded by classical communication and its precise role, such as the round complexity of classical communication\cite{Chitambar2017}, in entanglement-assisted circuit knitting.
Addressing these questions will be essential for understanding the interplay between entanglement, classical communication, and sampling complexity in distributed quantum computing.

\begin{algorithm}[h!]
\setcounter{algocf}{\value{theorem}}\addtocounter{theorem}{1}
\caption{Mixed QPD configuration under probabilistic entanglement distribution}
\label{algo:QPD_mixing}
\tcp{Estimate the success probability $p_{\mathrm{ENT}}$ of entanglement distribution from successful samples $N_{ENT}$ versus total samples $N_{\mathrm{total}}$:}
$p_{\mathrm{ENT}} \gets N_{\mathrm{ENT}}/N_{\mathrm{total}}$ \;
\tcp{Calculate the LUDs $\boldvec{\lambda}_{i}$ of the target black-box-composition components $\widehat{U}_{i}$:}
$L \gets \prod_{i}\|\boldvec{\lambda}_{i}\|_{1}^{2}$\;
\tcp{Determine the maximal Choi-Stretchable weight $c_{s}$ according to Corollary \ref{coro::one-Bell_CK_LOCC}:}
$c_{\max} \gets $ Corollary \ref{coro::one-Bell_CK_LOCC}\;
\tcp{Choose a Choi-Stretchable weight $c_{s}$ for the $\Phi_{2}$-LOCC QPD via the minimum of $c_{\max}$ and $1/2$}
$c_{s}\gets \min(c_{\max},1/2)$\;
\tcp{Determine the threshold probability $p_{c}$, under which one mixes the LO QPD and the $\Phi_{2}$-LOCC QPD with the Choi-Stretchable weight $c_{s}$:}
$p_{c} \gets (L-1)/(L - c_{s})$\;
\If{$p_{\mathrm{ENT}} \le p_{c}$}{
  $(p_{\mathrm{LO}},p_{\Phi_{2}-\mathrm{LOCC}}) \gets (1-p_{\mathrm{ENT}}/p_{c},p_{\mathrm{ENT}}/p_{c})$\;
  \tcp{Construct the total QPD via mixing the LO QPD and $\Phi_{2}$-LOCC QPD:}
  \begin{align}
    \gamma_{tot} & \gets
    p_{\mathrm{LO}}\,\gamma_{\mathrm{LO}} + p_{\Phi_{2}-\mathrm{LOCC}}\,\gamma_{\Phi_{2}-\mathrm{LOCC}}(c_{s})
    \nonumber\\
    & =
    (2L-1) - \left(L + \frac{(1-c_{s})c_{s}}{L-1}\right)p_{\mathrm{ENT}}.
  \end{align}
  \tcp{The $\Phi_{2}$-LOCC QPD is constructed with the Choi-stretchable weight $c_{s}$}
}
\Else{ 
  \tcp{Determine the threshold probability for the $\Phi_{2}$-LOCC QPD with the Choi-Stretchable weight $c_{\max}$.}
  $p_{\mathrm{thr}}\gets (L-1)/(L - c_{\max})$\;
  \If{$p_{c}<p_{\mathrm{ENT}}\le p_{\mathrm{thr}}$}{
    \tcp{Choose a Choi-stretchable $c_{s}$ according to $p_{\mathrm{ENT}}$.}
    $c_{s} \gets L - (L-1)/p_{\mathrm{ENT}}$\;
    \tcp{Construct the total QPD employing the $\Phi_{2}$-LOCC QPD with the Choi-stretchable weight $c_{s}$:}
    \begin{equation}
      \gamma_{tot} \gets \gamma_{\Phi_{2}-\mathrm{LOCC}}(c_{s}) = L-c_{s} = \frac{1}{p_{\mathrm{ENT}}}(L-1)
    \end{equation}
  }
  \Else{
    \tcp{Compare the optimal overhead of $\Phi_{2}$-LOCC QPD and $\Phi_{2}$-Telegate: }
    $\Delta_{\gamma}\gets \gamma_{\Phi_{2}-\mathrm{LOCC}}(c_{\max}) - \gamma_{\Phi_{2}-\mathrm{Tel}}$\;
    \If{$\Delta_{\gamma}>0$}{
      \tcp{Allocate all failure events of entanglement distribution to $\Phi_{2}$-LOCC QPD, and reduce the overhead via assigning the remaining entanglement to the $\Phi_{2}$-Telegate QPD}
      $p_{\Phi_{2}-\mathrm{LOCC}} \gets (1-p_{\mathrm{ENT}})/(1-p_{\max})$ \;
      $p_{\Phi_{2}-\mathrm{Tel}} \gets (1-p_{\Phi_{2}-\mathrm{LOCC}})$\;
      \tcp{Construct the total QPD via mixing the $\Phi_{2}$-LOCC and $\Phi_{2}$-Telegate QPDs:}
      \begin{align}
        \gamma_{\mathrm{tot}} & \gets
        p_{\Phi_{2}-\mathrm{LOCC}}\,\gamma_{\Phi_{2}-\mathrm{LOCC}}(c_{\max})
        +
        p_{\Phi_{2}-\mathrm{Tel}}\,\gamma_{\Phi_{2}-\mathrm{Tel}}
        \nonumber\\
        & =
        (1-p_{\mathrm{ENT}})\frac{L-c_{\max}}{1-c_{\max}}\Delta_{\gamma}
        +
        \left(
          \frac{2L}{\|\boldvec{\lambda}_{\max}\|^{2}}-1
        \right)
      \end{align}
    }
    \Else{
      \tcp{The optimum QPD is provided by $\Phi_{2}$-LOCC QPD with the Choi-stretchable weight $c_{\max}$}
      \begin{equation}
        \gamma_{tot} \gets L-c_{\max}.
      \end{equation}
    }
  }
}
\end{algorithm}

\acknowledgments
We thank Dr. Chi-Kwong Li, Dr. Chung-Yun Hsieh, Mr. Lukas Schmitt, Dr. Ray-Kuang Lee and Dr. Min-Hsiu Hsieh for the helpful and insightful discussion.
This work was supported by NSTC under the Grant No. 111-2923-M-032-002-MY5, 112-2112-M-032-008-MY3, 114-2119-M-008-008 and 115-2927-I-032-501.

\appendix

\section{Proof of the rules for entanglement-assisted QPD in Lemma \ref{lemma::gamma_properties}} \label{sec::proof_gamma_properties}
The rules for sampling overheads of resource-assisted QPDs in Lemma~\ref{lemma::gamma_properties} are proved as follows.
\begin{description}
    \item[Triangle submultiplicativity] $\gamma_{\mathbb{F}}(\Tilde{A}\rightarrow\Tilde{C})\leq \gamma_{\mathbb{F}}(\Tilde{A}\rightarrow\Tilde{B})\gamma_{\mathbb{F}}(\Tilde{B}\rightarrow\Tilde{C}).$
\end{description}
\begin{proof}
Based on Eq~\eqref{eq_QPD_with_r}, suppose the optimal overhead of the QPDs for $\Tilde{A}\rightarrow \Tilde{B}$ and $\Tilde{B}\rightarrow\Tilde{C}$ are achieved by the QPD configuration $\mathcal{Q}$ and $\mathcal{Q}'$, respectively,
\begin{align}
    \Tilde{B} &= \gamma_{\mathcal{Q}}\sum_{x}p_x \sum_{a} s_{x}(a)  \bbra{M_{a|x}}\circ\Tilde{F}^{(1)}_x \circ \Tilde{A} \circ \Tilde{F}^{(0)}_{x}
    \\
    \Tilde{C} &= \gamma_{\mathcal{Q}'}\sum_{y}p'_y \sum_{b} s'_{y}(b)\bbra{W_{b|y}}\circ\Tilde{L}^{(1)}_y \circ \Tilde{B} \circ \Tilde{L}^{(0)}_{y}.
\end{align}
One can then construct a QPD for $\Tilde{A}\rightarrow\Tilde{C}$ as
\begin{align}
   \Tilde{C} &= \gamma\gamma'\sum_{x,y}p_xp'_y \sum_{a,b}s_{x}(a) s'_{y}(b)  \bbra{M_{a|x}\otimes W_{b|y}}\circ (\Tilde{L}^{(1)}_y \circ \Tilde{F}^{(1)}_x) \circ \Tilde{A} \circ (\Tilde{F}^{(0)}_{x} \circ \Tilde{L}^{(0)}_{y}).
\end{align}
As one explicitly finds the QPD configuration, the minimum QPD of $\widetilde{A}\rightarrow \widetilde{C}$ is upper bounded by
\begin{equation}
  \gamma_{\mathbb{F}}(\Tilde{A}\rightarrow\Tilde{C})
  \le
  \gamma_{\mathcal{Q}}\gamma_{\mathcal{Q}'}
  =
  \gamma_{\mathbb{F}}(\Tilde{A}\rightarrow\Tilde{B})\gamma_{\mathbb{F}}(\Tilde{B}\rightarrow\Tilde{C}).
\end{equation}
\end{proof}

\begin{description}
\item[Ordering] $\gamma_{\mathbb{F}}(\Tilde{A}\rightarrow\Tilde{B})=1 \Rightarrow \gamma_{\mathbb{F}}(\Tilde{A}) \geq \gamma_{\mathbb{F}}(\Tilde{B}).$
\end{description}
\begin{proof}
With the triangle submultiplicativity, we can take $\gamma_{\mathbb{F}}(\Tilde{f}\rightarrow\Tilde{B})\leq \gamma_{\mathbb{F}}(\Tilde{f}\rightarrow\Tilde{A})\gamma_{\mathbb{F}}(\Tilde{A}\rightarrow\Tilde{B})$, in which $\Tilde{f}\in\mathbb{F}$.
If $\gamma_{\mathbb{F}}(\Tilde{A}\rightarrow\Tilde{B})=1$, the inequality becomes $\gamma_{\mathbb{F}}(\Tilde{B})=\gamma_{\mathbb{F}}(\Tilde{f}\rightarrow\Tilde{B})\leq \gamma_{\mathbb{F}}(\Tilde{f}\rightarrow\Tilde{A})=\gamma_{\mathbb{F}}(\Tilde{A})$.
Note that the "$\Leftarrow$" direction does not hold in general.
\end{proof}

\begin{description}
  \item[Composition submultiplicativity for output channels] $\gamma_{\mathbb{F}}(\widetilde{A}\otimes\widetilde{B} \rightarrow \widetilde{C}\circ\widetilde{D})
    \leq \gamma_{\mathbb{F}}(\widetilde{A} \rightarrow \widetilde{C}) \cdot \gamma_{\mathbb{F}}(\widetilde{B} \rightarrow \widetilde{D}).$
\end{description}
\begin{proof}
Suppose the optimal overhead of the QPDs for $\Tilde{A}\rightarrow \Tilde{B}$ and $\Tilde{B}\rightarrow\Tilde{D}$ are achieved by the QPD configuration $\mathcal{Q}$ and $\mathcal{Q}'$, respectively,
\begin{align}
    \Tilde{C} &= \gamma_{\mathcal{Q}}\sum_{x}p_x \sum_{a} s_{x}(a)  \bbra{M_{a|x}}\, \Tilde{F}^{(1)}_x \, \Tilde{A} \, \Tilde{F}^{(0)}_{x}
    \\
    \Tilde{D} &= \gamma_{\mathcal{Q}'}\sum_{y}p'_y \sum_{b} s'_{y}(b)\bbra{W_{b|y}} \, \Tilde{L}^{(1)}_y \, \Tilde{B} \, \Tilde{L}^{(0)}_{y}.
\end{align}
We can then construct a QPD for $\Tilde{A}\otimes \Tilde{B} \rightarrow \Tilde{C} \circ \Tilde{D}$ as
\begin{align}
   \Tilde{C} \circ \Tilde{D}
   &= \gamma_{\mathcal{Q}}\gamma_{\mathcal{Q}'}
   \sum_{x,y}p_x\,p'_y
   \sum_{a,b}s_{x}(a) s'_{y}(b)
   \bbra{M_{a|x}\otimes W_{b|y}} (\Tilde{L}^{(1)}_y \circ \Tilde{F}^{(1)}_x) \, (\Tilde{A} \otimes \Tilde{B}) \, (\Tilde{L}^{(0)}_{y} \circ \Tilde{F}^{(0)}_{x}).
\end{align}
This concrete QPD upper bounds the $\mathbb{F}$-QPD overhead for $\widetilde{A} \otimes \widetilde{B} \rightarrow \widetilde{C}\circ\widetilde{D}$
\begin{equation}
  \gamma_{\mathbb{F}}(\widetilde{A} \otimes \widetilde{B} \rightarrow \widetilde{C}\circ\widetilde{D}) \le
  \gamma_{\mathcal{Q}}\gamma_{\mathcal{Q}'}
  =
  \gamma_{\mathbb{F}}(\widetilde{A}\rightarrow\widetilde{C})
  \gamma_{\mathbb{F}}(\widetilde{B}\rightarrow\widetilde{D}).
\end{equation}
\end{proof}

\begin{description}
  \item[Parallel-composition submultiplicativity] $\gamma_{\mathbb{F}}(\widetilde{A}\otimes\widetilde{B} \rightarrow \widetilde{C}\otimes\widetilde{D})
    \leq \gamma_{\mathbb{F}}(\widetilde{A} \rightarrow \widetilde{C}) \cdot \gamma_{\mathbb{F}}(\widetilde{B} \rightarrow \widetilde{D}).$
\end{description}
\begin{proof}
  This is a special case of the composition submultiplicativity, where $\widetilde{C}\otimes\widetilde{D} = (\widetilde{C}\otimes\widetilde{\id})\circ(\widetilde{\id}\otimes\widetilde{D})$.
\end{proof}

\begin{description}
      \item[Output-channel convexity] $\gamma_{\mathbb{F}}(\widetilde{A} \rightarrow p\widetilde{B}+(1-p)\widetilde{C})
    \leq p\gamma_{\mathbb{F}}(\widetilde{A} \rightarrow \widetilde{B})+(1-p)\gamma_{\mathbb{F}}(\widetilde{A} \rightarrow \widetilde{C}) $ for all $ p\in [0,1]$.
\end{description}
\begin{proof}
Suppose the optimal overhead of the QPDs for $\widetilde{A}\rightarrow \widetilde{B}$ and $\widetilde{A}\rightarrow\widetilde{C}$ are achieved by the QPD configuration $\mathcal{Q}$ and $\mathcal{Q}'$, respectively,
\begin{align}
    \Tilde{B} &= \gamma_{\mathcal{Q}}\sum_{x}p_x \sum_{a} s_{x}(a)  \bbra{M_{a|x}}\, \Tilde{F}^{(1)}_x \, \Tilde{A} \, \Tilde{F}^{(0)}_{x}
    \\
    \Tilde{C} &= \gamma_{\mathcal{Q}'}\sum_{y}p'_y \sum_{b} s'_{y}(b)\bbra{W_{b|y}} \, \Tilde{L}^{(1)}_y \, \Tilde{A} \, \Tilde{L}^{(0)}_{y}.
\end{align}
One can obtain a QPD of $\left(\widetilde{A}\rightarrow(p\Tilde{B}+(1-p)\Tilde{C})\right)$ for all $p \in [0,1]$,
\begin{align}
    p\Tilde{B}+(1-p)\Tilde{C} =& p\left( \gamma_{\mathcal{Q}}\sum_x p_x \sum_{a} s_{x}(a) \bbra{M_{a|x}} \, \Tilde{F}^{(1)}_x  \, \Tilde{A} \, \Tilde{F}^{(0)}_x \right) \notag\\
    +&(1-p)\left(\gamma_{\mathcal{Q}'}\sum_{y} p'_y \sum_{b} s'_y(b) \bbra{W_{b|y}} \, \Tilde{G}^{(1)}_x  \,  \Tilde{A}  \,  \Tilde{G}^{(0)}_x\right).
\end{align}
This QPD gives us a sampling upper bound
\begin{equation}
  \gamma_{\mathbb{F}}\left(\widetilde{A} \rightarrow p\widetilde{B}+(1-p)\widetilde{C}\right)
  \le
  p\gamma_{\mathcal{Q}} + (1-p)\gamma_{\mathcal{Q}}
  =
  p\gamma_{\mathbb{F}}(\widetilde{A} \rightarrow \widetilde{B})+(1-p)\gamma_{\mathbb{F}}(\widetilde{A} \rightarrow \widetilde{C}).
\end{equation}
\end{proof}

\begin{description}
    \item[Input-resource concavity] $\gamma_{\mathbb{F}}(p\widetilde{A}+(1-p) \widetilde{B} \rightarrow \widetilde{C}) \ge p\gamma_{\mathbb{F}}(\widetilde{A}\rightarrow \widetilde{C}) + (1-p)\gamma_{\mathbb{F}}(\widetilde{B} \rightarrow \widetilde{C})$ for all $ p\in [0,1]$.
\end{description}
\begin{proof}
Suppose the optimal overhead of the QPDs for $\widetilde{A}\rightarrow \widetilde{B}$ and $\widetilde{B}\rightarrow\widetilde{C}$ are achieved by the QPD configuration $\mathcal{Q}$ and $\mathcal{Q}'$, respectively,
\begin{align}
    \Tilde{C} &= \gamma_{\mathcal{Q}}\sum_{x}p_x \sum_{a} s_{x}(a)  \bbra{M_{a|x}}\, \Tilde{F}^{(1)}_x \, \Tilde{A} \, \Tilde{F}^{(0)}_{x}
    \;\;\text{ and }\;\;
    \Tilde{C} &= \gamma_{\mathcal{Q}'}\sum_{y}p'_y \sum_{b} s'_{y}(b)\bbra{W_{b|y}} \, \Tilde{L}^{(1)}_y \, \Tilde{B} \, \Tilde{L}^{(0)}_{y}.
\end{align}
We can prepare the resource channel $\widetilde{A}$ and $\widetilde{B}$ with the probability $p$ and $(1-p)$, respectively.
A QPD of $\widetilde{C}$ via be then constructed via mixing the $\widetilde{A}$-assisted QPD and $\widetilde{B}$-assisted QPD
\begin{equation}
\label{eq::resource_convacity_proof}
  \widetilde{C} =
  p\left( \gamma_{\mathcal{Q}}\sum_x p_x \sum_{a} s_{x}(a) \bbra{M_{a|x}} \, \Tilde{F}^{(1)}_x  \, \Tilde{A} \, \Tilde{F}^{(0)}_x \right)
  +
  (1-p)\left( \gamma_{\mathcal{Q}'}\sum_y p_y \sum_{b} s'_{y}(b) \bbra{W_{b|y}} \, \Tilde{L}^{(1)}_y  \, \Tilde{B} \, \Tilde{L}^{(0)}_y \right)
\end{equation}
This QPD overhead is optimal when each instance of the ensemble of resourceful channels is known and one can only use the each resource instantly without storage.
Here, we denote such an ensemble by $\mathbb{R}:=\{(p_{r}, \widetilde{r}): \widetilde{r} \in\{\widetilde{A}, \widetilde{B}\}, p_{A}=p\}$.
The optimum overhead of implementing $\widetilde{C}$ with assistance of this ensemble is therefore given by
\begin{equation}
  \gamma_{\mathbb{F}}(\mathbb{R}\rightarrow\widetilde{C})
  =
  p\gamma_{\mathbb{F}}(\widetilde{A}\rightarrow\widetilde{C}) +
  (1-p)\gamma_{\mathbb{F}}(\widetilde{B}\rightarrow\widetilde{C}).
\end{equation}
Since the mixed resource $\widetilde{r}_{mix} := p\widetilde{A} + (1-p)\widetilde{B}$ can be obtained from the ensemble $\mathbb{R}$ by forgetting the label of each instance, i.e. $\gamma_{\mathbb{F}}(\mathbb{R}\rightarrow\widetilde{r}_{mix})=1$, the triangle submultiplicativity implies the following the resource concavity
\begin{equation}
  \gamma_{\mathbb{F}}(p\widetilde{A}+(1-p) \widetilde{B} \rightarrow \widetilde{C}) \ge \gamma_{\mathbb{F}}(\mathbb{R}\rightarrow\widehat{C})
  =
  p\gamma_{\mathbb{F}}(\widetilde{A}\rightarrow \widetilde{C}) + (1-p)\gamma_{\mathbb{F}}(\widetilde{B} \rightarrow \widetilde{C}).
\end{equation}
\end{proof}

\begin{description}
    \item[Free-map monotonicity] For any free map $\widetilde{f}\in \mathbb{F}$, it holds
       \begin{equation}
         \max\left\{ \gamma_{\mathbb{F}}(\widetilde{A}\rightarrow \widetilde{f}\circ \widetilde{B}),\gamma_{\mathbb{F}}(\widetilde{A}\rightarrow \widetilde{B}\circ\widetilde{f}) \right\}
         \le \gamma_{\mathbb{F}}(\widetilde{A}\rightarrow \widetilde{B}) \le
         \min\left\{ \gamma_{\mathbb{F}}(\widetilde{f}\circ\widetilde{A}\rightarrow \widetilde{B}), \gamma_{\mathbb{F}}(\widetilde{A}\circ \tilde{f}\rightarrow \widetilde{B}) \right\}
       \end{equation}
\end{description}

\begin{proof}
The first inequality is a direct result of the triangle submultiplicativity of $\widetilde{A}\rightarrow \widetilde{B} \rightarrow \widetilde{B}\circ \Tilde{f}$ or $\widetilde{A}\rightarrow \widetilde{B} \rightarrow \Tilde{f}\circ\widetilde{B}$; while the second inequality is a result of the following composition rule of input resource for $\widetilde{B}$ being free,
\begin{equation}
  \gamma_{\mathbb{F}}(\widetilde{A} \rightarrow \widetilde{C})
  \le
  \gamma_{\mathbb{F}}(\widetilde{B})
  \min\left(
    \gamma_{\mathbb{F}}(\widetilde{A} \circ \widetilde{B} \rightarrow \widetilde{C}),
    \gamma_{\mathbb{F}}(\widetilde{B} \circ \widetilde{A}\rightarrow \widetilde{C})
  \right).
\end{equation}
The composition rule for input resources is proved as follows.
Let $\mathcal{Q} = \{p_{x},\widetilde{F}_{x}^{(1)},\widetilde{F}_{x}^{(0)}; s_{x},\mathcal{M}_{x}\}_{x}$ be the optimal QPD of $(\widetilde{A}\circ\widetilde{B}\rightarrow\widetilde{C})$, i.e. $\gamma_{\mathbb{F}}(\widetilde{A}\circ\widetilde{B}\rightarrow\widetilde{C}) = \gamma_{\mathcal{Q}}(\widetilde{A}\circ\widetilde{B}\rightarrow\widetilde{C})$, it holds then
\begin{equation}
  \widetilde{C} = \gamma_{\mathcal{Q}}\sum_{x}p_{x}
  \sum_{m}s_{x}(m) \bbra{M_{m|x}} \widetilde{F}_{x}^{(1)}\,(\widetilde{A}\circ\widetilde{B})\,\widetilde{F}_{x}^{(0)}.
\end{equation}
Suppose $\mathcal{Q_{B}} = \{p_{y},\widetilde{G}_{x};s'_{y}, \mathcal{W}_{y}\}_{y}$ be the optimal resource-free QPD of $\widehat{B}$, i.e. $\gamma_{\mathbb{F}}(\widetilde{B}) = \gamma_{\mathcal{Q}_{B}}(\widetilde{B})$, one can than obtain an $\widetilde{A}$-assisted QPD of $\widetilde{C}$,
\begin{equation}
  \widetilde{C} = \gamma_{\mathcal{Q}}\gamma_{\mathcal{Q}_{B}}
  \sum_{x,y}p_{x}p_{y}
  \sum_{m,b}s_{x}(m) s'_{y}(b)\bbra{M_{m|x}}\otimes \bbra{W_{b|y}} \widetilde{F}_{x}^{(1)}\;\widetilde{A}\;(\widetilde{G}_{y}\widetilde{F}_{x}^{(0)}).
\end{equation}
It determines an upper bound on $\widetilde{A}\rightarrow\widetilde{C}$,
\begin{equation}
  \gamma_{\mathbb{F}}(\widetilde{A}\rightarrow\widetilde{C})
  \le
  \gamma_{\mathcal{Q}}\gamma_{\mathcal{Q}_{B}}
  =
  \gamma_{\mathbb{F}}(\widetilde{A}\circ\widetilde{B}\rightarrow\widetilde{C})
  \gamma_{\mathbb{F}}(\widetilde{B}).
\end{equation}

\end{proof}

\section{Proofs for entanglement-assisted circuit knitting in Section \ref{sec::ent-assisted_circknit}}
\label{sec::proof_EACK}

Here, we provide a proof for Proposition~\ref{prop::r-qpd_lower_bound_by_ent}.
\begin{customprop}{\ref{prop::r-qpd_lower_bound_by_ent}}
Let $\hat{U}$ be a bipartite unitary, for any pure state $\kett{\psi}$ with the entanglement entropy $\mathscr{E}({\Phi_U}) \geq \mathscr{E}({\psi} )$, the $\psi$-assisted QPD overhead over LOCC is lower bounded by
\begin{align}
    2^{\mathscr{E}({\Phi_U})-\mathscr{E}(\psi )} \le \gamma_{\mathrm{LOCC}}(\kett{\psi}\rightarrow \Tilde{U}).
\end{align}

\begin{proof}
As a result of Corollary \ref{coro::Choi_state_lower_bound}, the lower bound is determined by the overhead for the Choi state $\gamma_{LOCC}(\kett{\psi}\rightarrow\kett{J_{U}})$.
First of all, according to the parallel-composition submultiplicativity, it holds that
\begin{equation}
\label{eq::r-qpd_lower_bound_proof_1}
  \sqrt[N]{\gamma_{\mathrm{LOCC}}\left(\kett{\psi}^{\otimes N}\rightarrow \kett{J_{U}}^{\otimes N}\right) }
  \le
  \gamma_{\mathrm{LOCC}}\left(\kett{\psi}\rightarrow \kett{J_{U}}\right)
\end{equation}
Based on the transition $\kett{\Phi_{2}}^{\otimes \lceil N \mathscr{E}(\psi)\rceil} \rightarrow \kett{\psi}^{\otimes N} \rightarrow \kett{J_{U}}^{\otimes N} \rightarrow \kett{\Phi_{2}}^{\otimes \lfloor N \mathscr{E}(J_{U})\rfloor} $, one obtains the following inequality via the triangle submultiplicativity,
\begin{align}
\label{eq::r-qpd_lower_bound_proof_2}
  & \gamma_{\mathrm{LOCC}}\left( \kett{\Phi_{2}}^{\otimes \lceil N \mathscr{E}(\psi)\rceil} \rightarrow \kett{\Phi_{2}}^{\otimes \lfloor N \mathscr{E}(J_{U})\rfloor}\right)
  \nonumber\\
  \le&
  \gamma_{\mathrm{LOCC}}\left( \kett{\Phi_{2}}^{\otimes \lceil N \mathscr{E}(\psi)\rceil} \rightarrow \kett{\psi}^{\otimes N} \right)
  \gamma_{\mathrm{LOCC}}\left(\kett{\psi}^{\otimes N} \rightarrow \kett{J_{U}}^{\otimes N} \right)
  \gamma_{\mathrm{LOCC}}\left(\kett{J_{U}}^{\otimes N} \rightarrow \kett{\Phi_{2}}^{\otimes \lfloor N \mathscr{E}(J_{U})\rfloor }\right)
\end{align}
Since entanglement entropy is the asymptotic conversion ratio between pure entangled state and the Bell states using LOCC, one has the following  asymptotic overheads,
\begin{align}
    \lim_{N\rightarrow \infty} \gamma_{\mathrm{LOCC}}\left( \kett{\Phi_2}^{\otimes \lceil N\mathscr{E}({\psi}) \rceil}\rightarrow \kett{\psi}^{\otimes N} \right)
    = \lim_{N\rightarrow \infty} \gamma_{\mathrm{LOCC}}\left( \kett{J_{U}}^{\otimes N} \rightarrow \kett{\Phi_2}^{\otimes \lfloor N\mathscr{E}({\psi}) \rfloor} \right) = 1.
\end{align}
By taking the asymptotic limit in Eq.~\eqref{eq::r-qpd_lower_bound_proof_1} and ~\eqref{eq::r-qpd_lower_bound_proof_2} , one achieves the lower bound
\begin{align}
    \gamma_{\mathrm{LOCC}}\left({\kett{\psi}} \rightarrow {\kett{{J_U}}} \right)
    &\geq \lim_{N\rightarrow \infty}\sqrt[N]{\gamma_{\mathrm{LOCC}}\left(\kett{\psi}^{\otimes N} \rightarrow \kett{J_U}^{\otimes N} \right)}
    \geq
    \lim_{N\rightarrow \infty}
    \sqrt[N]{\gamma_{\mathrm{LOCC}}\left( \kett{\Phi_{2}}^{\otimes \lceil N \mathscr{E}(\psi)\rceil} \rightarrow \kett{\Phi_{2}}^{\otimes \lfloor N \mathscr{E}(J_{U})\rfloor}\right)}
    \nonumber\\
    & =
    \lim_{N\rightarrow \infty}
    \sqrt[N]{2^{1+\lceil N \mathscr{E}(J_U)\rceil - \lfloor N \mathscr{E}(\psi) \rfloor }-1}
    = 2^{\mathscr{E}(J_U) -\mathscr{E}(\psi) }.
\end{align}
Together with Corollary \ref{coro::Choi_state_lower_bound}, one completes the proof.
\end{proof}
\end{customprop}

\bigskip

The proof of Theorem \ref{thm::lBound_one-Bell_CK} works as follows.
\begin{customthm}{\ref{thm::lBound_one-Bell_CK}}
The overhead of a one-Bell-state-assisted gate cutting of a unitary over LOCC is lower bounded by
\begin{align}
    \gamma_{\mathrm{LOCC}}(\kett{ {\Phi}_2} \rightarrow \widetilde{U}) \geq ||\boldvec{s}||_{1}^2 - 1 + \left(\max\{0, 2s_1 - ||\boldvec{s}||_{1}\}\right)^2,
\end{align}
where $s_i$ is the Schmidt coefficient of the Choi state $\kett{J_{U}}$ with decreasing order.
\begin{proof}
As a result of majorization condition for quantum state transform, it holds that $\gamma_{\mathrm{LOCC}}(\kett{\Phi_2} \rightarrow \kett{\varphi}) = 1$ for all the states $\kett{\varphi}$ that have a Schmidt rank smaller or equal to $2$. It implies that for any state $\kett{\rho_{\le 2}}$ as a mixture of rank-$r$ pure states with $r\le 2$, the transform $(\kett{\Phi_{2}}\rightarrow\kett{\rho_{2}})$ is free under LOCC, i.e. $\gamma_{\mathrm{LOCC}}(\kett{\Phi_2} \rightarrow \kett{\rho_{\le 2}}) = 1$.
The optimum overhead of $(\kett{\Phi_2} \rightarrow \kett{J_{U}})$ is therefore determined by the optimum QPD of $\kett{J_{U}}$ over all possible states $\kett{\rho_{\le 2}}$,
\begin{equation}
  \gamma_{\mathrm{LOCC}}(\kett{ {\Phi}_2} \rightarrow \kett{ J_U} )
  =
  \min_{\kett{\rho_{\le 2}^{(\pm)}}} \left\{
    (q_{+}+q_{-}):
    \kett{J_{U}} = q_{+}\kett{\rho_{\le 2}^{(+)}} - q_{-}\kett{\rho_{\le 2}^{(-)}}
  \;\;\text{ and }\;\;
  q_{+} - q_{-} =1
  \right\},
\end{equation}
which can be determined by the robustness $\mathcal{R}_{2}(\kett{J_{U}})$ of Schmidt rank $r=2$
\begin{equation}
  \gamma_{\mathrm{LOCC}}(\kett{ {\Phi}_2} \rightarrow \kett{ J_U} )
  = 1+2\mathcal{R}_{2}(\kett{J_{U}}),
\end{equation}
where $\mathcal{R}_{2}(\kett{J_{U}})$ is defined as
\begin{equation}
  \mathcal{R}_{2}(\kett{J_{U}})
  :=
  \min_{\kett{\rho_{\le 2}^{(\pm)}}} \left\{t: \kett{\rho_{\le2}^{(+)}} = \frac{1}{1+t}(\kett{J_{U}} + t \kett{\rho_{\le2}^{(-)}})\right\}.
\end{equation}
The the robustness of Schmidt rank $r=2$ of the Choi state $\kett{J_{U}}$ is explicitly determined by its Schmidt coefficient \cite{}
\begin{equation}
  \mathcal{R}_{2}(\kett{J_{U}})
  =
  \frac{1}{2}(||\boldvec{s}||_{1}^{2}-2)
  +
  \frac{1}{2}\left(\max\{0, 2s_1 - ||\boldvec{s}||_{1}\}\right)^2.
\end{equation}
Altogether, we obtain
\begin{equation}
  \gamma_{\mathrm{LOCC}}\left(\kett{\Phi_{2}}\rightarrow\widetilde{U}\right)
  \ge
  \gamma_{\mathrm{LOCC}}\left(\kett{\Phi_{2}}\rightarrow\kett{J_{U}}\right)
  \ge
  ||\boldvec{s}||_{1}^2 - 1 + \left(\max\{0, 2s_1 - ||\boldvec{s}||_{1}\}\right)^2.
\end{equation}
This completes the proof.
\end{proof}
\end{customthm}

\begin{customthm}{\ref{thm::one-Bell_CK_LO}}
Let $\hat{U}$ be a bipartite unitary with the LUD $\hat{U} = \sum_i \lambda_i \hat{A}_i\otimes \hat{B}_i$, and $\kett{\Phi_2}$ be a two-qubit maximally entangled state.
There exists a $\Phi_{2}$-assisted QPD for $\widetilde{U}$ over LO
\begin{align}
  \widetilde{U} =
  \gamma_{\mathcal{Q}} \left(
    \sum_{i}p_{i,i}(\widetilde{A}_{i}\otimes\widetilde{B}_{i})
    +
    \sum_{i,j} p_{i,j}\sum_{\bm{m} \in \{0,1\}^{\otimes2}} (-1)^{|\bm{m}|} \bbra{\boldvec{m}_{X}^{(anc.)}}\tilde{F}_{i,j}  \kett{\Phi_2},
  \right)
\end{align}
where $p_{i,j} = \frac{\lambda_i\lambda_j}{||\boldvec{\lambda}||_{1}^2}$, and $\widetilde{F}_{i,j}$ is given in Eq.~\eqref{eq::free_LO_for_Bell_QPD} and illustrated in Fig. \ref{fig::circuit-2LO}.
The overhead of the $\Phi_2$-assisted QPD over LO is then upper bounded by $\gamma_{\mathcal{Q}}$
\begin{align}
  \gamma_{\mathrm{LO}}(\kett{{\Phi}_2} \rightarrow \Tilde{U} )
  \leq
  \gamma_{\mathcal{Q}}
  =
  \|\boldvec{\lambda}\|_{1}^2.
\end{align}
\begin{proof}
  Suppose the unitary $\widehat{U}$ has a LUD of $\widehat{U} = \sum_{i=1}^{n}\lambda_{i}\widehat{\Lambda}_{i}$, where $\hat{\Lambda}_i = \hat{A}_i\otimes \hat{B}_i$ are local unitaries.
  A unitary can be expressed as
  \begin{align}
    \widetilde{U}
    & =
    \sum_{i,j}\lambda_i \lambda_j(\hat{\Lambda}_i\otimes\hat{\Lambda}_j^\ast)
    =
    \sum_{i}\lambda_i^2\widetilde{\Lambda}_i+
    \sum_{i\neq j}\lambda_i\lambda_j \widetilde{C}_{ij},
  \end{align}
  where $\widetilde{\Lambda}_i$ are the diagonal term, and $\widetilde{C}_{ij}$ are the cross terms
  \begin{equation}
    \widetilde{C}_{ij} := \frac{1}{2}(\widehat{\Lambda}_i\otimes \widehat{\Lambda}^*_j + \widehat{\Lambda}_j\otimes \widehat{\Lambda}^*_i).
  \end{equation}
  We construct two quantum channels $\Tilde{r}_{ij}^{\pm}$ from the sum of the operators $\widehat{\Lambda}_{i}$ and $\widehat{\Lambda}_{j}$
  \begin{equation}
    \Tilde{R}_{ij}^{(\pm)} :=
    \frac{1}{2}(\widehat{\Lambda}_{i} \pm \widehat{\Lambda}_{j})\otimes(\widehat{\Lambda}_{i} \pm \widehat{\Lambda}_{j})^{\ast}.
  \end{equation}
  The cross terms can expressed by
  \begin{align}
    \widetilde{C}_{ij} = \frac{1}{2}\Tilde{R}_{ij}^{(+)} - \frac{1}{2}\Tilde{R}_{ij}^{(-)}
  \end{align}
  An explicit $\Phi_{2}$-assisted QPD over LO can be then constructed in Fig. \ref{fig::circuit-2LO}.
  Fig. \ref{fig::circuit-2LO} (b) implement the diagonal term, while Fig. \ref{fig::circuit-2LO} (c) implement the cross terms.
  Under such the construction in Fig. \ref{fig::circuit-2LO} (c), we have the LO
  \begin{align}
    \widehat{F} =
    (\widehat{\pi}_{0}^{(a)}\otimes\widehat{A}_{i} + \widehat{\pi}_{1}^{(a)}\otimes \widehat{A}_{j})
    \otimes
    (\widehat{\pi}_{0}^{(b)}\otimes\widehat{B}_{i} + \widehat{\pi}_{1}^{(b)}\otimes \widehat{B}_{j}),
  \end{align}
  where $\widehat{\pi}_{0}=\projector{0}$, $\widehat{\pi}_{1}=\projector{1}$, and the local measurement operator in $X$ basis is
  \begin{align}
    \bbra{M_{+}} := \bbra{+}\otimes\bbra{+} + \bbra{-}\otimes\bbra{-}
    \;\;\text{ and }\;\;
    \bbra{M_{-}} := \bbra{+}\otimes\bbra{-} + \bbra{+}\otimes\bbra{-},
  \end{align}
  and the entanglement resource is
  \begin{equation}
    \kett{\Phi_{2}} = \frac{1}{\sqrt{2}}(\ket{0,0}+\ket{1,1}).
  \end{equation}
  The Kraus operators obtained in the $\kett{\Phi_{2}^{(ab)}}$-assisted LO is therefore
  \begin{equation}
    \frac{1}{2}\widetilde{K}_{\pm}
    = \bbra{M_{\pm}} \widehat{F} \kett{\Phi_{2}^{(ab)}}
    = \frac{1}{2}\Tilde{R}^{(\pm)}_{ij}
  \end{equation}
  As a result, one obtain the QPD for the cross terms
  \begin{equation}
    \widetilde{C}_{ij} = \frac{1}{2}\widetilde{K}_{+} - \frac{1}{2}\widetilde{K}_{-}
  \end{equation}
  As a whole, the $\Phi_{2}$-QPD of $\widetilde{U}$ over LO is then given by
  \begin{equation}
    \widetilde{U} =
    \sum_{i}\lambda_{i}^{2}\widetilde{\Lambda}_{i}
    +
    \sum_{i\neq j}\lambda_{i}\lambda_{j}(\frac{1}{2}\widetilde{K}_{+}-\frac{1}{2}\widetilde{K}_{-}),
  \end{equation}
  which result in an overhead of
  \begin{equation}
    \gamma =||\boldvec{\lambda}||_{1}^{2}.
  \end{equation}
\end{proof}
\end{customthm}

The upper bound on the sampling overhead of one-Bell-state-assisted LOCC-QPD is proved as follows
\begin{customcoro}{\ref{coro::one-Bell_CK_LOCC}}
Let $\hat{U}$ be a bipartite unitary with the LUD $\widehat{U} = \sum_i \lambda_i \widehat{\Lambda}_{i}$ with $\widehat{\Lambda}_{i} = \hat{A}_i\otimes \hat{B}_i$, and $\kett{\Phi_2}$ be a two-qubit maximally entangled state.
Suppose $\mathbb{S}$ is the set of index subset associated with all Choi-stretchable unitaries $\widetilde{S}_{L}$ constructed in Eq.~\eqref{eq::CS_construction},
\begin{equation}
  \mathbb{S} := \{L: \widetilde{S}_{L} \text{ is Choi-stretchable}\}.
\end{equation}
One can then construct a $\Phi_{2}$-assisted QPD over LOCC given in Eq.~\eqref{eq::CS_Non-CS_decomposition} with an overhead
\begin{equation}
  \gamma_{\mathrm{LOCC}}(\kett{{\Phi}_2} \rightarrow \Tilde{U} ) \leq
  ||\boldvec{\lambda}||_{1}^2-c_{S},
\end{equation}
where $c_{S}$ denotes the maximum fraction of the Choi-stretchable quantum channel $\widetilde{\mathcal{E}}_{S}$ in $\widetilde{U}$ that satisfies the following condition,
\begin{align}
  c_{S} = \max_{\{p_{L}\}_{L\in\mathbb{S}}} c,
  \;\;\text{ s.t. }\;\;
  c \sum_{L}\frac{p_{L}}{|L|}\delta_{l\in L}
  \le
  \lambda_{l}^{2}.
\end{align}
\begin{proof}
  According to Eq.~\eqref{eq::CS_Non-CS_decomposition}, the target unitary admits a decomposition into a Choi-stretchable component $\widetilde{\mathcal{E}}_{S}$ and a complementary entanglement-assisted LO component $\widetilde{\mathcal{E}}_{\mathrm{ELO}}$.
  \begin{equation}
  \label{eq::UBOund_LOCC_CK_proof_1}
    \widetilde{U} = c \widetilde{\mathcal{E}}_{S} + (1-c)\widetilde{\mathcal{E}}_{\mathrm{ELO}}.
  \end{equation}
  The explicit expression of $\widetilde{\mathcal{E}}_{S}$ is given by
  \begin{align}
    \widetilde{\mathcal{E}}_{S}
    =
    \sum_{L\in\mathbb{S}}p_{L} (\widetilde{D}_{L} + \widetilde{C}_{L})
  \end{align}
  where $\widetilde{D}_{L}$ and $\widetilde{C}_{L}$ are given in Eq.~\eqref{eq::CS_UL}.
  The overhead of $\widetilde{\mathcal{E}}_{S}$ is given by
  \begin{equation}
  \label{eq::UBOund_LOCC_CK_proof_2}
    \gamma_{\mathrm{LOCC}}(\kett{\Phi_{2}}\rightarrow\widetilde{\mathcal{E}}_{S})
    =
    \sum_{L\in\mathbb{S}}p_{L}|L|-1.
  \end{equation}
  On the other hand, the entanglement-assisted LO part $\widetilde{\mathcal{E}}_{\mathrm{ELO}}$ is given by
  \begin{equation}
    (1-c)\widetilde{\mathcal{E}}_{\mathrm{ELO}}
    =
    \sum_{i}\left(
      \lambda_{i}^{2} - c\sum_{L\in\mathbb{S}}\frac{p_{L}}{|L|}\delta_{i\in L}
    \right)\widetilde{\Lambda}_{i}
    +
    \sum_{i\neq j}\left(
      \lambda_{i}\lambda_{j} - c\sum_{L\in\mathbb{S}}\frac{p_{L}}{|L|}\delta_{i\in L}\delta_{j\in L}
    \right)
    \frac{1}{2}(\widehat{\Lambda}_{i}\otimes\widehat{\Lambda}_{j}^{\ast}+\widehat{\Lambda}_{j}\otimes\widehat{\Lambda}_{i}^{\ast})
  \end{equation}
  The first term can be implemented using LO alone, while the second term is implemented via $\Phi_{2}$-assisted LO, as shown in Fig.~\ref{fig::circuit-2LO}. Under the condition in Eq.~\eqref{eq::one-Bell_LOCC-QPD_cond}, all coefficients are positive. Moreover, the overhead contributions from $\widetilde{\Lambda}_{i}$ via LO and from $\frac{1}{2}(\widehat{\Lambda}_{i}\otimes\widehat{\Lambda}_{j}^{\ast}+\widehat{\Lambda}_{j}\otimes\widehat{\Lambda}_{i}^{\ast})$ via $\Phi_{2}$-assisted LO are both equal to $1$. Therefore, the overhead of $\widetilde{\mathcal{E}}_{\mathrm{ELO}}$ is simply given by the sum of the corresponding coefficients.
  \begin{equation}
  \label{eq::UBOund_LOCC_CK_proof_3}
    \gamma_{\mathrm{LO}}(\kett{\Phi_{2}}\rightarrow\widetilde{\mathcal{E}}_{\mathrm{ELO}})
    =
    \frac{1}{1-c}\left(\sum_{i,j}\lambda_{i}\lambda_{j} -
    c\sum_{i,j}\sum_{L\in\mathbb{S}}\frac{p_{L}}{|L|}\delta_{i\in L}\delta_{j\in L}\right)
    =
    \frac{1}{1-c}\left(\sum_{i,j}\lambda_{i}\lambda_{j} -
    c\sum_{L\in\mathbb{S}}p_{L}|L|\right).
  \end{equation}
  As a result of Eq.~\eqref{eq::UBOund_LOCC_CK_proof_1}, \eqref{eq::UBOund_LOCC_CK_proof_2} and \eqref{eq::UBOund_LOCC_CK_proof_3}, the total overhead is given by
  \begin{equation}
    \gamma_{\mathrm{LOCC}}(\kett{\Phi_{2}}\rightarrow\widetilde{U})
    \le
    \sum_{i,j}\lambda_{i}\lambda_{j}-c.
  \end{equation}
  The maximum $c_{S}:=\max c$ under the condition in Eq.~\eqref{eq::one-Bell_LOCC-QPD_cond}, therefore provides the smallest overhead under the combination of $\Phi_{2}$-LOCC-QPD $\widetilde{\mathcal{E}}_{S}$ and $\Phi_{2}$-LO-QPD $\widetilde{\mathcal{E}}_{\mathrm{ELO}}$,
  \begin{equation}
    \gamma_{\mathrm{LOCC}}(\kett{\Phi_{2}}\rightarrow\widetilde{U})
    \le
    \|\boldvec{\lambda}\|_{1}^{2}-c_{S}.
  \end{equation}
\end{proof}
\end{customcoro}

\section{Proofs in for entanglement-assisted black-box circuit knitting in Section \ref{sec::ent-assist_bb_circknit}}
\label{sec::proof_BB-EACK}

The optimal overhead of black-box QPD for Choi-stretchable unitaries is proved as follows.
\begin{customthm}{\ref{thm_BB-with-Clifford}}
Let $ \{\tilde{U}^{(t)}\}_t$ be a sequence of Choi-stretchable unitaries, i.e. $\gamma_{\mathrm{LOCC}}\left(\kett{J_{U^{(t)}}}\rightarrow \widetilde{U}^{(t)}\right) = 1$ for all $t$.
For any initial resourceful operation $\Tilde{r}$, the optimal black-box sampling overhead over LOCC is equivalent to the overhead of the parallel preparation of their Choi states,
\begin{align}
\label{eq::BB_QPD_Choi-stretchable}
  \gamma_{\mathrm{LOCC}} \left(\Tilde{r}\rightarrow \bigotimes_{t}\kett{J_{U^{(t)}}}  \right)
  =
  \gamma_{ \mathrm{LOCC}}\left( \Tilde{r} \rightarrow \bbcompo_{t}\Tilde{U}^{(t)} \right).
\end{align}
If $\Tilde{r} = \kett{\rho}$ is a state preparation, the optimal overhead is given by the fully entangled fraction $F_{D}(\rho)$ of $\rho$, where $D = \prod_{t} d_{U^{(t)}}$ and $d_{U^{(t)}}$ is the operator Schmidt rank of $\widetilde{U}^{(t)}$,
\begin{align}
    \gamma_{ \mathrm{LOCC} }\left( \kett{\rho}\rightarrow \bbcompo_{t}\widetilde{U}^{(t)} \right)
    = \frac{2}{F_D(\kett{\rho})} - 1.
\end{align}
\begin{proof}
According to the composition submultiplicativity for output channels in Lemma \ref{lemma::gamma_properties}, for any interleaving quantum channels $\{\widetilde{\mathcal{E}}_{t}\}_{t}$, the composition $\bigcirc_{t}(\widetilde{\mathcal{E}}^{(t)}\widetilde{U}^{(t)})$ has a Choi-state-assisted QPD upper bounded by
\begin{align}
  \gamma_{LOCC}\left(
    \bigotimes_{t}(\widetilde{\mathcal{E}}^{(t)}\otimes\kett{J_{U^{(t)}}})
    \rightarrow
    \bigcirc_{t}(\widetilde{\mathcal{E}}^{(t)}\widetilde{U}^{(t)})
  \right)
  & \le
  \prod_{t} \gamma_{LOCC}\left(
    (\widetilde{\mathcal{E}}^{(t)}\otimes\kett{J_{U^{(t)}}})
    \rightarrow
    (\widetilde{\mathcal{E}}^{(t)}\widetilde{U}^{(t)})
  \right)
  \nonumber \\
  & \le
  \prod_{t} \gamma_{LOCC}\left(
    \kett{J_{U^{(t)}}}
    \rightarrow
    \widetilde{U}^{(t)}
  \right)
\end{align}
Since this inequality holds for all interleaving quantum channels and  $\widetilde{U}^{(t)}$ are all Choi-stretchable, i.e. $\gamma_{LOCC}(    \kett{J_{U^{(t)}}} \rightarrow \widetilde{U}^{(t)} )=1$ for all $t$, it leads to
\begin{equation}
  \gamma_{LOCC}\left(
    \bigotimes_{t} \kett{J_{U^{(t)}}}
    \rightarrow
    \bbcompo_{t} \widetilde{U}^{(t)}
  \right)
  =1
\end{equation}
Given an input resource $\Tilde{r}$, as a result of the triangle inequality of the transition $\left(\Tilde{r}\rightarrow \bigotimes_{t}\kett{J_{U}^{(t)}}\rightarrow\bbcompo_{t}\widetilde{U}^{(t)}\right)$, it holds
\begin{equation}
  \gamma_{\mathrm{LOCC}}\left(\Tilde{r}\rightarrow \bbcompo_{t}\widetilde{U}^{(t)}\right)
  \le
  \gamma_{\mathrm{LOCC}}\left(\Tilde{r}\rightarrow \bigotimes_{t}\kett{J_{U}^{(t)}}\right).
\end{equation}
This upper bound coincides with the lower bound derived in Corollary \ref{coro::lower-bound-BB}, which leads to the equality in Eq.~\eqref{eq::BB_QPD_Choi-stretchable}. Consequently, as a result of Theorem \ref{thm::gamma-LOCC-Choi-stretchable}, if $\Tilde{r} = \kett{\rho}$ is a state preparation, the optimal overhead is given by
\begin{equation}
  \gamma_{\mathrm{LOCC}}(\kett{\rho}\rightarrow\kett{\Phi_{D}}) = \frac{2}{F_{D}( \rho )}-1.
\end{equation}
This completed the proof.
\end{proof}
\end{customthm}

\bigskip

The upper bound on the overhead of $\Phi_{2}$-assisted black-box QPD over LO is shown as follows.
\begin{customthm}{\ref{thm::Bell_BBCK_LO}}
Let $\{\tilde{U}^{(t)}\}_{t=1}^{T}$ be a sequence of untiaries, which have the LUD coefficients $\{\boldvec{\lambda}^{(t)}\}_{t}$.
Suppose $\ket{\Phi_{2}}$ is a two-qubit Bell state.
One can construct a $\Phi_{2}$-assisted black-box QPD over LO for $\{\tilde{U}^{(t)}\}_t$ as follows,
\begin{equation}
  \bbcompo_{t}\widetilde{U}^{(t)}
  =
  \gamma_{\mathcal{Q}}
  \left(
    \sum_{\boldvec{i}}p_{\boldvec{i},\boldvec{i}}
    \;\bbcompo_{t}(\widetilde{A}_{i^{(t)}}^{(t)}\otimes\widetilde{B}_{i^{(t)}}^{(t)})
    +
    \sum_{\boldvec{i}\neq\boldvec{j}}p_{\boldvec{i},\boldvec{j}}
    \sum_{\boldvec{m}\in\mathbb{Z}_{2}^{\otimes 2}}
    (-1)^{|\boldvec{m}|}
    \bbra{\boldvec{m}} (\bbcompo_{t} \widetilde{F}_{i^{(t)},j^{(t)}}^{(t)}) \kett{\Phi_{2}}
  \right),
\end{equation}
where $\widetilde{F}_{i^{(t)},j^{(t)}}^{(t)}$ is given in Eq.~\eqref{eq::free_LO_for_Bell_BB_QPD} and illustrated in Fig.~\eqref{fig::black-box-with-Bell}, and the classical labels $(\boldvec{i},\boldvec{j})$ are sampled from the set in Eq.~\eqref{eq::cl_labels_BB_QPD} with the probability
\begin{equation}
  p_{\boldvec{i},\boldvec{j}} =
  \prod_{t=1}^{T}   \frac{\lambda_{i^{(t)}}^{(t)}\lambda_{j^{(t)}}^{(t)}}{\|\boldvec{\lambda}^{(t)}\|_{1}^{2}}.
\end{equation}
The optimum overhead over LO is upper bounded by $\gamma_{\mathcal{Q}}$, which is determined by the 1-norm of the LUD coefficients,
\begin{equation}
  \gamma_{ \mathrm{LO} }\left(
    \kett{\Phi_2} \rightarrow \bbcompo_{t}\widetilde{U}^{(t)}
  \right)
  \le
  \gamma_{\mathcal{Q}}
  =
  \prod_{t=1}^{T}\|\boldvec{\lambda}^{(t)}\|_{1}^{2}.
\end{equation}
\begin{proof}
The black-box composition of $\{\tilde{U}^{(t)}\}_{t=1}^{T}$ can be decomposed to the diagonal and cross terms, respectively
\begin{equation}
  \bbcompo_{t}\widetilde{U}^{(t)}
  =
  \widetilde{D}_{\mathrm{BB}} + \widetilde{C}_{\mathrm{BB}},
\end{equation}
where $\widetilde{D}_{\mathrm{BB}}$ and $\widetilde{C}_{\mathrm{BB}}$ are the diagonal and cross terms, respectively
\begin{align}
  \widetilde{D}_{\mathrm{BB}} :=
  \sum_{\boldvec{i}} (\prod_{t}\lambda_{i^{(t)}}^{2})\;\bbcompo_{t}\widetilde{\Lambda}_{i}^{(t)}\;
  \;\;\text{ and }\;\;
  \widetilde{C}_{\mathrm{BB}} :=
  \sum_{\boldvec{i}\neq\boldvec{j}}
  (\prod_{t}\lambda_{i^{(t)}}\lambda_{j^{(t)}})\;
  \widetilde{C}_{\boldvec{i},\boldvec{j}},
\end{align}
where each cross term is given by
\begin{equation}
  \widetilde{C}_{\boldvec{i},\boldvec{j}} =
  \frac{1}{2}
  \left(
    \bigcirc_{t} (\widetilde{\mathcal{B}}^{(t)}
    (\widehat{\Lambda}_{i^{(t)}}^{(t)}\otimes\widehat{\Lambda}_{j^{(t)}}^{(t)\ast}))
    +
    \bigcirc_{t} (\widetilde{\mathcal{B}}^{(t)}
    (\widehat{\Lambda}_{j^{(t)}}^{(t)}\otimes\widehat{\Lambda}_{i^{(t)}}^{(t)\ast}))
  \right)
\end{equation}

Without loss of generality, we can describe the unknown channel as
\(
  \widetilde{\mathcal{B}}^{(t)} = \sum_{k^{(t)}}\widetilde{K}^{(t)}_{k^{(t)}}.
\)
The cross term can be then described by
\begin{equation}
  \widetilde{C}_{\boldvec{i},\boldvec{j}}
  =
  \frac{1}{2}
  \sum_{\boldvec{k}}
  \left(
    \bigcirc_{t} (
      \widehat{K}_{k^{(t)}}^{(t)}\widehat{\Lambda}_{i^{(t)}}^{(t)}\otimes
      \widehat{K}_{k^{(t)}}^{(t)\ast}\widehat{\Lambda}_{j^{(t)}}^{(t)\ast}
    )
    +
    \bigcirc_{t} (
      \widehat{K}_{k^{(t)}}^{(t)}\widehat{\Lambda}_{j^{(t)}}^{(t)}\otimes
      \widehat{K}_{k^{(t)}}^{(t)\ast}\widehat{\Lambda}_{i^{(t)}}^{(t)\ast}
    )
  \right)
\end{equation}
We can now construct two quantum channels $\widetilde{R}_{\boldvec{i},\boldvec{j}}^{\pm}$
from the sum of the operators $\bigcirc_{t}\widehat{K}_{k^{t}}^{(t)}\widehat{\Lambda}_{i^{(t)}}^{(t)}$ and $\bigcirc_{t}\widehat{K}_{k^{t}}^{(t)}\widehat{\Lambda}_{j^{(t)}}^{(t)}$
  \begin{equation}
    \widetilde{R}_{\boldvec{i},\boldvec{j}}^{(\pm)} :=
    \frac{1}{2}\sum_{\boldvec{k}}
    \left(
      \bigcirc_{t}
      (\widehat{K}_{k^{(t)}}^{(t)}\widehat{\Lambda}_{i^{(t)}}^{(t)})
      \pm
      \bigcirc_{t}
      (\widehat{K}_{k^{(t)}}^{(t)}\widehat{\Lambda}_{j^{(t)}}^{(t)})
    \right)
    \otimes
    \left(
      \bigcirc_{t}
      (\widehat{K}_{k^{(t)}}^{(t)}\widehat{\Lambda}_{i^{(t)}}^{(t)})
      \pm
      \bigcirc_{t}
      (\widehat{K}_{k^{(t)}}^{(t)}\widehat{\Lambda}_{j^{(t)}}^{(t)})
    \right)^{\ast}.
  \end{equation}
  The cross terms can then constructed from the following decomposition
  \begin{align}
    \widetilde{C}_{ij} = \frac{1}{2}\widetilde{R}_{\boldvec{i},\boldvec{j}}^{(+)} - \frac{1}{2}\widetilde{R}_{\boldvec{i},\boldvec{j}}^{(-)}.
  \end{align}
  The quantum operation $\widetilde{R}_{\boldvec{i},\boldvec{j}}^{(+)}$ can be implemented via the $\Phi_{2}$-assisted LO extended from Eq.~\eqref{eq::cross_term_BB_QPD}
  \begin{equation}
    \frac{1}{2}\widetilde{R}_{\boldvec{i},\boldvec{j}}^{(\pm)}
    =
    \sum_{\boldvec{m}:(-1)^{|\boldvec{m}|} = \pm 1}
    \bbra{\boldvec{m}_{X}}
      \bigcirc_{t}(\widetilde{B}^{(t)}\widetilde{F}_{i^{(t)},j^{(t)}}^{(t)})
    \kett{\Phi_{2}}
    =
    \sum_{\boldvec{m}:(-1)^{|\boldvec{m}|} = \pm 1}
    \bbra{\boldvec{m}_{X}}
      \bbcompo_{t}\widetilde{F}_{i^{(t)},j^{(t)}}^{(t)}
    \kett{\Phi_{2}},
  \end{equation}
  where the LO $\widetilde{F}_{i^{(t)},j^{(t)}}^{(t)}$ is a local control unitary given in Eq.~\eqref{eq::free_LO_for_Bell_BB_QPD} and illustrated in Fig. \ref{fig::black-box-with-Bell} for $T=2$,
  \begin{align}
    \widetilde{F}_{i^{(t)},j^{(t)}}^{(t)} :=
    (\widehat{\pi}_{0}^{(e_{a})}\otimes\widehat{A}_{i^{(t)}}^{(t)} + \widehat{\pi}_{1}^{(e_{a})}\otimes \widehat{A}_{j^{(t)}}^{(t)})
    \otimes
    (\widehat{\pi}_{0}^{(e_{b})}\otimes\widehat{B}_{i^{(t)}}^{(t)} + \widehat{\pi}_{1}^{(e_{b})}\otimes \widehat{B}_{j^{(t)}}^{(t)}).
  \end{align}
  As a result, the $\Phi_{2}$-assisted black-box QPD is
  \begin{equation}
    \bbcompo_{t}\widetilde{U}^{(t)}
    =
    \widetilde{D}_{\mathrm{BB}}
    +
    \sum_{\boldvec{i}\neq \boldvec{j}}\prod_{t}\lambda_{i^{(t)}}\lambda_{j^{(t)}}
    (\frac{1}{2}\widetilde{R}_{\boldvec{i},\boldvec{j}}^{(+)} - \frac{1}{2}\widetilde{R}_{\boldvec{i},\boldvec{j}}^{(-)})
  \end{equation}
  The overhead contributed from the diagonal and cross terms are
  \begin{equation}
    \gamma_{D} = \sum_{\boldvec{i}}\prod_{t}\lambda_{i^{(t)}}^{2}
    ,\;\;\text{ and }\;\;
    \gamma_{C} = \sum_{\boldvec{i}\neq\boldvec{j}}\prod_{t}\lambda_{i^{(t)}}\lambda_{j^{(t)}},
  \end{equation}
  respectively, which sum up to
  \begin{equation}
    \gamma_{\mathcal{Q}} = \gamma_{D} + \gamma_{C} = \prod_{t}\|\boldvec{\lambda}\|_{1}^{2}.
  \end{equation}
\end{proof}
\end{customthm}


\newpage
\myprintglossary

\myprintbibliography

\end{document}